\documentclass[a4paper,fleqn,usenatbib]{mnras}

\usepackage[T1]{fontenc}
\usepackage{ae,aecompl}

\usepackage{graphicx}	% Including figure files
\usepackage{amsmath}	% Advanced maths commands
\usepackage{amssymb}	% Extra maths symbols
\usepackage[english]{babel}
\usepackage{graphicx}
\usepackage{epstopdf}
\usepackage{longtable}
\usepackage{supertabular}
\usepackage{color}
\graphicspath{{./plots/}}
\title[Bolometric light curves of SE-SNe]{The bolometric light curves and physical parameters of stripped-envelope supernovae}

\author[S. J. Prentice]{S. J. Prentice$^{1}$\thanks{E-mail: S.J.Prentice@2014.ljmu.ac.uk}, P. A. Mazzali$^{1,2}$, E. Pian$^{3,4}$,  A. Gal-Yam$^{5}$, S. R. Kulkarni$^{6}$, \newauthor A. Rubin$^{5}$, A. Corsi$^{7}$, C. Fremling$^{8}$, J. Sollerman$^{8}$, O. Yaron$^{5}$, I. Arcavi$^{9,10}$, \newauthor  W. Zheng$^{11}$, M. M. Kasliwal$^{6,12}$, A. V. Filippenko$^{11}$, S. B. Cenko$^{13}$, Y. Cao$^{6}$, \newauthor P. E. Nugent$^{14,11}$\\
$^{1}$Astrophysics Research Institute, Liverpool John Moores University, IC2, Liverpool Science Park, 146 Brownlow Hill, \\  Liverpool L3 5RF, UK\\
$^{2}$Max-Planck-Institut f{\"u}r Astrophysik, Karl-Schwarzschild-Str. 1, D-85748 Garching, Germany\\
$^{3}$Institute of Space Astrophysics and Cosmic Physics, via P. Gobetti 101, I-40129 Bologna, Italy\\
$^{4}$Scuola Normale Superiore, Piazza dei Cavalieri 7, I-56126 Pisa, Italy\\
$^{5}$Department of Particle Physics and Astrophysics, Weizmann Institute of Science, 76100 Rehovot, Israel\\
$^{6}$Division of Physics, Mathematics, and Astronomy, California Institute of Technology, 1200 E California Blvd., Pasadena, CA 91125, USA \\
$^{7}$Texas Tech University, Physics Department, Box 41051, Lubbock, TX 79409-1051, USA\\
$^{8}$The Oskar Klein Centre, Department of Astronomy, Stockholm University, 10691 Stockholm, Sweden\\
$^{9}$Las Cumbres Observatory Global Telescope Network, 6740 Cortona Dr, Suite 102, Goleta, CA 93111, USA\\
$^{10}$Kavli Institute for Theoretical Physics, University of California, Santa Barbara, CA 93106, USA\\
$^{11}$Department of Astronomy, University of California, Berkeley, CA 94720-3411, USA\\
$^{12}$The Observatories, Carnegie Institution for Science, 813 Santa Barbara st, Pasadena, CA 91101, USA\\
$^{13}$Astrophysics Science Division, NASA Goddard Space Flight Center, Mail Code 661, Greenbelt, MD 20771, USA\\
$^{14}$Lawrence Berkeley National Laboratory, 1 Cyclotron Rd., Berkeley, CA 94720, USA\\
}
\date{Accepted XXX. Received YYY; in original form ZZZ}

\pubyear{2015}

\begin{document}
\label{firstpage}
\pagerange{\pageref{firstpage}--\pageref{lastpage}}
\maketitle

\begin{abstract}
The optical and optical/near-infrared pseudobolometric light curves of 85 stripped-envelope supernovae (SNe) are constructed using a consistent method and a standard cosmology. The light curves are analysed to derive temporal characteristics and peak luminosity $L_{\mathrm{p}}$, enabling the construction of a luminosity function. Subsequently, the mass of $^{56}$Ni synthesised in the explosion, along with the ratio of ejecta mass to ejecta kinetic energy, are found. Analysis shows that host-galaxy extinction is an important factor in accurately determining luminosity values as it is significantly greater than Galactic extinction in most cases. It is found that broad-lined SNe~Ic (SNe~Ic-BL) and gamma-ray burst SNe are the most luminous subtypes with a combined median $L_{\mathrm{p}}$, in erg s$^{-1}$, of log($L_{\mathrm{p}})=43.00$ compared to $42.51$ for SNe Ic, $42.50$ for SNe~Ib, and $42.36$ for SNe~IIb. It is also found that SNe~Ic-BL synthesise approximately twice the amount of $^{56}$Ni compared with SNe~Ic, Ib, and IIb, with median $M_{\mathrm{Ni}} = 0.34$, 0.16, 0.14, and 0.11 M$_{\odot}$, respectively. SNe~Ic-BL, and to a lesser extent SNe~Ic, typically rise from $L_{\mathrm{p}}/2$ to $L_{\mathrm{p}}$ more quickly than SNe~Ib/IIb; consequently, their light curves are not as broad.
\end{abstract}


%%%%%%%%%%%%%%%%%%%%%%%%%%%%%%%%%%%%%%%%%%%%%%%%%%

%%%%%%%%%%%%%%%%% BODY OF PAPER %%%%%%%%%%%%%%%%%%

\section{Introduction}
Stars with zero-age main sequence (ZAMS) mass $\ga 8$ M$_{\odot}$ have short lives that end with the catastrophic gravitational collapse of the stellar core \citep[e.g.,][]{Smartt2009}. The structure of the resulting electromagnetic emission observed due to the explosion depends upon the evolution of the progenitor star. It is possible for the star to be stripped of its outer envelope by either strong winds during the Wolf-Rayet phase (e.g., \citealt{Galyam2014}; \citealt{Smith2006}), interaction with a binary companion \citep[e.g.,][]{Eldridge2013}, or some combination of the two. The result is an absence of H in the spectra, which characterises Type Ib/c supernovae (SNe~Ib, SNe~Ic). If the envelope stripping is highly efficient, then the He shell is removed as well, and it is the presence or absence of He which differentiates between Type Ib and Type Ic SNe, respectively. (For extensive reviews of the optical spectra of SNe~Ib and Ic, see \citealt{Filippenko1997} and  \citealt{Matheson2001}.) In addition, some SNe show H lines around maximum light which rapidly disappear \citep{Filippenko1988,Filippenko1993}. They are canonically classified as Type IIb SNe \citep[e.g.,][]{Nomoto1993} but are included here, along with SNe~Ib/c, in the broader category of stripped-envelope supernovae \citep[SE-SNe;][]{Filippenko1997,Clocchiatti1997}.

SE-SNe are typically characterised by heterogeneous light-curve (LC) shapes, and luminosities where the LC $M_{\mathrm{peak}}$ is typically between $-16$ mag and $-18.5$ mag. The decay of $^{56}$Ni to $^{56}$Co, and then to $^{56}$Fe, powers the light curve \citep{Tominaga2005} in the vast majority of cases. A notable exception to this is the magnetar powered SN 2011kl associated with the ultralong gamma-ray burst (GRB) 111209A \citep{Greiner2015}. The spectra of SE-SNe show photospheric velocities ($v_{\mathrm{ph}}$) ranging between $\sim 5000$ km s$^{-1}$ and 25,000 km s$^{-1}$ at maximum light. A significant amount of ejecta at higher velocities, $\ga 18,000$ km s$^{-1}$, results in broad absorption lines that led to significant line blending in the spectra and reduces the number of visible features. SNe with high $v_{\mathrm{ph}}$ are almost exclusively Type Ic; for exceptions, see the Type IIb SN 2003bg \citep{Hamuy2009} and the Type IIn SN 1997cy \citep{Turatto2000}. They possess kinetic energies $E_{\mathrm{k}} > 10^{52}$ erg, an order of magnitude greater than typical SE-SNe. SNe with $E_{\mathrm{k}}$ in this extreme range have sometimes been labeled ``hypernovae'' \citep[HNe;][]{Iwamoto1998}. Type Ic HNe form a subpopulation known as broad-lined SNe~Ic \citep[SNe~Ic-BL; e.g., SN 2002ap, SN 1997ef;][]{Mazzali2002,Mazzali2000}.

Since 1998, some SNe~Ic have been associated with GRBs when SN 1998bw was discovered in the error box of GRB 980425 \citep{Galama1998}. The GRB/SN connection was later confirmed by the spectroscopic association of SN 2003dh with GRB 030329 \citep[e.g.,][]{Matheson2003,Hjorth2003}. To date, more than two dozen GRB-SNe have been discovered, either photometrically or spectroscopically. Of those SNe that are spectroscopically confirmed, all are Type Ic HNe showing characteristic broad lines in their spectra. 
In addition, the Type Ib SN 2008D \citep{Mazzali2008,Soderberg2008} was associated with the weak X-ray flash (XRF) 080109. The energetics of the GRBs associated with SNe are varied. It was initially thought that these GRBs represented a different population as they were often underluminous in $\gamma$-rays. With more data, it was found that luminosities in this photon range were typical but showed variations of a few orders of magnitude. For example, GRB 060218 associated with SN 2006aj is more commonly considered to be an XRF given the low $\gamma$-ray luminosity. The diversity of GRB energetics associated with the GRB SNe has implications for the progenitor stars and the powering mechanism of the SN itself. It is unfortunate that observational circumstances, such as distance and competition with the GRB afterglow, conspire to make GRB-SNe generally difficult to observe. However, their unambiguous explosion dates make them an important baseline for this study.

There have been attempts in the last few years to investigate the bulk properties of core-collapse SNe. \cite{Cano2013} presented a method of estimating explosion properties of 61 SNe~Ib/c using SN 1998bw as a template. \cite{Taddia2015} used data from the Sloan Digital Sky Survey SN survey II (SDSS-II) to build a set of 20 SN~Ib/c bolometric light curves and their corresponding properties. \cite{Drout2011} took 25 SNe~Ibc/II and used multiband light curves, as well as taking $R$-band data as a proxy for the bolometric emission, to estimate the characteristics of the SNe. \cite{Pritchard2014} used \textit{Swift} data to build bolometric light curves for core-collapse SNe but were limited to a redward limit of the \textit{Swift}-UVOT $V$ band. \cite{Lyman2014b} utilised a method of deriving bolometric values for 38 SNe from colours. However, as yet no single study of the individually derived bulk properties of SE-SNe found within the literature has been published. Motivated by this gap, the aims of this study are to build a self-consistent set of bolometric light curves using the available photometry, investigate the derived explosion parameters, determine the temporal characteristics of each SN, and compare with the light curves of multiband photometry.

\section{Database}
We compiled a list of data for SE-SNe which are publicly available in the literature. Over 100 were found, typically covering a period of $\sim 20$ years. In most cases, the data were found in studies involving single objects. However, larger datasets have recently been published. \cite{Bianco2014} presented 64 SNe~Ib/c observed by the Harvard-Smithsonian Center for Astrophysics (CfA) SN group in the period 2001 to 2009. 20 SNe~Ib/c from the SDSS-II were analysed by \cite{Taddia2015} using data from \cite{Sako2014}. Consequently, the number of SNe available more than doubled in less than a year. In addition to this, we gained access to the Palomar Transient Factory (PTF)\footnote{www.ptf.caltech.edu} \citep{Law2009} and the intermediate Palomar Transient Factory (iPTF) archives, increasing the number of SNe available. Thus, the total number of SNe in the initial database is as follows: 
\begin{itemize}
	\item{Public --- single object --- $\sim 50$}
	\item{Public --- CfA --- 64}
	\item{Public --- SDSS II --- 20}
	\item{Public --- \textit{Swift} --- 15}
	\item{iPTF/PTF --- 128.}
\end{itemize}

\subsection{Selection criteria}
With a large database of SNe, the next step was to consider what was required from the dataset. To build a consistent group of bolometric LCs it was necessary that, as best as possible, the photometry from SN to SN was well sampled over the same rest-frame wavelength range. Ideally, the sample should have shown good coverage in the wavelength range corresponding to the bulk of the SN light (i.e., between the $V$ and $R$ bands), which also corresponds to the turnover in the spectral energy distribution (SED) around bolometric peak. It should also have sufficient coverage in adjacent bands in order to build SEDs across a uniform wavelength range and, as we wanted to derive time-dependent parameters from the SN peak, it was essential that the coverage included this epoch. This led to the following two criteria for inclusion in our sample:
	\begin{itemize}
		\item{The peak of the SN must be observed in the $BVR$ bands or equivalents.}					
		\item{The temporal coverage must have been sufficient that the rise and/or decay time from half-maximum luminosity to maximum luminosity could be derived or the explosion date well constrained.}
		\end{itemize}	

These restrictions immediately ruled out more than 70$\%$ of the literature SNe, with the majority lacking observations at peak or sufficient coverage in multiple filters. This affected the CfA and \textit{Swift} set considerably. Only half of the CfA sample showed a clear photometric peak, and for SNe observed solely by \textit{Swift} the lack of observations redward of the UVOT $V$ band means that the peak of the SED, and hence the bulk of the light, was missed in every case. Of the PTF/iPTF SNe, most had to be rejected because they were observed only in the $r$ band. 

Table~\ref{database} lists the 85 SNe that fulfilled the criteria. The sample consists of 25 SNe~Ib, 21 SNe~Ic, 12 SNe~Ic-BL, 10~GRB-SNe, 15 SNe~IIb, and 2 SNe~Ibc. (The SNe~Ibc have an ambiguous classification; the data were not of sufficient quality to distinguish between the Ib and Ic subclasses.) 
 SNe with full optical and near-infrared (NIR) coverage allowed the construction of an effective bolometric light curve as opposed to a pseudobolometric optical variant. 

% Database
\begin{table*}
 \centering
 \begin{minipage}{125mm}
  \caption{The database of 85 SNe included in this work.}
  \begin{tabular}{lcccccc}%{@{}llcccclclcc@{}}
  \hline
 SN & Type & $\mu$ (mag) & $z$ & $E(B-V)_{\mathrm{MW}}$ (mag) & $E(B-V)_{\mathrm{host}}$ (mag) & references \\
  \hline
1993J&IIb&27.8&-0.000113&0.071&0.1&(1)\\
1994I&Ic&29.6&0.0015&0.03&0.3&(2)\\
1996cb&IIb&29.95&0.0024&0.12&negligible&(3)\\
1998bw&GRB-SN&32.76&0.0087&0.052&negligible&(4),(5)\\
1999dn&Ib&32.93&0.0093&0.052&0.1&(6)\\
1999ex&Ib&33.44&0.0114&0.02&0.28&(7)\\
2002ap&Ic-BL&29.5&0.0022&0.071&0.008&(8),(9),(10)\\
2003bg&IIb&31.68&0.0046&0.02&negligible&(11)\\
2003dh&GRB-SN&39.21&0.168&0.025&negligible&(12)\\
2003jd&Ic-BL&34.43&0.019&0.06&0.09&(13)\\
2004aw&Ic&34.31&0.016&0.021&0.35&(14)\\
2004fe&Ic&34.28&0.018&0.0210&-&(15)\\
2004gq&Ib&32.09&0.0065&0.0627&0.095&(15)\\
2005az&Ic&33.14&0.0085&0.0097&-&(15)\\
2005bf&Ib&34.62&0.019&0.045&negligible&(16)\\
2005hg&Ib&34.68&0.021&0.0901&-&(15)\\
2005hl&Ib&34.92&0.023&0.073&-&(17)\\
2005hm&Ib&35.85&0.035&0.048&-&(17)\\
2005kl&Ic&31.64&0.0035&0.0219&-&(15)\\
2005kr&Ic-BL&38.91&0.134&0.087&-&(17)\\
2005ks&Ic-BL&38.21&0.099&0.05&-&(17)\\
2005kz&Ic&35.31&0.027&0.046&-&(15)\\
2005mf&Ic&35.27&0.027&0.0153&-&(15)\\
2006T&IIb&32.68&0.0080&0.0647&-&(15)\\
2006aj&GRB-SN&35.61&0.033&0.097&negligible&(18),(19),(20)\\
2006el&IIb&34.25&0.017&0.0973&-&(15)\\
2006ep&Ib&33.93&0.015&0.036&-&(15)\\
2006fe&Ic&37.41&0.07&0.098&-&(17)\\
2006fo&Ib&34.58&0.021&0.025&-&(15)\\
14475$^{\mathrm{a}}$&Ic-BL&39.17&0.149&0.072&-&(17)\\
2006jo&Ib&37.63&0.077&0.032&-&(17)\\
2006lc&Ib&34.13&0.016&0.057&-&(17)\\
2006nx&Ic-BL&38.97&0.137&0.108&-&(17)\\
2007C&Ib&32.15&0.0059&0.0363&0.73&(15)\\
2007D&Ic-BL&34.84&0.023&0.2881&-&(15)\\
2007Y&Ib&31.29&0.0046&0.022&0.09&(24)\\
2007ag&Ib&34.78&0.020&0.025&-&(15)\\
2007cl&Ic&34.84&0.022&0.02&-&(15)\\
2007gr&Ic&29.84&0.0017&0.055&0.03&(21)\\
2007kj&Ib&34.3&0.018&0.0691&-&(15)\\
2007ms&Ic&36.09&0.039&0.184&-&(17)\\
2007nc&Ib&37.91&0.087&0.025&-&(17)\\
2007qv&Ic&38.11&0.095&0.048&-&(17)\\
2007qx&Ic&37.71&0.08&0.023&-&(17)\\
2007ru&Ic-BL&34.04&0.016&0.27&negligible&(22)\\
2007sj&Ic&36.09&0.039&0.032&-&(17)\\
2007uy&Ib&32.48&0.0065&0.022&0.63&(23)\\
2008D&Ib&32.48&0.0065&0.02&0.63&(27),(15),(28)\\
2008ax&IIb&29.82&0.0019&0.022&0.278&(25),(26)\\
2008bo&IIb&32.06&0.005&0.0513&0.0325&(15)\\
2008hw&GRB-SN&42.35&0.53&0.42&negligible&(15)\\
2009bb&Ic&33&0.00988&0.098&0.482&(29)\\
2009er&Ib&35.9&0.035&0.0389&-&(15)\\
2009iz&Ib&33.8&0.014&0.0729&-&(15)\\
2009jf&Ib&32.64&0.0079&0.112&0.05&(30)\\
2010as&IIb&32.17&0.0073&0.15&0.42&(31)\\
2010bh&GRB-SN&36.94&0.059&0.12&0.14&(32)\\
2010ma&GRB-SN&42.40&0.552&0.019&0.04&(32)\\
2011bm&Ic&34.95&0.022&0.032&0.032&(33)\\
2011dh&IIb&29.48&0.0020&0.035&0.05&(34)\\
2011ei&IIb&33.09&0.0093&0.059&0.18&(35)\\
2011fu&IIb&34.36&0.019&0.068&0.015&(36)\\
2011hs&IIb&31.91&0.0057&0.011&0.16&(37)\\
2011kl&GRB-SN&43.09&0.677&0.019&0.038&(38)\\
  \hline
\label{database}
\end{tabular}
\end{minipage}
\end{table*}

\begin{table*}
 \centering
 \begin{minipage}{125mm}
  \contcaption{The database of 85 SNe included in this work}
  \begin{tabular}{lcccccc}
  \hline
 SN & Type & $\mu$ (mag) & $z$ & $E(B-V)_{\mathrm{MW}}$ (mag) & $E(B-V)_{\mathrm{host}}$ (mag) & references \\
  \hline
2012ap&Ic-BL&33.45&0.012&0.045&0.4&(39)\\
2012bz&GRB-SN&40.31&0.28&0.037&negligible&(40)\\
2013cq&GRB-SN&41.19&0.34&0.02&0.05&(41)\\
2013cu&IIb&35.23&0.026&0.011&negligible&(42)\\
2013df&IIb&31.65&0.0024&0.017&0.08&(43)\\
2013dx&GRB-SN&39.04&0.145&0.04&0.10&(44)\\
2013ge&Ibc&31.87&0.0044&0.02&0.047&(45)\\
PTF09dh/2009dr$^{\mathrm{b}}$&Ic-BL&37.60&0.076&0.022&-&-\\
PTF10gvb&Ic-BL&38.26&0.098&0.022&-&-\\
PTF10inj&Ib&37.31&0.066&0.01&-&-\\
PTF10qif&Ib&37.26&0.064&0.0587&-&-\\
PTF10vgv&Ic&34.01&0.015&0.145&-&(49)\\
PTF11bli&Ibc&35.81&0.034&0.013&-&-\\
PTF11jgj&Ic&36.15&0.04&0.027&-&-\\
PTF11klg&Ic&35.26&0.027&0.03&-&-\\
PTF11qiq&Ib&35.66&0.032&0.066&-&-\\
PTF11rka&Ic&37.61&0.074&0.03&-&-\\
PTF12gzk&Ic&33.8&0.014&0.14&negligible&(48)\\
PTF12os&IIb&31.89&0.0045&0.045&-& (50)\\
iPTF13bvn&Ib&31.89&0.0045&0.0278&0.0437&(46),(47)\\
iPTF14dby&Ic-BL&37.54&0.074&0.048&-&(51)\\
  \hline
\multicolumn{7}{p{\textwidth}}{References: (1) \cite{Richmond1994}, (2) \cite{Richmond1996}, (3) \cite{Qiu1999}, (4) \cite{Clocchiatti2011}, (5) \cite{Patat2001}, (6) \cite{Benetti2011}, (7) \cite{Stritzinger2002}, (8) \cite{Foley2003}, (9) \cite{Galyam2002}, (10) \cite{Tomita2006}, (11) \cite{Hamuy2009}, (12) \cite{Deng2005}, (13) \cite{Valenti2008}, (14) \cite{Taubenberger2006}, (15) \cite{Bianco2014}, (16) \cite{Anupama2005}, (17) \cite{Taddia2015},  (18) \cite{Pian2006},(19) \cite{Mirabal2006}, (20) \cite{Kocevski2007}, (21) \cite{Hunter2009}, (22) \cite{Sahu2009}, (23) \cite{Roy2013}, (24) \cite{Stritzinger2009}, (25) \cite{Pastorello2008}, (26) \cite{Taubenberger2011}, (27) \cite{Mazzali2008}, (28) \cite{Modjaz2009}, (29) \cite{Pignata2011}, (30) \cite{Valenti2011}, (31) \cite{Folatelli2014}, (32) \cite{Bufano2012}, (33) \cite{Valenti2012}, (34) \cite{Marion2014}, (35) \cite{Mili2013B}, (36) \cite{Kumar2013}, (37) \cite{Bufano2014}, (38) \cite{Greiner2015}, (39) \cite{Mili2015}, (40) \cite{Melandri2012}, (41) \cite{Melandri2014}, (42) \cite{Galyam2014}, (43) \cite{MG2014}, (44) \cite{DElia2015}, (45) \cite{Drout2015}, (46) \cite{Fremling2014}, (47) \cite{Srivastav2014}, (48) \cite{BenAmi2012}, (49) \cite{Corsi2012}, (50) Fremling et al. (2016, in prep.), (51) Corsi et al. (2016, in prep.).}\\
\multicolumn{7}{p{\textwidth}}{$^{\mathrm{a}}$SN 14475 was discovered in 2006 as part of the SDSS-II SN survey.}\\
\multicolumn{7}{p{\textwidth}}{$^{\mathrm{b}}$PTF09dh/SN 2009dr was originally classified at a Type Ia SN but later reclassified as a Ic-BL.}
\end{tabular}
\end{minipage}
\end{table*}

\section{Constructing the bolometric light curve}
\subsection{Missing data}
In order to derive the bolometric luminosity of a SN at a particular date, an SED constructed from the photometry is required. Thus, it was essential that there was photometry in the relevant bands at that time, which was not always the case. In order to attain temporal uniformity, the worst-sampled band was chosen to be a reference point and the remaining bands were fit with a linear spline. The magnitudes were interpolated on the dates of the reference band. Early-time data points are especially important as they help determine the rise time of the SN and constrain $t_0$, the time of explosion. However, these epochs also tended to be sparsely sampled, often with observations in just a couple of bands. 

To obtain estimates of the early-time bolometric data points, two methods were used to extrapolate missing photometry provided that at least two bands were available and one of them was a $V$-band equivalent (e.g., an effective wavelength around 4000--5000~\AA). If the temporal gap between the first date in two adjacent bands was no greater than a few days, then either a constant colour was assumed or the mean colour evolution of similar SN types was adopted. If it was not possible to use this method, but there were sufficient pre-peak data, then extrapolations were done via a low-order polynomial fit to the data. As per the previous method, this technique was limited to time periods on the order of a few days. Care was taken to avoid extrapolating early-time data points based upon the behaviour of the light curves near peak, as this would have underestimated the rate of change. Given the uncertain nature of the shape of the light curve outside the observed dates, large errors of $\sim 1$ mag on the extrapolations were assumed.

\subsection{GRB-SNe and afterglow subtraction}
The desire to include as many GRB-SNe as possible in the database is compromised by the difficulty in deconvolving the SN light from that of the GRB afterglow and host galaxy. For some SNe (e.g., SN 1998bw), the afterglow is negligible, so optical emission is dominated by SN photons; however, this is not usually the case. To calculate the afterglow component of the light curve, the spectrum is considered to follow a simple relation given by $F_{\nu}(t,\nu)\propto t^{-\alpha}\nu^{-\beta}$. We defer to the literature for the numerical values of the temporal and spectral indices and subtract the afterglow flux from the SEDs as required. Additionally, it is common in the literature to fit the afterglow-SN-host light curve with a template SN based on SN 1998bw \citep{Cano2013}, but this method is not adopted here to avoid biasing any temporal characteristics that may be extracted from the light curve. 

There are approximately 20 GRB-SNe given in the literature. Most are photometrically associated with the GRB (e.g., they show a late-time bump in the light curve), but unfortunately the majority of these are poorly observed, with few data points in only a few bands and showing large afterglow contamination. Consequently, the number of usable GRB-SNe is greatly restricted.  It is an unfortunate irony that GRB-SNe in general have the best-known explosion date yet the most poorly constrained photometry. 

\subsubsection{SN 2003dh / GRB 030329}
To form as complete a sample as possible, the light curve of SN 2003dh from \cite{Deng2005} is included in the database. The ultraviolet-optical-infrared (UVOIR) light curve is the result of synthetic photometry from spectra and a mean bolometric correction, derived from SN 1998bw, added to the photometric data. A consequence of this is that we do not desconstruct the light curve to obtain a {\it BVRI} variant and use the data as is.

\subsection{Distance}

A primary goal of this study was to present a self-consistent set of bolometric light curves. In order to achieve this, it was necessary (where possible) to obtain a distance modulus for each SN adopting a standard cosmological model. Thus, the distance modulus of each SN was taken from the NASA/IPAC Extragalactic Database (NED)\footnote{https://ned.ipac.caltech.edu/} for the host galaxy using the standard NED cosmological model (H$_0$ = 73.0 km s$^{-1}$ Mpc$^{-1}$, $\Omega_m$ = 0.27, $\Omega_\Lambda$ = 0.73) and corrected for Galactic motion toward Virgo, the Great Attractor, and the Shapley Supercluster. The standard uncertainty of 0.15 mag in the NED values was adopted throughout.

\begin{figure*}
\centering
\includegraphics[scale=0.9]{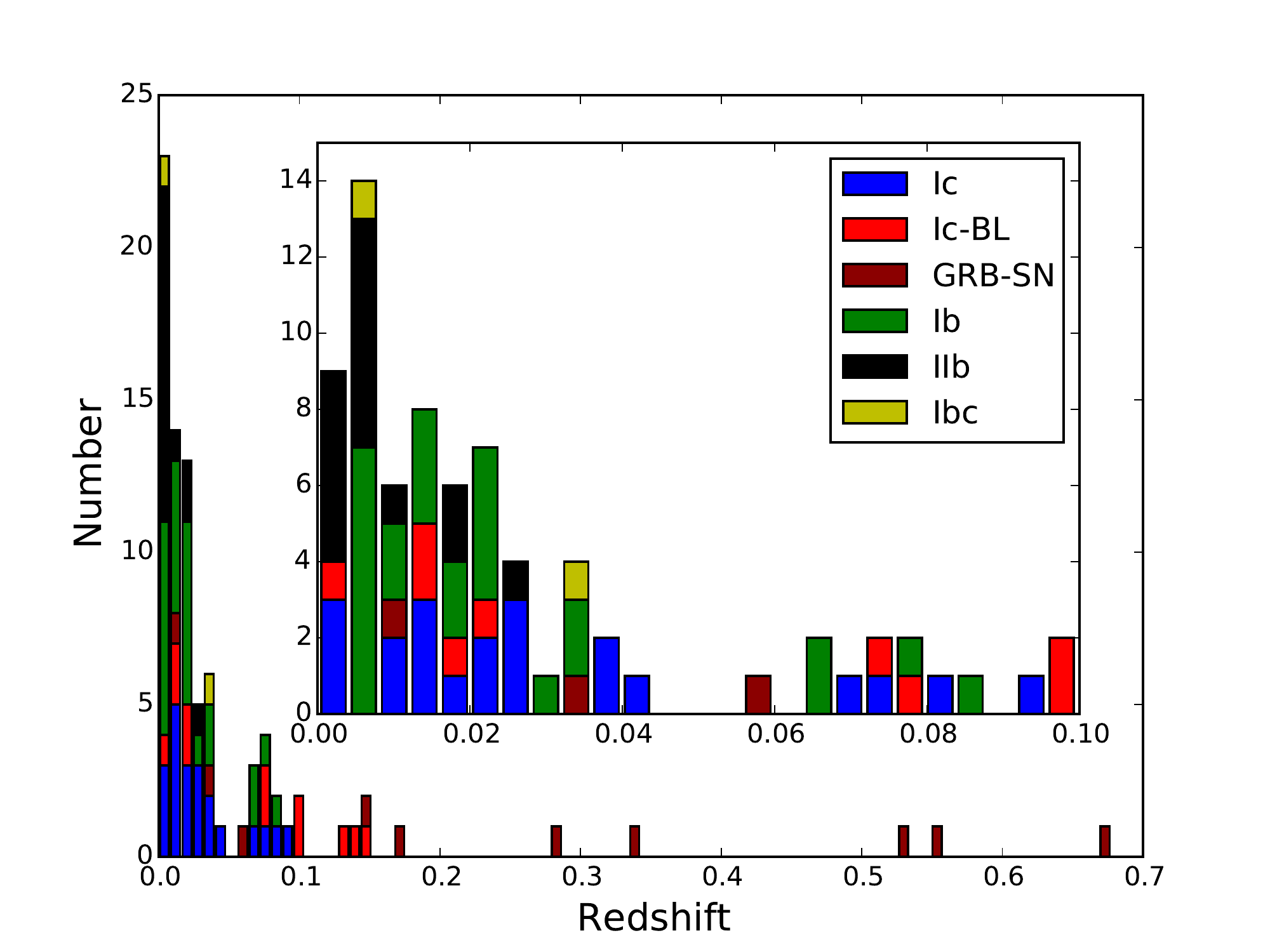}
\caption{Redshift distribution of the SNe in the sample. The vast majority of SNe are found at $z < 0.1$, as demonstrated in the inset. It is clear that the high-$z$ regime is dominated by GRB-SNe.}
\label{fig:redshift}
\end{figure*}

Figure~\ref{fig:redshift} shows the redshift distribution of SNe within the database. $88\%$ lie at $z < 0.1$, with the median being $z=0.0189$. Statistics for the redshift distribution by spectral type are given in Table~\ref{zstats} 

% redshift table
\begin{table}
 %\centering
 \begin{minipage}{100mm}
  \caption{Redshift statistics of sample SNe by spectral type}
 \begin{tabular}{@{}lccccccclcc@{}}
  \hline
 Type & median $z$ & mean $z$&  range & Number  \\
  \hline
  GRB-SNe & 0.224 & 0.279 & 0.0087--0.677& 10\\
Ic-BL & 0.075 & 0.069 & 0.0022--0.149&12\\
Ic & 0.022 & 0.031 &0.0015--0.095&21\\
Ibc & 0.019 & 0.019 & 0.0044--0.034 & 2 \\
Ib & 0.018 & 0.025 & 0.0045--0.087& 25 \\
IIb & 0.0050 & 0.0076& $<0.0258$&15\\
\hline
 \label{zstats}
\end{tabular}
\end{minipage}
\end{table}

% redshift figure

\subsection{Extinction}
Extinction values were taken from the literature, and the reddening correction was done using the extinction law given by \cite{CCM}. However, it was apparent that not all SNe have spectroscopically derived host-galaxy extinction, an issue which affects half the SNe used here. We find that in cases where both extinction values are known, the database has a mean $E(B-V)_{\mathrm{MW}}$ = 0.059 mag, significantly less than the mean host-galaxy extinction of $E(B-V)_{\mathrm{host}}$ = 0.135 mag. Thus, failure to include the effect of host reddening results in less-luminous bolometric LCs for those SNe and places the values derived as lower limits. 

It is desirable to make as few assumptions as possible while constructing the set of bolometric light curves. However, it is acknowledged that taking the host-galaxy reddening to be zero for some SNe and then attempting to include them in population statistics  biases the sample by selecting a value at a boundary of a distribution. We also want to maximise the number of SNe in the sample, so we adopt the following procedure. When calculating the properties of individual SNe, no correction was made when the host extinction was unknown. The values derived are taken as lower limits in this case. For the bulk sample we dereddened the flux for a median host extinction of that SN type. This works on the basis that the extinction would be typically around the mean and the number of SNe overcorrected would balance the number of SNe undercorrected. This increases the spread of the distribution but provides a better estimate of the median and mean than using lower limits.  

To examine this in more detail, we searched the literature for core-collapse SNe, including all SNe~II, and built a set of host-galaxy extinction functions by type. The host galaxies of observed GRB-SNe are known to be different from those of other core-collapse SNe in terms of metallicity \citep[e.g.,][]{Modjaz2008,GF2012} --- so if the extinction is dependent on host, host inclination, and SN type, then the distributions should reflect this. The results of this work are presented in the Appendix, with the results given in Table~\ref{redstats}. 

Finally, we note that only the spectroscopically derived host-galaxy extinction, such as from the equivalent width of \ion{Na}{I}~D absorption lines \citep{Poznanski2012}, is used here.

\subsection{Constructing the SED}

To construct the SED the photometry was corrected for Galactic extinction in the observer frame and, if possible, corrected for host extinction in the rest frame. Uncertainties in reddening were included in quadrature. When the monochromatic flux, calculated from the photometry (see \citealt{BCP1998}; \citealt{Fukugita1995}), was shifted to the rest frame wavelength it was also multiplied by (1+$z$), which is a useful approximation in the absence of spectroscopically derived K-corrections. The process of flux conversion depends upon the filter system used. Data in the literature cover more than twenty years of observations and over this period there has been a shift in the usage of filters from Johnson-Cousins (J-C) \textit{UBVRIJHK} (and minor variations) to SDSS type filters ($u'g'r'i'z'$) or some combination of these, as well as space telescope specific filters such as those on the \textit{Swift}-UVOT. The two main standards use different flux units with the J-C system being based on Vega and SDSS on AB. As default all SEDs are constructed in units of erg s$^{-1}$ cm$^{-2}$ $\mathrm{\AA}^{-1}$. This required conversion of the SDSS filters which was achieved by using the relation $\lambda f_\lambda$ = $\nu f_\nu$.

The rest frame flux was fit with a linear spline to create SEDs over the range 3000 $\mathrm{\AA}$ to 10000 $\mathrm{\AA}$ for \textit{UBVRI}-equivalents, 4000 $\mathrm{\AA}$ to 10000 $\mathrm{\AA}$ for \textit{BVRI}-equivalents, and 10000 $\mathrm{\AA}$  to 24000 $\mathrm{\AA}$ for NIR.

\subsection{From SED to pseudobolometric luminosity}
The rest-frame SEDs were then integrated over the wavelength range, assuming zero flux outside the limits. 
The effect due to redshift and the subsequent blueshifting of the effective wavelengths of the photometry is small (a few percent) out to $z \approx 0.1$. Beyond this, the bluest bands start to be shifted outside the integration range and the reddest effective wavelength shifts to a more central position; in this regime, the largest uncertainty comes from the behaviour of the tail of the SED. If there is photometry in bands with effective wavelengths longer than $I/i'$, we incorporate these into the optical SEDs, as they are blueshifted into the optical wavelength range or close to it.

We note that one might be tempted to fit a blackbody emission curve to the SEDs, but this is erroneous as the spectrum of a SN is not a blackbody. During the photospheric phase, the UV suffers line blanketing, the severity of which is related to the amount of iron-group elements \citep{Mazzali2000UV} and the velocity of the ejecta, which causes line broadening. The photons scattered in this process eventually escape at redder wavelengths creating a flux excess at these points (e.g., \citealt{Mazzali1993}). In the nebular phase ($> 60$ days), the spectrum is dominated by line emission as the optically thin ejecta are excited via energy deposition from the $\gamma$-rays and positrons emitted during the decay of $^{56}$Ni to $^{56}$Fe \citep{Mazzali2005}. 

The uncertainties were carried through the integration by evaluating the integral at the upper and lower errors of the flux. Once the bolometric flux was determined it was converted to bolometric luminosity using the distance.

\section{Pseudobolometric light curves}
In order to compare bolometric light curves effectively, a set of \textit{BVRI}, \textit{UBVRI}, and NIR (or equivalent) LCs were produced, where the data allowed. Table~\ref{bands} lists the bands, total wavelength range, and reference name throughout the paper for the different light curves. Within the sample, 84 SNe have a \textit{BVRI} LC, 44 have a \textit{UBVRI} LC, and 24 have a NIR LC. Various extensions of the LCs can be made by combining the data (e.g., \textit{UBVRI}NIR). In most cases, the NIR is less well sampled than the optical, so to construct an optical-NIR LC the ratio between the two datasets was calculated for coincident dates. This was subsequently fit with a linear spline, and the NIR flux was deduced by interpolation of the ratio and the value of the optical flux on the dates when NIR observations were absent. When the optical LC extended beyond the NIR, the flux ratio was kept constant beyond the boundaries.  

\begin{table}
	\centering
	\caption{SED integration wavelength range and terminology.}
	\begin{tabular}{lccc}
	\hline
	Constituent bands & Wavelength range (\AA) & Nomenclature\\ 
	\hline
	$BVRI/g'r'i'z'$ & 4000--10,000 & $BVRI$ \\
	$UBVRI/u'g'r'i'z'$ & 3000--10,000 & $UBVRI$ \\
	$JHK$ & 10,000--24,000 & NIR \\
	\hline
	\end{tabular}
	\label{bands}
\end{table}

Finally, the temporal evolution of the light curve was corrected to the rest frame and set with a fiducial $t(0)$ at the time of maximum luminosity with the following caveat: any peak caused by early-time shock breakout \citep[e.g., SN IIb 2013df][]{Vandyk2014} is ignored, and the later peak powered by radioactive decay is selected. In the case of peculiar SN~Ib 2005bf, with its double-peaked light curve \citep{Tominaga2005}, we took the first peak to be the result of the decay of $^{56}$Ni and the second peak a different energy-injection process, specifically a magnetar \citep{Maeda2007}. For a different interpretation, see \cite{Folatelli2006}. 

\subsection{The contribution of $U$ to $UBVRI$}
% plot of U contribution
\begin{figure}
\centering
\includegraphics[scale=0.4]{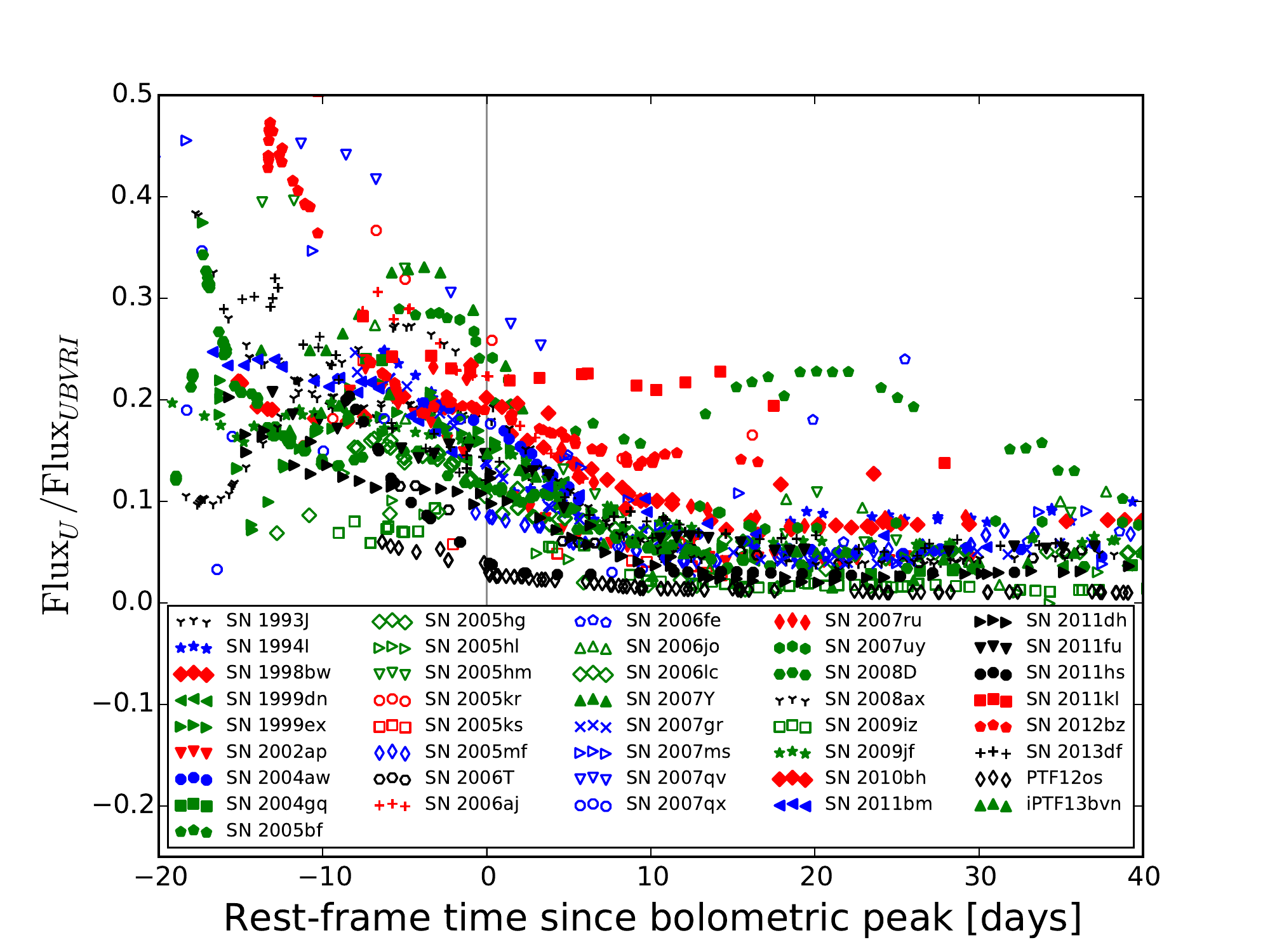}
\caption{The ratio of the $U/u'$-band flux to the $UBVRI$ flux as a function of time. Open symbols represent SNe that have not had host-galaxy extinction corrections applied. The high host extinction of PTF12os is apparent in this diagram as the $U/u'$ flux is negligible. The temporal evolution of peculiar SN~Ib 2005bf is a noteworthy feature}
\label{fig:Uper}
\end{figure}

Figure~\ref{fig:Uper} gives the time-dependent contribution of the $U/u'$-band flux to the total $UBVRI$ flux. As expected, the SNe are bluer at earlier times than later. Table~\ref{UNIR} shows the statistics for the sample at bolometric peak by spectral type and with/without host-extinction corrections applied. Note that the statistics are derived from two overlapping populations and there is a bias in the non-host-corrected SN~Ic values caused by a few SNe with poorly constrained $u'$ photometry. The SNe~Ic-BL are demonstrably bluer at this epoch than other SN types with a $U$ flux contribution of $\sim 20$\% as opposed to $\sim 16$\% for the other types (host-extinction correction included). We do note the small sample size in some cases.

\subsection{NIR contribution}
% plot of NIR contribution
\begin{figure}
\centering
\includegraphics[scale=0.4]{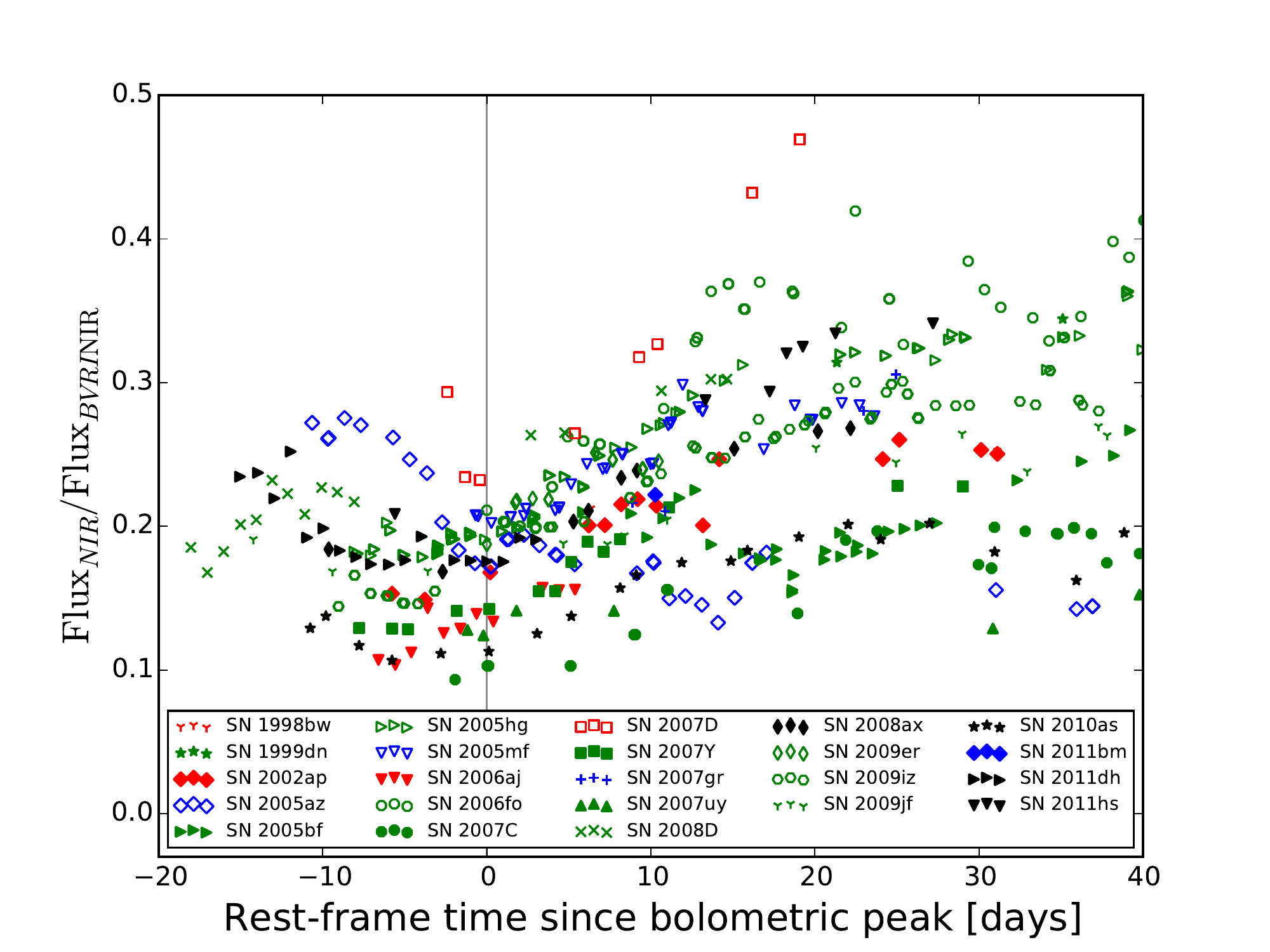}
\caption{The ratio of NIR flux to \textit{BVRI}NIR flux. Only dates when the optical and NIR observations were coincident are included.}
\label{fig:BVRINIR}
\end{figure}
% plot of U contribution
\begin{figure}
\centering
\includegraphics[scale=0.4]{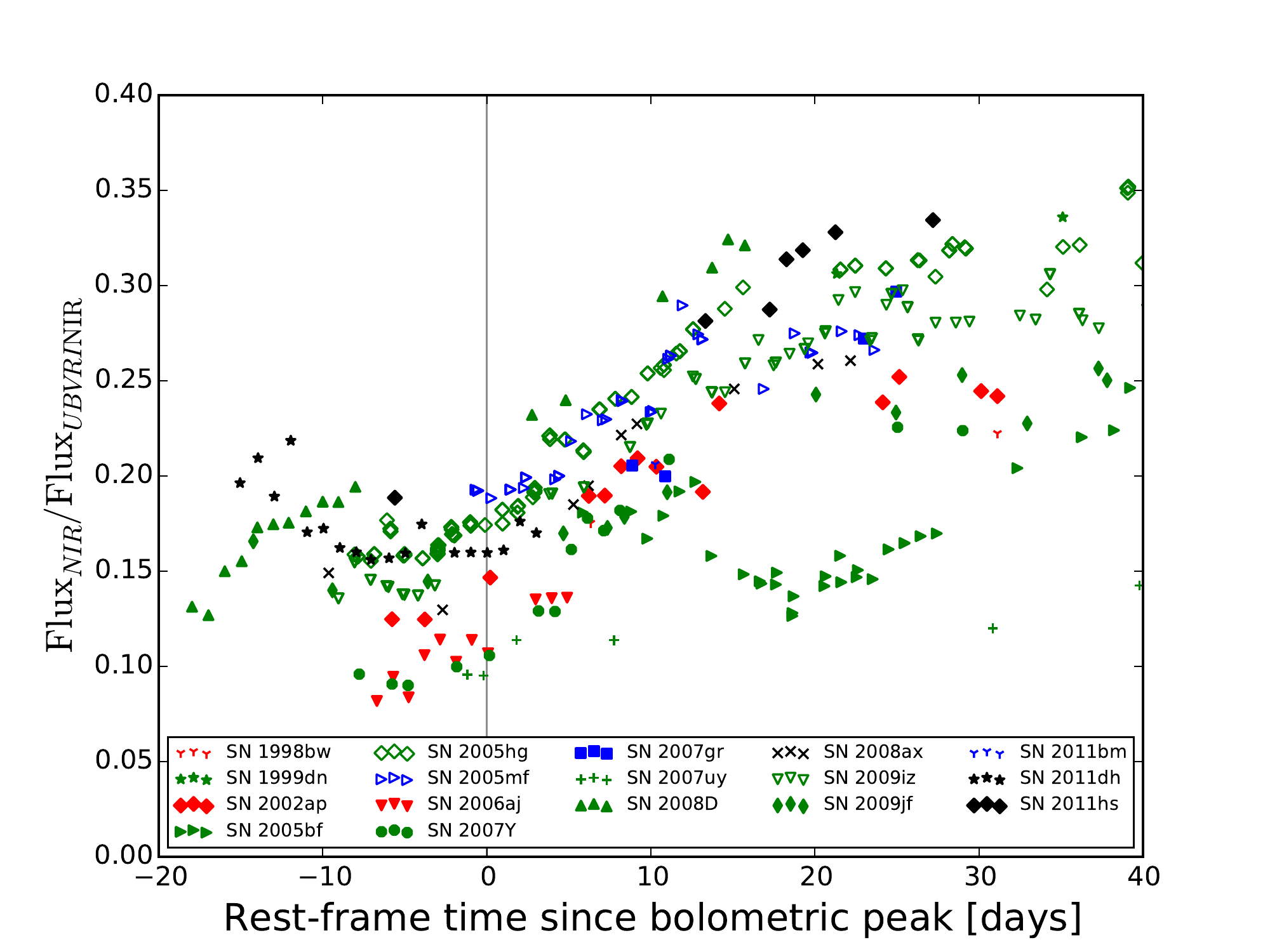}
\caption{The ratio of NIR flux to \textit{UBVRI}NIR flux. Only dates when the optical and NIR observations were coincident are included.}
\label{fig:UBVRINIR}
\end{figure}

Figures \ref{fig:BVRINIR} and \ref{fig:UBVRINIR} show the contribution of the NIR to the \textit{BVRI}NIR and \textit{UBVRI}NIR light curves, respectively. The median, mean, and standard-deviation statistics are given in Table \ref{UNIR}. We have insufficient numbers to split the sample into subtypes, so we perform the analysis on all available SNe, regardless of whether there is a host-galaxy extinction value. We find that around peak, mean $\approx$ median, with a standard deviation of $\sim 4$\% in both cases. It is possible that we have a bias in the sample given that the total number in each case is 18 and 12, which represents 21\% and 14\% of the total database, respectively. To test the probability of the statistical values being returned by chance, we run a Monte-Carlo simulation in which we randomly place 85 SNe with a uniform distribution in a NIR/ONIR ``box'' of varying width. From this box, 18 SNe (12 for NIR/\textit{UBVRI}NIR) are randomly selected and their bulk median/mean ratio and standard deviation measured. We define an acceptable parameter set as $0.9 <$ ratio $<1.1$ and standard deviation $<0.4$; there is no constraint on the values of median or mean, only their ratio. If the returned values fulfil these criteria then it is considered a hit. We perform 5000 runs for each window and return $P$(ratio) $=$ hit/runs. The results are sensitive to the allowed variation in the ratio and the standard deviation, both of which we choose to be in excess of the measured values. We find that for NIR/\textit{BVRI}NIR, a window of $> 0.16$ gives $P<0.025$, while for NIR/\textit{UBVRI}NIR, a window of $> 0.18$ gives $P<0.064$. We conclude that the observed median and mean are not a sampling bias caused by too few SNe; they represent typical values.

The contribution of the NIR is generally small throughout with the exception of SN 2005kl, which was extremely red and has a NIR to optical ratio $> 1$; consequently, it has been omitted from Figure~\ref{fig:BVRINIR}. \cite{Bianco2014} noted that this SN occurred in an \ion{H}{II} region of NGC 4369, a galaxy with a high continuum gradient. They also do not attribute the red colour to high intrinsic extinction. Spectra of the object show that it is dominated by galactic emission and displays a red continuum. Accordingly, we do attribute the red colour to host-galaxy extinction. On this basis it has not been included in the process of determining the NIR flux fraction.

\begin{table}
	\centering
	\caption{Flux-ratio statistics at bolometric peak.}
	\begin{tabular}{lccc}
	\hline
	\multicolumn{4}{c}{\textit{U}-band contribution to \textit{UBVRI} with host extinction}\\
	Type & median & mean & standard deviation \\
	\hline
	SNe~Ic-BL/GRB-SNe & 0.20 & 0.20 & 0.03 \\
	SNe~Ic & 0.15 & 0.15 & 0.02 \\
    SNe~Ib & 0.16 & 0.20 & 0.05 \\
    SNe~IIb & 0.16 & 0.17 & 0.06 \\
    \hline
	\multicolumn{4}{c}{\textit{U}-band contribution to \textit{UBVRI} without host extinction}\\
	Type & median & mean & standard deviation \\
	\hline
	SNe~Ic-BL/GRB-SNe & 0.18 & 0.19 & 0.07 \\
	SNe~Ic & 0.16 & 0.19 & 0.09 \\
    SNe~Ib & 0.13 & 0.15 & 0.07 \\
    SNe~IIb & 0.11 & 0.14 & 0.07 \\
    \hline
	\multicolumn{4}{c}{NIR contribution to optical/NIR}\\
	Type & median & mean & standard deviation \\
	\hline
	NIR/\textit{BVRI}NIR & 0.17 & 0.17 & 0.04 \\
	NIR/\textit{UBVRI}NIR & 0.14 & 0.15 & 0.03 \\
    \hline
	\end{tabular}
	\label{UNIR}
\end{table}

\section{Light-curve statistics}
\subsection{Light curves}
\begin{figure}
\centering
\includegraphics[scale=0.4]{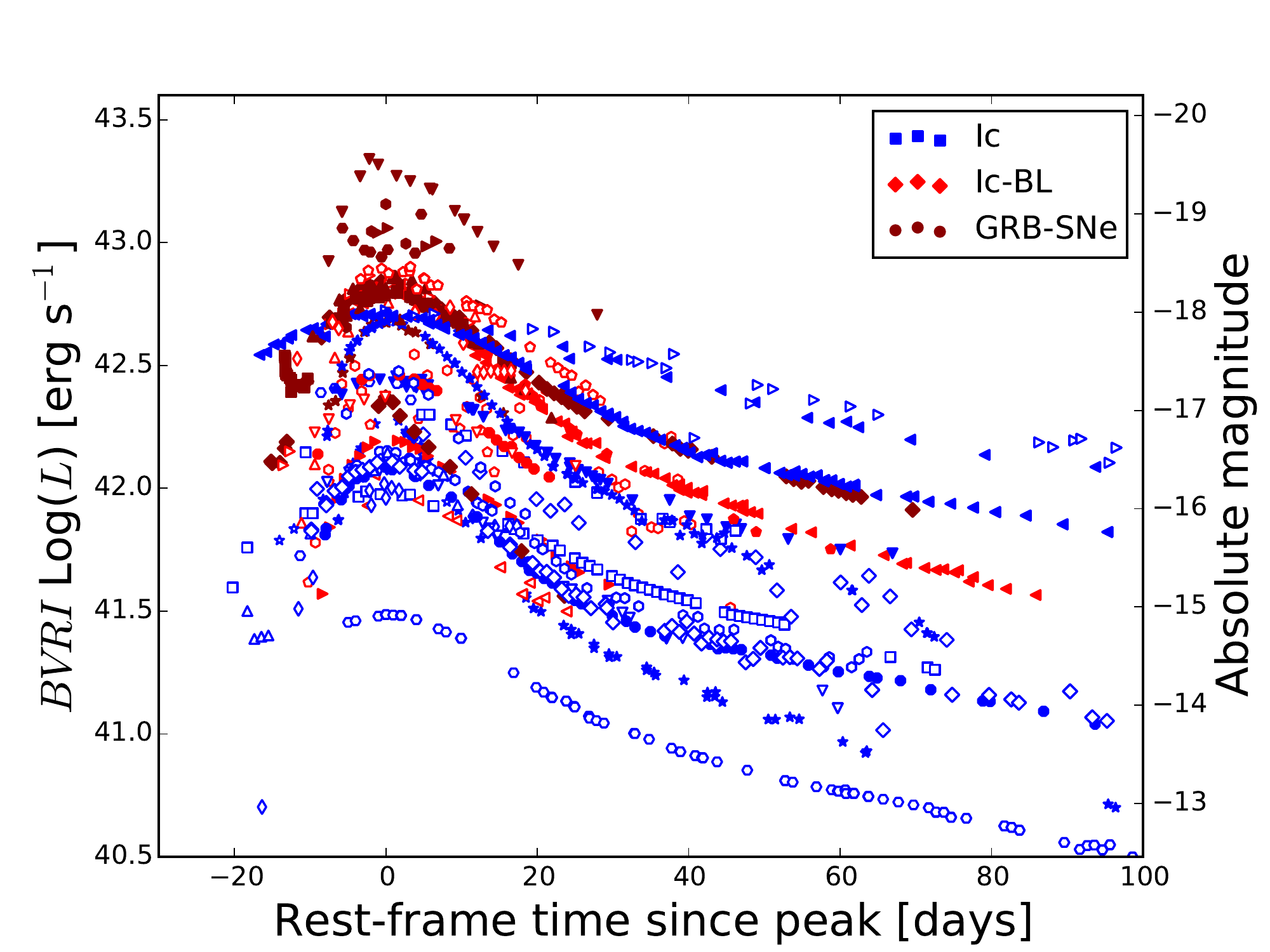}
\caption{\textit{BVRI} light curves of all Type Ic variants in the sample. SNe denoted by open markers have not been corrected for host-galaxy extinction. Markers may differ from the legend to aid the eye.}
\label{fig:IcBVRILCs}
\end{figure}

\begin{figure}
\centering
\includegraphics[scale=0.4]{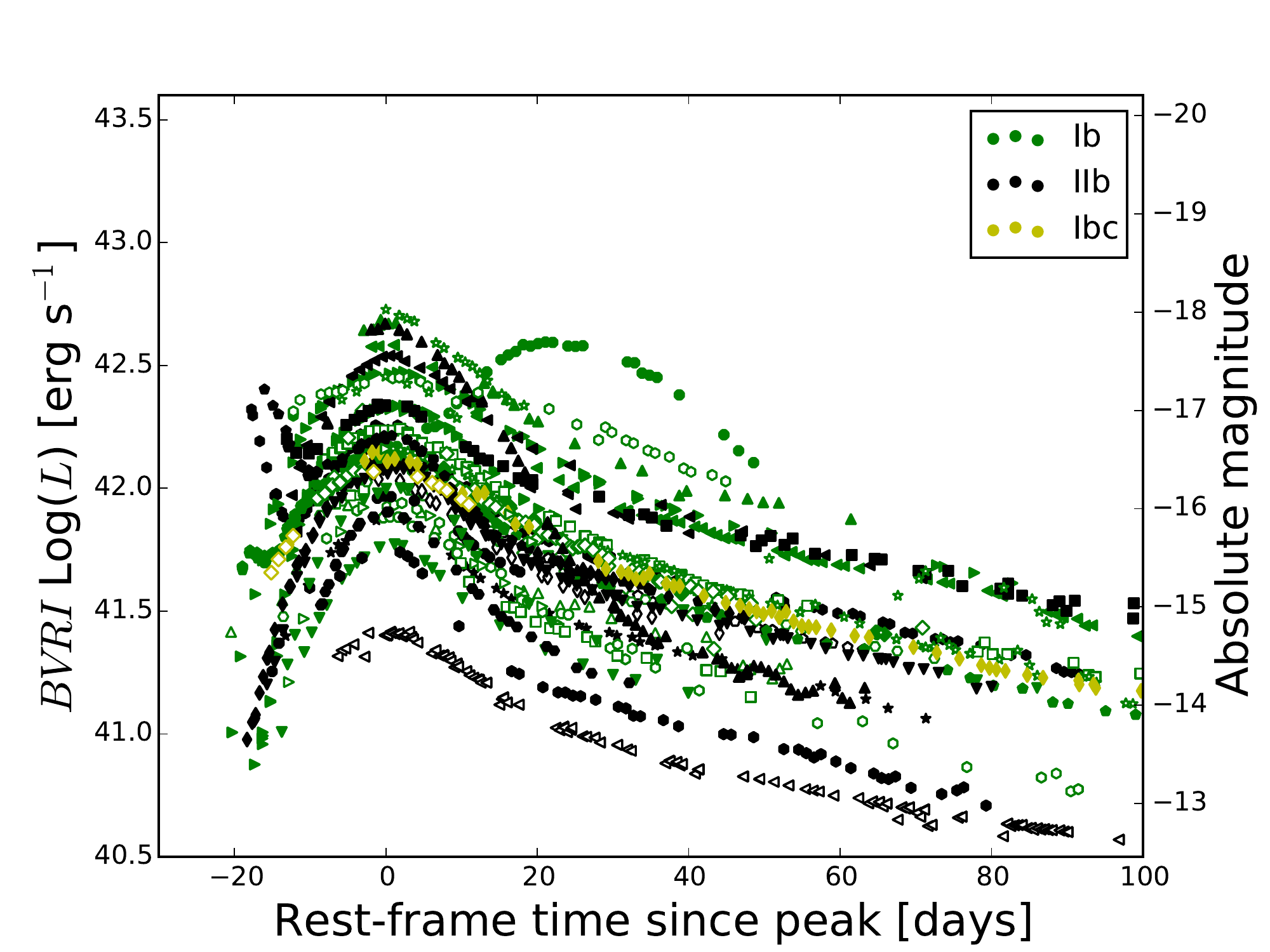}
\caption{\textit{BVRI} light curves of SNe~Ib and SNe~IIb in the sample. Open symbols represent SNe without a correction for host-galaxy extinction. Markers may differ from the legend to aid the eye.}
\label{fig:IbBVRILCs}
\end{figure}

\begin{figure}
\centering
\includegraphics[scale=0.4]{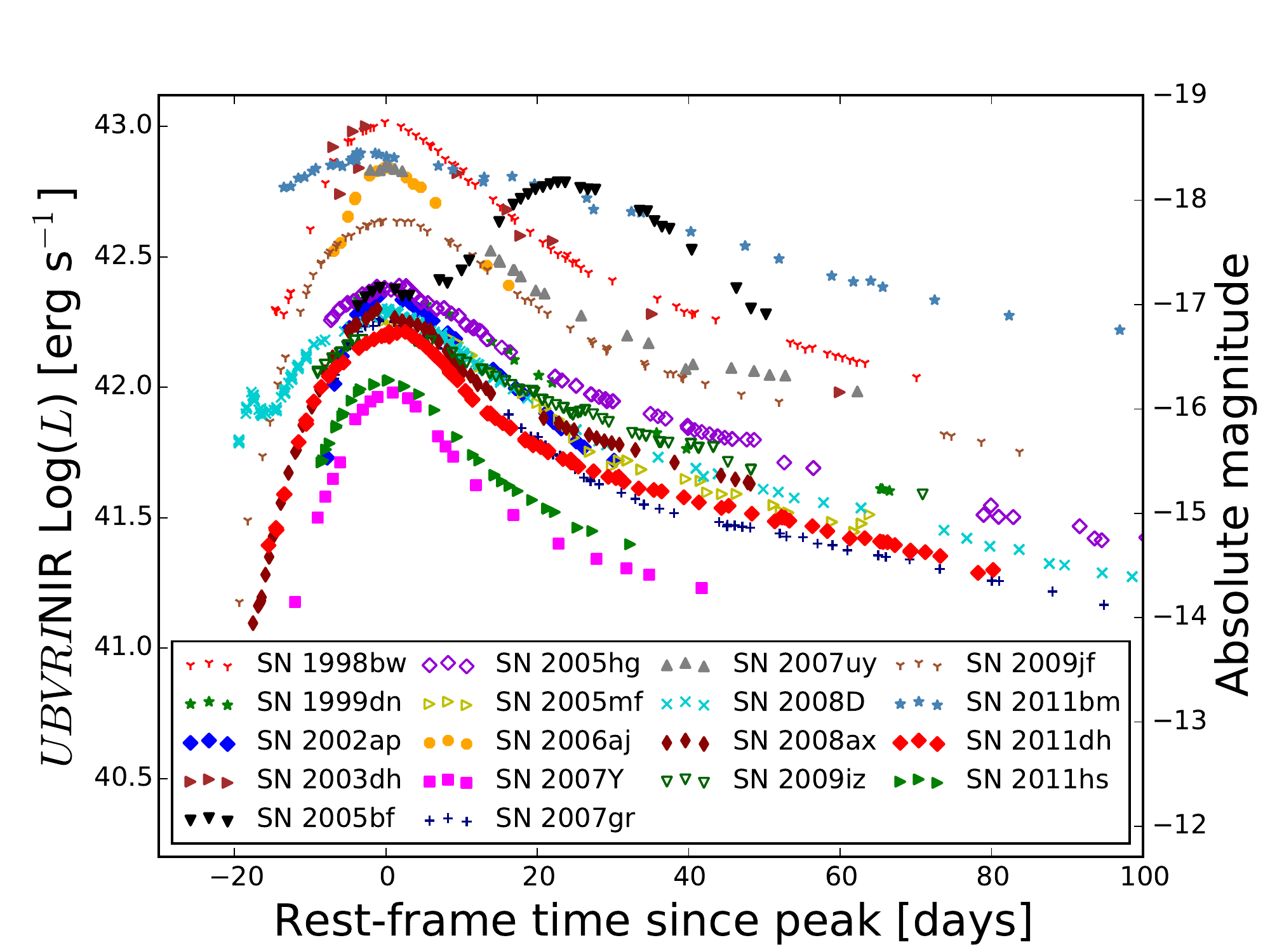}
\caption{18 \textit{UBVRI}+ NIR pseudobolometric light curves constructed from optical and NIR photometry. Open symbols represent SNe without correction for host-galaxy extinction}
\label{fig:ONIRLCs}
\end{figure}

The $BVRI$ LCs are shown by spectral type in Figures~\ref{fig:IcBVRILCs} and \ref{fig:IbBVRILCs}. The peak luminosity ranges from $5.6\times10^{41}$ erg s$^{-1}$ to $\sim 2.1 \times10^{43}$ erg s$^{-1}$, a factor of $\sim 100$. Note that this range does not include PTF12os or SN 2005kl, both of which suffer from significant but unquantified host-galaxy extinction. 18 of the SNe available have sufficient photometry to be able to construct \textit{UBVRI}NIR light curves. These SNe represent our ``gold'' sample and are shown in Figure~\ref{fig:ONIRLCs}. In some cases, the light curves are lower limits owing to the lack of information on host extinction. 

\begin{figure*}
\centering
\includegraphics[scale=0.7]{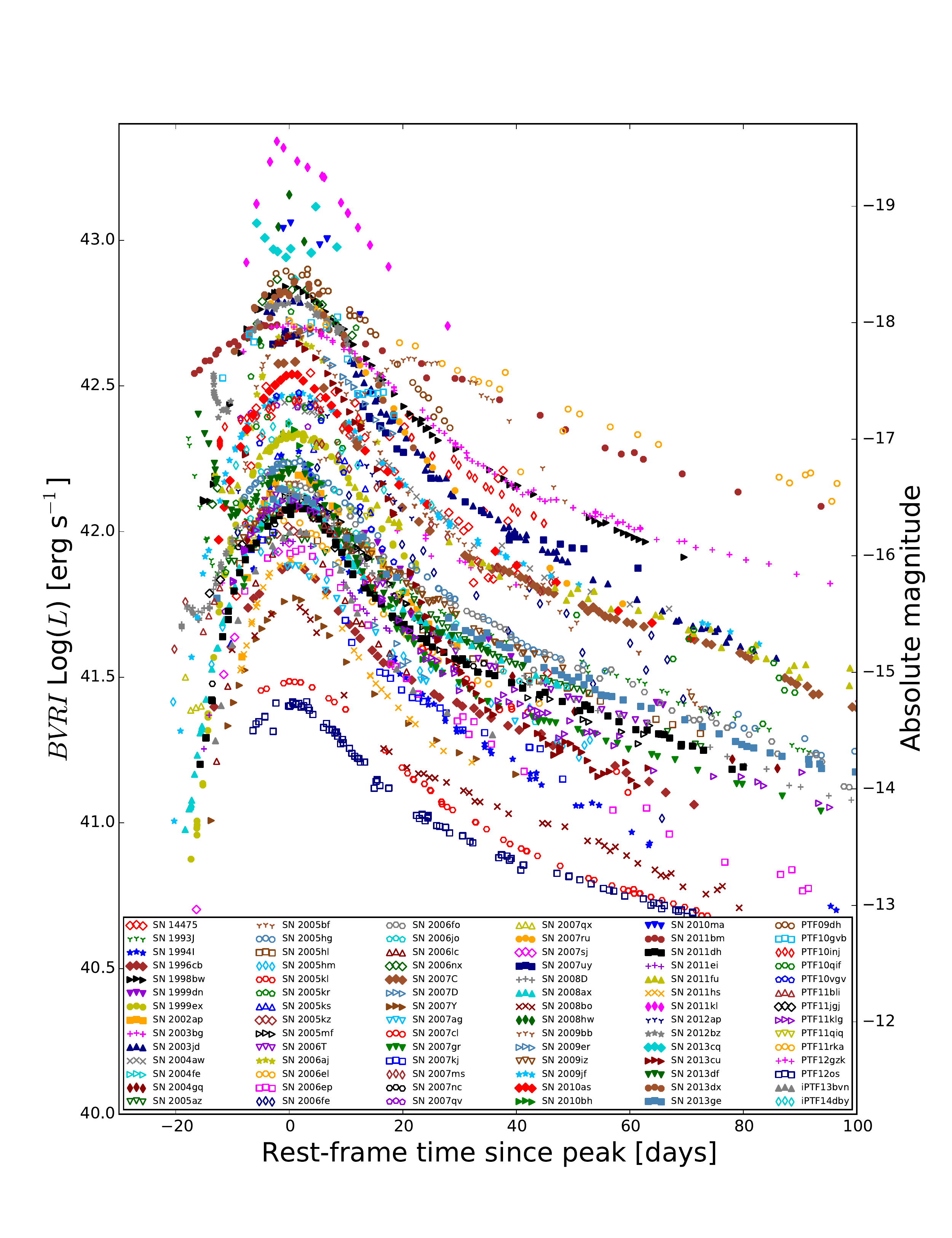}
\caption{84 \textit{BVRI} light curves of all SN types in the sample. GRB-SN 2003dh is not included because it lacks a \textit{BVRI} light curve. Note that GRB-SN 2013cq is extremely noisy, and its peak luminosity is constrained by a {\it Hubble Space Telescope} observation \citep{Melandri2014}. Open symbols represent SNe without corrections for host-galaxy reddening.}
\label{fig:BVRILCs}
\end{figure*}

The \textit{BVRI} pseudobolometric light curves are shown en-masse in Figure~\ref{fig:BVRILCs}.
The final light curves will be uploaded to WISeREP\footnote{http://wiserep.weizmann.ac.il/} \citep{Yaron2012}.

% BVRI LCs of all Ic

% BVRI Ib/IIb LCs

% plot of ONIR LCs
 
% All BVRI LCs

\subsection{Luminosity functions}
\begin{figure}
\centering
\includegraphics[scale=0.4]{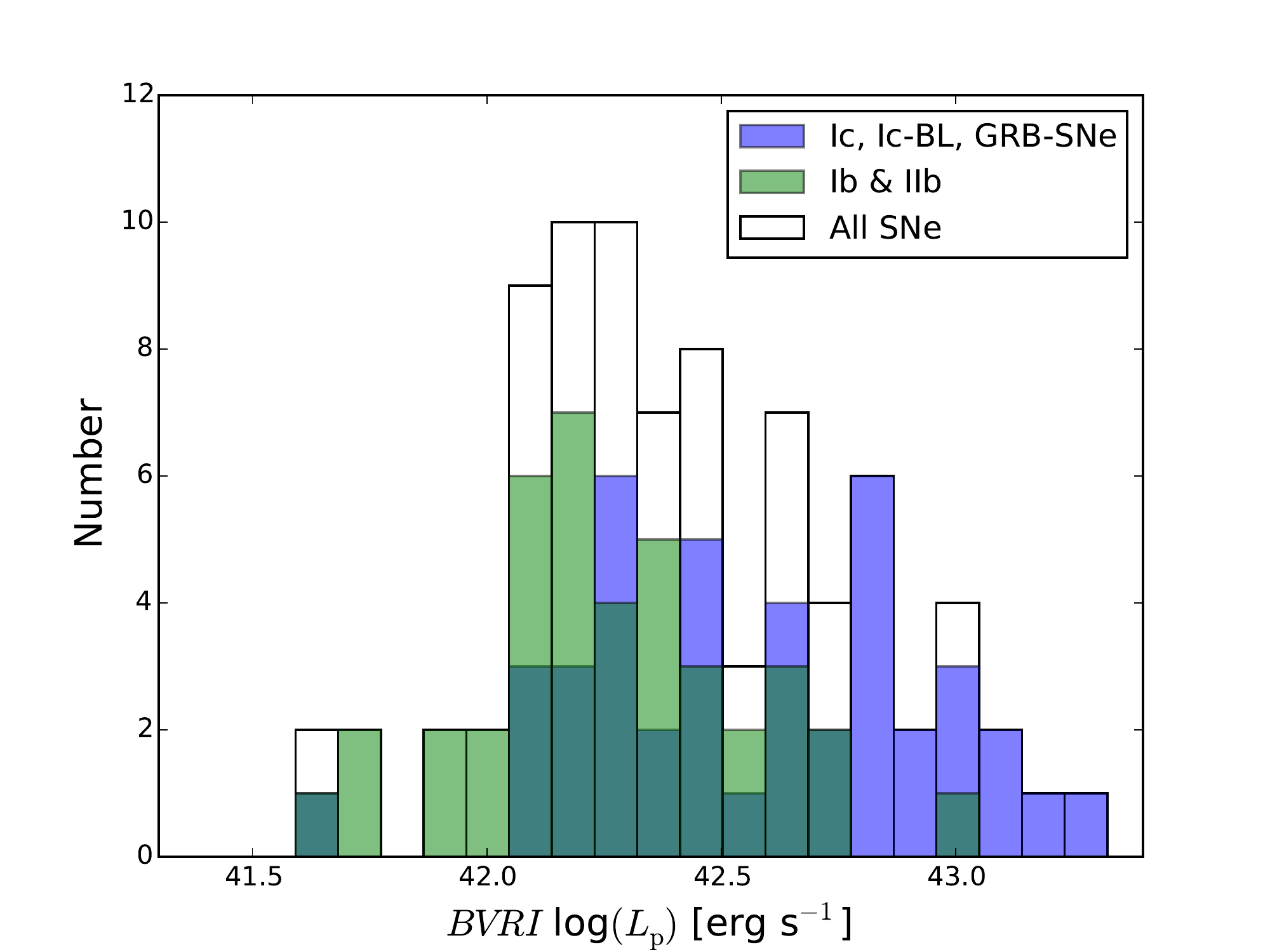}
\caption{The \textit{BVRI} pseudobolometric luminosity function of 82 SNe. The two SNe of Type Ibc are not included owing to ambiguity in their classification, nor is SN 2003dh owing to the lack of a \textit{BVRI} LC. This distribution includes SNe with a correction for the median host-galaxy extinction of that type applied.  The dark-green region represents the overlap of the blue and green distributions. The luminosity function for the combined green and blue distributions is shown in white.}
\label{fig:BVRIlumfunc}
\end{figure}

The luminosity function \citep[see, for example,][]{Li2011} for the \textit{BVRI} sample, which includes those SNe where the median host-galaxy extinction has been included, is shown in Figure~\ref{fig:BVRIlumfunc}. Table~\ref{bollumfunc} gives the statistics of the distribution. The collective SN~Ic-BL/GRB-SN group is most luminous, and while this is somewhat driven by the GRB-SNe, it can be seen from Figure~\ref{fig:IcBVRILCs} that SNe~Ic-BL are typically more luminous than other SNe~Ic. The least-luminous subpopulation is SNe~IIb. The standard deviation derived for each subset shows that there is considerable overlap between SNe~Ic, SNe~Ib, and SNe~IIb. 

% Luminosity function (BVRI) image
 
% Luminosity function (ONIR) image
\begin{figure}
\centering
\includegraphics[scale=0.4]{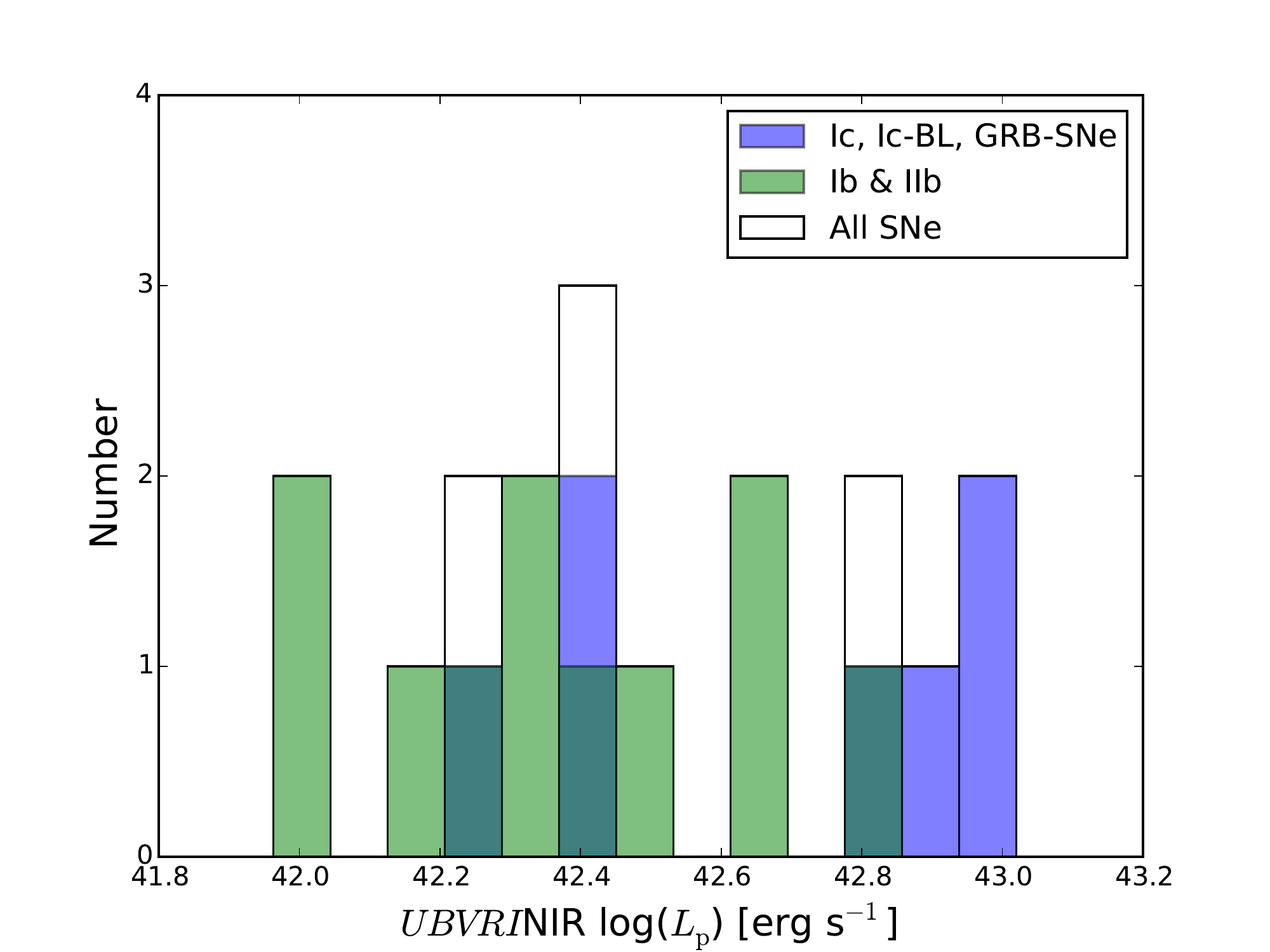}
\caption{The luminosity function derived from the peak luminosity of the 18 \textit{UBVRI}NIR light curves in the sample. Median host-galaxy extinction is assumed in three cases for statistical purposes. Colours are as described in Figure~\ref{fig:BVRIlumfunc}. }
\label{fig:fulllumfunc}
\end{figure}

% Fully bolometric luminosity function values
\begin{table}
 \centering
 \begin{minipage}{90mm}
  \caption{The \textit{BVRI} luminosity-function statistics.}
 \begin{tabular}{lccc}
  \hline
 Type & Median &Mean & Standard deviation \\
  \hline
SNe~Ic-BL/GRB-SNe & 42.81 &  42.78 & 0.29\\
SNe~Ic & 42.29 & 42.36 & 0.28 \\
SNe~Ib & 42.33 & 42.34 & 0.27 \\
SNe~IIb & 42.19 & 42.14 & 0.26\\
\hline
 \label{bollumfunc}
\end{tabular}
\end{minipage}
\end{table}

\subsection{Parameter values and statistics}
With the bolometric light curves complete, we are now in a position to begin determining their properties. The following parameters are of interest:
\begin{itemize}
	\item{Peak luminosity --- $L_{\mathrm{p}}$}
	\item{Rise time from explosion to $L_{\mathrm{p}}$ --- $t_{\mathrm{p}}$}
	\item{Rise time from $L_{\mathrm{p}}$/2 to $L_{\mathrm{p}}$ --- $t_{-1/2}$}
	\item{Decay time from $L_{\mathrm{p}}$ to $L_{\mathrm{p}}$/2 --- $t_{+1/2}$}
	\item{Light-curve width --- $t_{-1/2} + t_{+1/2}$}
	\item{Nickel mass --- $M_{\mathrm{Ni}}$}
	\item{Ratio of ejecta mass to kinetic energy --- $M_{\mathrm{ej}}^3/E_{\mathrm{k}}$}
\end{itemize}

The statistics were found using the set of \textit{BVRI} and \textit{UBVRI}NIR-equivalent LCs. The first step was to determine the values of $L_{\mathrm{p}}$, $t_{-1/2}$, and $t_{+1/2}$. A fourth-order polynomial was fit to each light curve around the peak using \textsc{curve\_fit} from the \textsc{python} \textsc{scipy}\footnote{www.scipy.org} package. If the photometric coverage was sufficient, then the three parameters could be determined. However, in most cases only $L_{\mathrm{p}}$ plus one other of the temporal values was directly measurable. In the instances where the photometric observations did not extend sufficiently far before or after $L_{\mathrm{p}}$ to return  $t_{-1/2}$ or $t_{+1/2}$, it may have been possible to extrapolate to this time if the initial/final luminosity was sufficiently close to $L_{\mathrm{p}}/2$. In an attempt to derive $t_{-1/2}$, a second-order polynomial was fit to the early-time observations up to the time of peak luminosity, provided a sufficient number of observations was available. If the fit could converge to a solution within two days of the boundary data point it was accepted. Conversely, on the rare occasion when $t_{+1/2}$ could not be taken directly from the observations, the late-time data were fit with a linear function and extrapolated out to five days to find a solution. The extrapolations were inspected visually for irregularities and  accepted if they seemed reasonable. To derive estimates of the uncertainty, the upper and lower photometric errors were fitted in a similar fashion as previously described. In some cases, it was not possible to determine one or more of these parameters, and so these SNe are omitted from the initial studies involving this parameter. 

Table~\ref{BVRIgroup} gives the median temporal characteristics as derived from the {\it BVRI} group; we note that these values remain consistent with those derived for the {\it UBVRI}, {\it BVRI}NIR, and {\it UBVRI}NIR samples. We caution the reader that the median width is not the sum of the median values of $t_{-1/2}$ and $t_{+1/2}$ but is the median of the sum of the two parameters for SNe where both have been derived. This means that the median width is drawn from a smaller sample size than either $t_{-1/2}$ or $t_{+1/2}$. In particular the large median value of $t_{+1/2}$ for SNe Ic is largely driven by the extremely broad light curves of SNe 2011bm \citep{Valenti2012} and PTF11rka, neither of which can be included in the calculation of median width because they lack a value for $t_{-1/2}$. The relationship between $t_{-1/2}$ and $t_{+1/2}$ is considered in more detail in section 7.2.

In the Appendix, Table~\ref{BVRIstats} and Table~\ref{intONIRstats} give the values derived for individual SNe from the \textit{BVRI} sample and the \textit{UBVRI}NIR-equivalent sample, respectively.

\subsection{Determining Errors}
The error in any particular value derived is related to the uncertainties in the photometry and the extinction. In certain situations the photometric errors are very small, perhaps unjustly so, which is the cause of very small uncertainties in the values determined in this section. Very small uncertainties lead to a higher degree of certainty in a value than is justified, and they have the effect of biasing possible correlations between parameters by increasing the weighting of the values for a particular SN. Ideally, a well-observed SN with good Galactic and host-galaxy extinction would have such an effect, but we see that some SNe which fall short of this standard display extremely small uncertainties in their photometry. It can be seen in the LCs of these SNe that the uncertainties are unjustified, as they show variability which is greater than the photometric errors bars. 
\subsection{Rise time} 
The parameter with the largest uncertainty is rise time, $t_{\mathrm{p}}$, principally because it requires well-covered photometry pre-peak and because of the uncertain behaviour of the light curve at very early times. Generally, only GRB/XRF SNe have an observed value of $t_0$ and hence a well-constrained $t_{\mathrm{p}}$, although the explosion time of SNe~IIb with prominent early-time emission (e.g., SN 1993J, \citealt{Matheson2000}; SN 2013df, \citealt{Vandyk2014}; SN 2011dh, \citealt{Arcavi2011}; SN 2013cu, \citealt{Galyam2014}) caused by the cooling of the stellar surface following shock breakout \citep{Woosley1994} can be estimated to within a day or two of explosion. If the SN is found in a host galaxy that is regularly observed, then $t_0$ can be constrained between the detection and nondetection dates assuming that $\left\langle \mathrm{d}L/\mathrm{d}t\right\rangle$ is sufficiently large at early epochs so as to minimise the ``dark time'' (the period between explosion and detectability) of the SN \citep[e.g., PTF10vgv;][]{Corsi2012}. Of course, this method is limited to the time interval between the two dates, which ideally should be on the order of a few days. Values for $t_{\mathrm{p}}$ from the literature are given in Table~\ref{tpall}.

\begin{table}
\renewcommand{\arraystretch}{1.5}
 \centering
 \begin{minipage}{85mm}
  \caption{Median temporal values derived from the \textit{BVRI} data.}
 \begin{tabular}{lccc}
  \hline
 Type &  $t_{-1/2}$ (days) & $t_{+1/2}$ (days) & Width (days) \\
  \hline
SNe~Ic-BL/GRB-SNe & 8.6$\pm^{1.9}_{1.1}$ & 15.1$\pm^{1.0}_{2.0}$ & 24.7$\pm^{2.7}_{2.3}$\\
SNe~Ic & 9.3$\pm^{2.6}_{1.1}$ & 19.2$\pm^{4.7}_{5.4}$ & 23.8$\pm^{7.3}_{5.4}$\\
SNe~Ib & 11.2$\pm^{2.2}_{1.4}$ & 17.0$\pm^{2.8}_{2.9}$ & 26.4$\pm^{3.6}_{3.9}$\\
SNe~IIb & 10.1$\pm^{1.2}_{0.4}$ & 15.3$\pm^{2.8}_{1.6}$ & 25.4$\pm^{2.3}_{0.8}$\\
\hline
 \label{BVRIgroup}
\end{tabular}
\end{minipage}
\end{table}

\subsection{Comparison between optical and optical/NIR light curves at peak}
\begin{figure}
\centering
\includegraphics[scale=0.4]{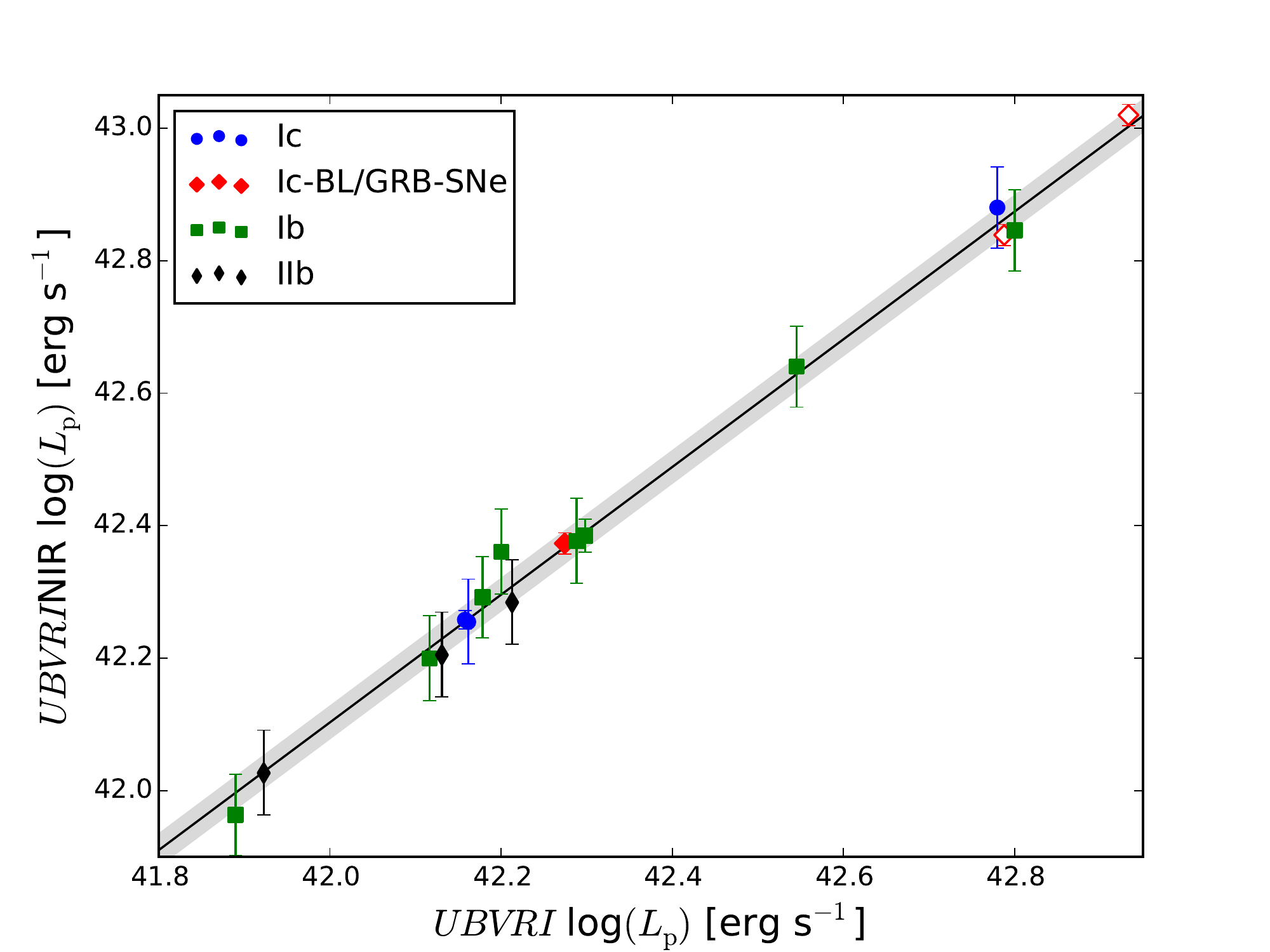}
\caption{\textit{UBVRI}+NIR peak luminosity as a function of \textit{UBVRI} peak luminosity. The correlation appears to be independent of SN type. The grey area represents the $1\sigma$ uncertainty in the fit determined by the standard deviation of the residuals. The error bars are representative of the uncertainty only in the NIR contribution. Open symbols represent GRB-SNe.}
\label{fig:luminc}
\end{figure}

We took the step of plotting the \textit{UBVRI} luminosity at maximum versus \textit{UBVRI}NIR maximum luminosity to investigate the possibility of determining some form of ``bolometric correction'' for the much larger \textit{BVRI}-only sample. The results of this test are shown in Figure~\ref{fig:luminc}. It is apparent, at a glance, that all SNe in this set follow a very tight correlation between the two values, and while caution must be exercised when dealing with values in log-log space, it appears that a simple linear fit would produce the desired results. The data were fit with a simple linear polynomial of the form given by the equation 
\begin{equation}
\mathrm{log}(L_{UBVRI\mathrm{NIR}})=a_{1}+a_{2}\,\mathrm{log}(L_{UBVRI})
\label{lumincfit}
\end{equation}
The best fit gives $a_{1}=1.62$ and $a_{2}=0.96$ with the standard deviation of the residual distribution equal to 0.02, which is adopted as the uncertainty in the fit. The SN with maximum displacement from the best-fit line shows a difference of $< 20$\% between the value returned from the polynomial and the photometric value.

Overall, it appears that the variation in spectral shape is less important outside the optical wavelengths. This correlation is also very strong for \textit{BVRI} against \textit{BVRI}NIR. Thus, the conversion to fully bolometric luminosity requires little more than a simple multiplicative factor that is proportional to the optical flux. This implies that, around peak luminosity, the absorption and re-emission of photons as they diffuse through the ejecta happens primarily in the optical regime. Consequently, the NIR region is effectively the Rayleigh-Jeans tail of a blackbody with a temperature that gives an optical integrated flux similar that of the SN optical SED. 

\section{Pseudobolometric to fully bolometric}
The disparity in number of SNe with \textit{BVRI} data compared with \textit{UBVRI}NIR data presents a challenge to efforts to reveal the statistics of the explosion, especially the luminosity function and the mass of $^{56}$Ni synthesised in the first seconds after core collapse. Here we detail how we construct fully bolometric peak luminosity values for all the SNe in the sample using the results from earlier sections.
\subsection{The conversion method}

To minimise the amount of uncertainty in each SN, the shortest method of obtaining the fully bolometric peak luminosity was used. The details of this procedure are outlined here.
\subsubsection{Conversion from \textit{BVRI} to \textit{UBVRI} at peak}
Approximately half of the SNe in the database lack $U/u'$ photometry. To compensate for this, we use the results of Section 3.7 and apply a correction to account for the missing photometric band to the \textit{BVRI} flux of these SNe. We use Table~\ref{UNIR} to assign a correction value depending on SN type and whether there is a value for host-galaxy extinction. We assume the errors are commensurate with the standard deviation of each distribution. As a justification, we note that from Table~\ref{UNIR} the median \textit{U} fraction is always between 15\% and 20\%.
\subsubsection{Conversion from \textit{UBVRI} to \textit{UBVRI}NIR}
There are two pathways to convert the \textit{UBVRI} flux to \textit{UBVRI}NIR flux. In the first instance we can use the NIR data where they exist and combine their peak value with that of the \textit{UBVRI} flux. The second method utilises Equation~\ref{lumincfit} and the values for the \textit{UBVRI} to \textit{UBVRI}NIR conversion as given in Section 5.6.

\subsubsection{\textit{UBVRI}NIR to fully bolometric}
We finalise the conversion of \textit{UBVRI}NIR $L_{\mathrm{p}}$ to a fully bolometric value by assuming a 10\% contribution from unobserved wavelengths. We justify this value by integrating a Planck function at temperatures between 4000~K and 8000~K, which is typical for SE-SNe at peak, and comparing the flux ratio of our {\it UBVRI}NIR wavelength range to that of the flux outside it. We find that the UV and far-infrared can account for between $\sim 7$\% and $\sim 20$\% of the total flux but is typically $\sim 10$\%. If we assume some reprocessing of UV photons into the optical regime then this value can be lower. Thus, we take the error to be $+10$\%--5\%. We do not attribute any particular fraction of this amount to IR or UV, the latter of which is small for SE-SNe \citep{Pritchard2014}.

\subsection{Luminosity function for the bolometric sample}

\begin{figure}
\centering
\includegraphics[scale=0.4]{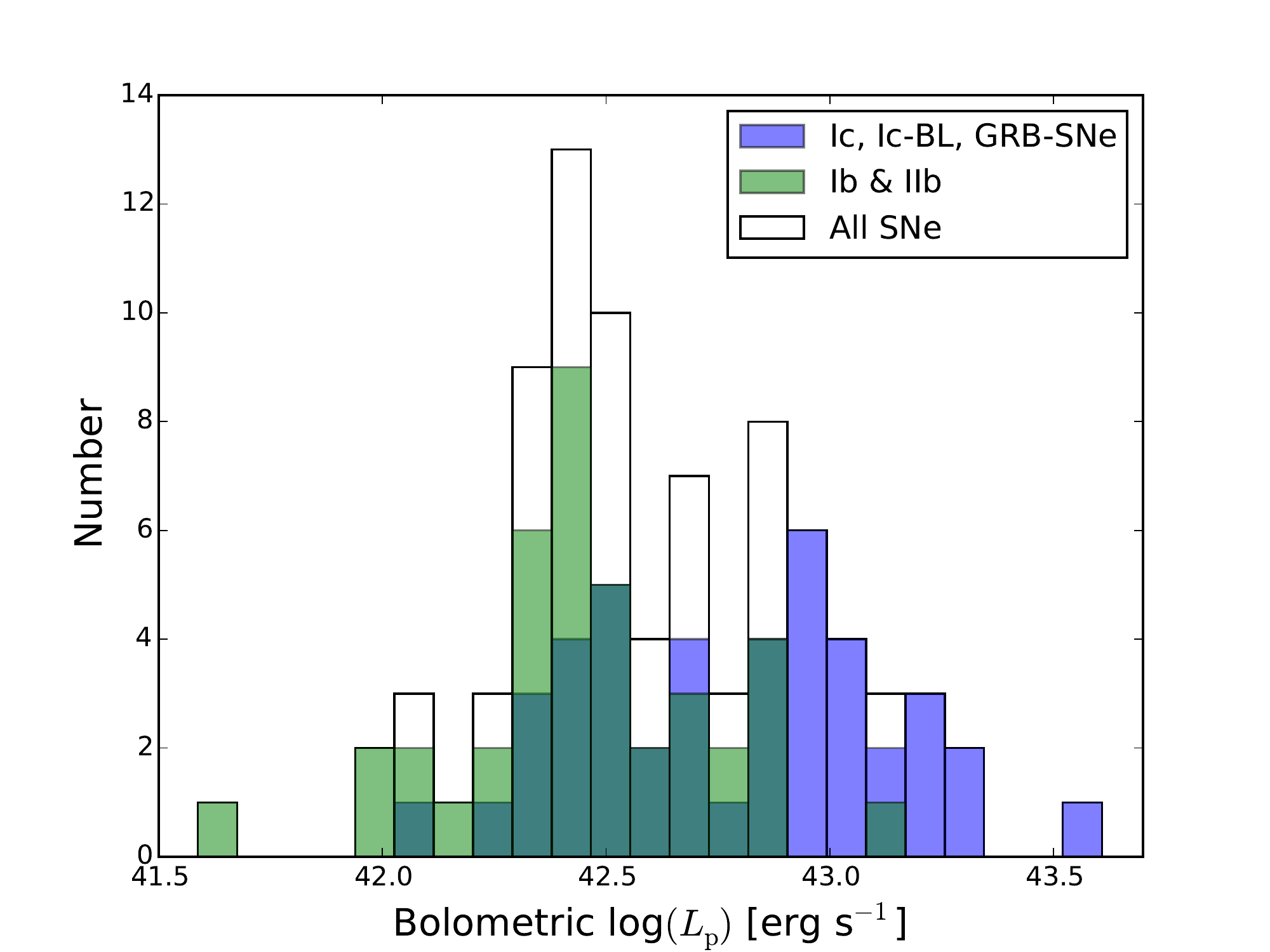}
\caption{Fully bolometric luminosity function of the entire sample. Median host-galaxy extinction is adopted when the actual value is absent. Colours are as described in Figure~\ref{fig:BVRIlumfunc}.}
\label{fig:ONIRlumfunc}
\end{figure}

The resulting fully bolometric luminosity function for the entire SN sample is shown in Figure~\ref{fig:ONIRlumfunc} and the statistics are shown in Table~\ref{meanstats}. The SN~Ic/Ic-BL/GRB-SN group remains the most luminous population. SNe~Ic and SNe~Ic-BL can be found throughout this distribution while GRB-SNe occupy the upper end.

% Luminosity function of bolometric sample

\section{temporal properties}
\subsection{$L_{\mathrm{p}}$ as a function of $t_{+1/2}$}
\begin{figure}
\centering
\includegraphics[scale=0.4]{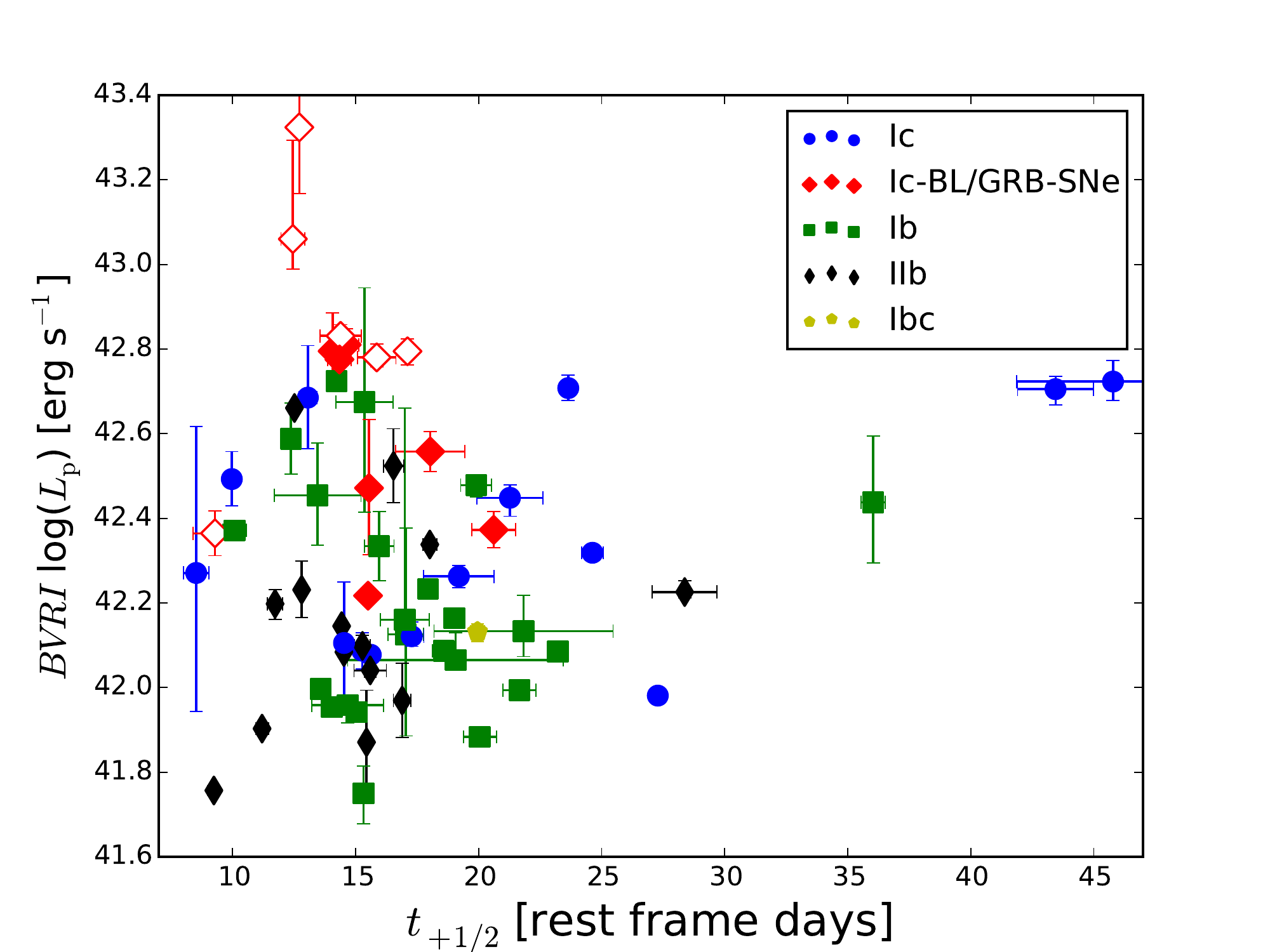}
\caption{$L_{\mathrm{p}}$ as a function of $t_{+1/2}$ for the \textit{BVRI}-equivalent light curves. There is no correlation between either parameter. Open symbols represent GRB-SNe.}
\label{fig:Lpvdecay}
\end{figure}

The value of Type Ia SNe for cosmological studies owing to the correlation between \textit{B}-band peak and light-curve decline over 15 days \citep{Phillips1993} is well known. All previous attempts to find a similar relation for core-collapse SNe have returned negative results \citep{Drout2011,Lyman2014}. Figure \ref{fig:Lpvdecay} shows how the {\it BVRI} pseudobolometric peak of our sample compares with $t_{+1/2}$. It is apparent that there is no equivalent ``Phillips relation'' for any SN type used here, thus confirming earlier studies and indicating that the dynamics of the explosion mechanism of SE-SNe and the relationship to the ejecta is nonuniform. 

% Lpeak v t+1/2

\subsection{Is there a relationship between $t_{-1/2}$ and $t_{+1/2}$?}
Photometric coverage in not consistent across the whole sample, so a full set of temporal parameters is available only for a small number of SNe. Motivated by the desire to maximise the statistics derived from our sample, we examined temporal parameter values as a function of other temporal characteristics where the values were known. A plot of $t_{-1/2}$ and $t_{+1/2}$ indicated that a loose correlation exists. We removed SNe that showed an excess of luminosity caused by shock breakout and nonradioactive power, further strengthening the apparent correlation between the two properties. However, this is at odds with the results of \cite{Taddia2015}, where the Kolmogorov-Smirnov (K-S) tests on the value of $\Delta m_{15}$ for the $r$ band for 40 SNe indicated that they were drawn from the same population. 

\begin{figure}
	\centering
	\includegraphics[scale=0.4]{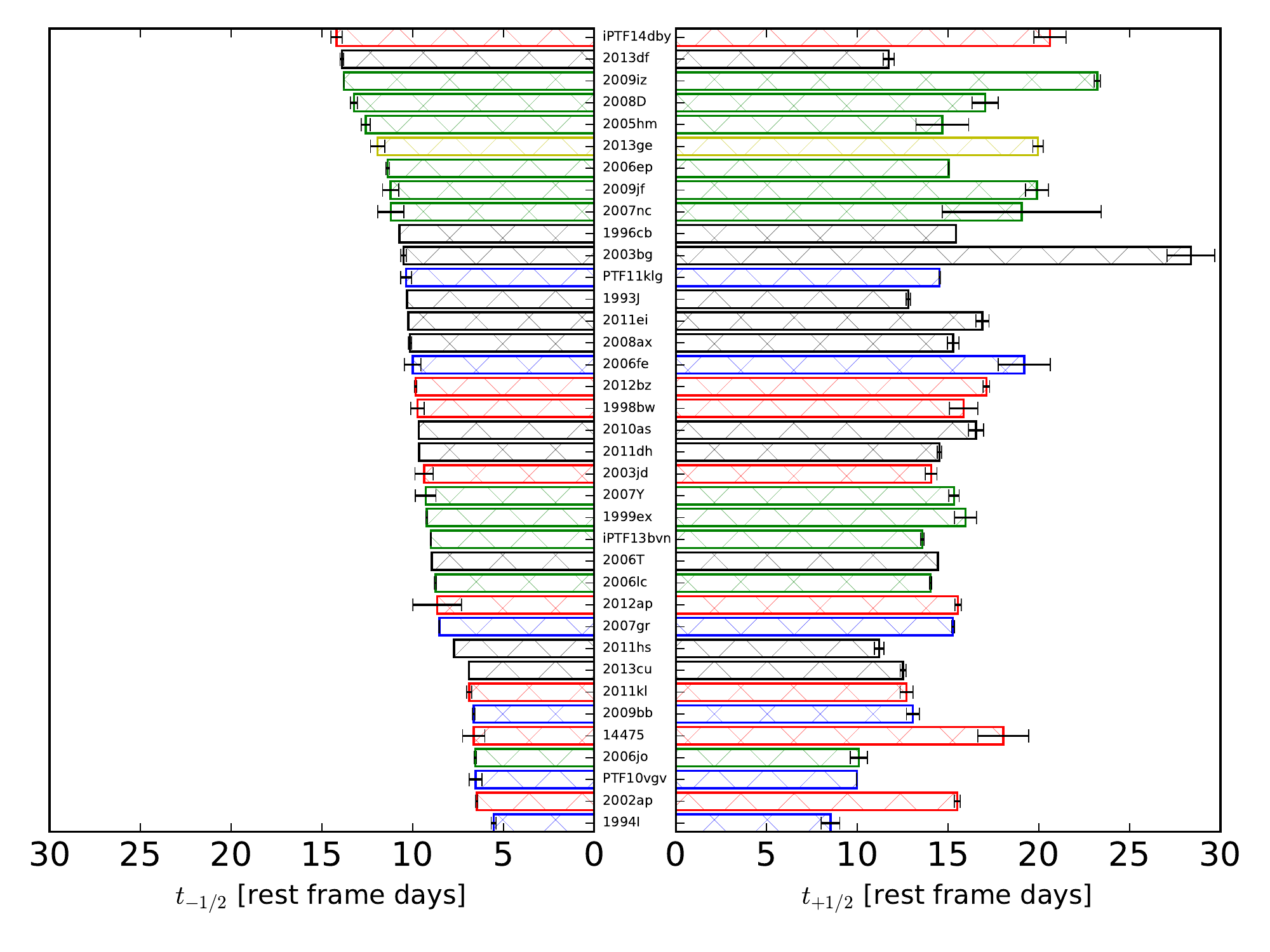}
	\caption{The $t_{-1/2}$ and $t_{+1/2}$ values for the SNe in the sample with both parameters. The range of values in $t_{+1/2}$ is readily apparent despite a trend of the decay time to increase as the rise time increases. SNe Ib are shown in green, Ic-BL/GRB-SNe are red, Ic are blue, IIb are black, and Ibc are yellow. }
	\label{fig:risedecay}
\end{figure}

\begin{figure}
	\centering
	\includegraphics[scale=0.4]{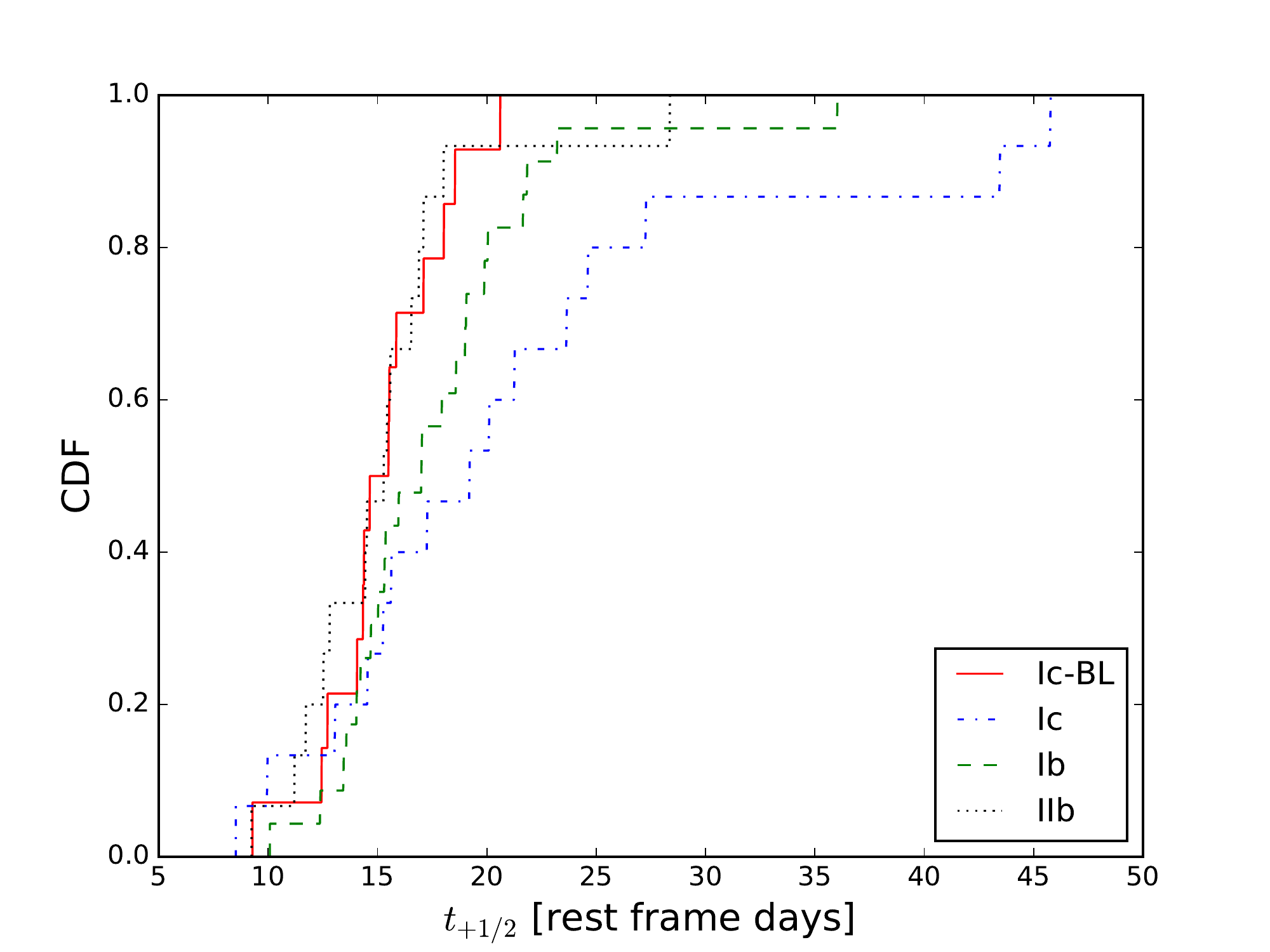}
	\caption{The cumulative distribution function of all the $t_{+1/2}$ values in the sample sorted by spectral type. The effect of the extremely slowly declining Type Ic SNe 2011bm and PTF11rka on the SN~Ic CDF is clearly seen.}
	\label{fig:decaycumu}
\end{figure}

Consequently, we performed a similar analysis on our $t_{+1/2}$ values. Figure~\ref{fig:risedecay} shows $t_{+1/2}$ as a function of $t_{-1/2}$. It can be seen that although there is a general trend to a slower decline for a slower rise, there is a considerable spread in the values. We plotted cumulative distribution functions (CDFs) for all $t_{+1/2}$ values derived in the sample and split them by type, as seen in Figure~\ref{fig:decaycumu}. K-S tests indicate that most of the distributions are drawn from the same population ($P>0.05$) with the exception of the SN~Ic/IIb group ($P=0.048$), although the SN~Ic/Ic-BL group returns $P=0.061$ which is marginally over the threshold. We can attribute this entirely to the presence of broad light curve SNe~Ic 2011bm and PTF11rka in conjunction with the relatively small number of SNe throughout: 12, 14, 23, and 13 for SNe~Ic-BL, Ic, Ib, and IIb, respectively. It is clear that these two SNe disproportionately affect the sample, because without their presence the SN distributions are very similar, sharing a similar median $t_{+1/2}$ of $\sim 15$ days. With this result, we cannot conclude that there is a significant correlation between $t_{-1/2}$ and $t_{+1/2}$, and while the general trend is toward a longer decay for a longer rise, the variance along the trend makes it unreliable as a method for converting one to the other.

%Fishbone

\subsection{Correlation between $t_{\mathrm{p}}$ and $t_{-1/2}$}
\begin{figure}
\centering
\includegraphics[scale=0.4]{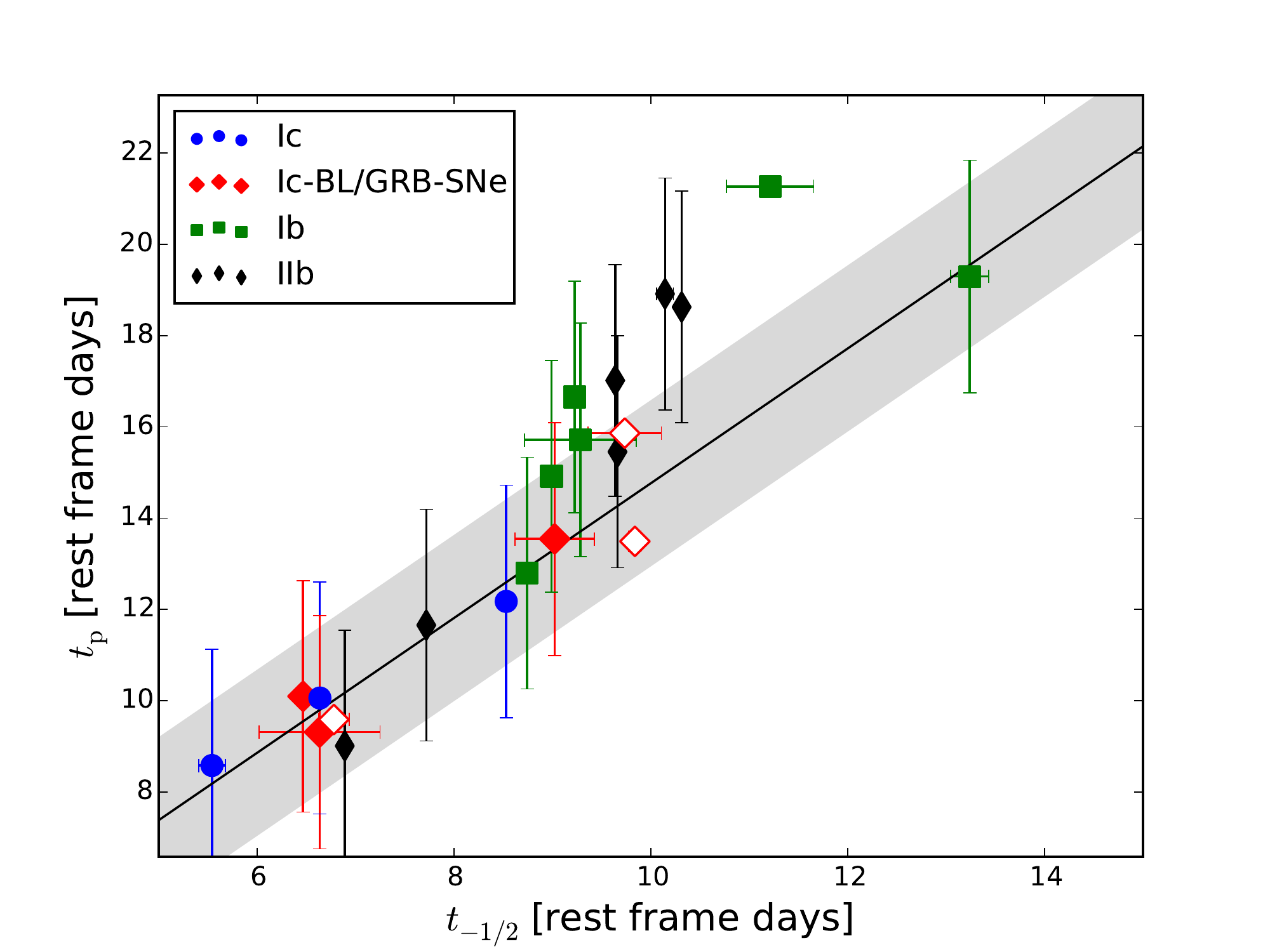}
\caption{$t_{\mathrm{p}}$ as a function of $t_{-1/2}$ for SNe in the sample having both parameters. The black line represents the best fit to the data incorporating errors in both the ordinate and abscissa. The grey region represents the standard deviation of the residuals from this line, which is taken to be the uncertainty in the fit. Open symbols represent GRB-SNe.}
\label{fig:risetp}
\end{figure}

It is pertinent to see if the time for rise to peak of the SN is in some way correlated with the rise from $L_{\mathrm{p}}/2$ to $L_{\mathrm{p}}$, as it would lead to a method to estimate the SN rise time from $t_{-1/2}$. The limitations of such an assessment are apparent, as few SNe have known explosion times. Thus, we take the GRB-SNe with their known $t_0$, some of the SN~IIb set (because their explosion times can be constrained from the initial shock-breakout peak), plus any SN with well-constrained explosion time owing to the short interval between nondetection and detection. Figure \ref{fig:risetp} shows $t_{\mathrm{p}}$ against $t_{-1/2}$; it is apparent that there is some correlation which appears to be independent of SN type. Using a linear fit as before, defined as
\begin{equation}
t_{\mathrm{p}}=\alpha t_{-1/2},
\label{fit2}
\end{equation}
we find $\alpha=1.5$. To determine uncertainties in the fit we took the standard deviation of the distribution of the residuals as described previously, which we find to be 1.68 days. SNe were omitted from the fitting procedure if the early-time emission was dominated by shock breakout. These were typically SNe~IIb or GRB-SNe for which $t_{-1/2}$ was poorly constrained owing to two components, and well-defined explosion times heavily influenced the result.

% rise/decay relation plot

\subsection{Inferred rise time $t_{\mathrm{p}}$}
Table~\ref{tpall} shows the inferred rise time for the sample as derived by the previously described correlation. Note that the rise times for GRB-SN 1998bw ($\sim 16$ days) and GRB-SN 2006aj ($\sim 10$ days) are recovered within the uncertainties of the extrapolated explosion time. Notable exceptions are as follows. (i) SNe that show relatively slowly declining shock break-out emission so that it is a nonnegligible contributor to the flux of the SN during the rise (e.g., SN 2008D, SN 2011hs). (ii) The case of SN 2011kl, associated with ultralong GRB 111209A \citep{Greiner2015}; this SN is interesting with regard to the explosion mechanism. 

First, we must consider an explanation for the correlation between $t_{-1/2}$ and $t_{\mathrm{p}}$. This can be explained by the interplay with $^{56}$Ni $\rightarrow$ $^{56}$Co $\rightarrow$ $^{56}$Fe energy injection and the light-curve rise time, which is itself a function of opacity, ejecta mass, and photospheric velocity \citep{Arnett1982}. Because the derivative of the energy injection rate is constant, the shape of the light curve is determined solely by the diffusion time. Altering the diffusion time means the peak may move, but the way the light curve rises to the peak retains the same relative shape provided the nickel distribution remains located centrally. This is not necessarily the case for an alternate energy source such as a magnetar, where the magnetar rotational energy can be deposited into the ejecta on a timescale much less, or much greater, than the diffusion time, or the case where shock breakout (particularly for SNe~IIb) or GRB afterglow contribute to the optical flux. For all GRB-SNe, except SN 1998bw where the GRB afterglow was negligible, afterglow subtraction is a major issue for accurately determining $t_{-1/2}$; fortunately, the need for this value is made redundant by a known explosion date.

%% Estimated tp
\begin{table}
 \centering
 \begin{minipage}{90mm}
  \caption{$t_{\mathrm{p}}$ values.}
 \begin{tabular}{lcc}
  \hline
SN & Literature $t_{\mathrm{p}}$ & $t_{\mathrm{p}}$ from $t_{-1/2}$ \\
\hline
1993J & 19.15$\pm$0.03 & 16.32$\pm$1.65 \\
1994I & 12.25$\pm$0.21 & 8.77$\pm$1.66 \\
1996cb & - &16.99$\pm$1.65 \\
1998bw & 15.86$\pm$0.18 & 15.41$\pm$1.69 \\
1999dn & 13.92$\pm$2.84 & - \\
1999ex & 18.35$\pm$0.04 & 14.60$\pm$1.65 \\
2002ap & 13.01$\pm$0.00 & 10.23$\pm$1.65 \\
2003bg & - &16.62$\pm$1.66 \\
2003dh & 12.65$\pm{1.66}$ & - \\
2003jd & 12.50$\pm$0.18 & 14.83$\pm$1.73 \\
2004fe & - &14.68$\pm$1.65 \\
2005bf & 18.32$\pm$0.35 & - \\
2005hm & - &19.93$\pm$1.67 \\
2005kr & - &11.41$\pm$1.67 \\
2005ks & - &12.37$\pm$1.66 \\
2006T & - &14.15$\pm$1.65 \\
2006aj & 9.59$\pm$0.04 & 10.73$\pm$1.66 \\
2006ep & - &18.00$\pm$1.65 \\
2006fe & - &15.83$\pm$1.71 \\
2006jo & - &10.37$\pm$1.65 \\
14475 & - &10.50$\pm$1.76 \\
2006lc & - &13.84$\pm$1.65 \\
2006nx & - &14.28$\pm$1.70 \\
2007Y & 18.76$\pm$0.35 & 14.69$\pm$1.75 \\
2007gr & 13.15$\pm$0.23 & 13.50$\pm$1.65 \\
2007ms & - &22.23$\pm$1.67 \\
2007nc & - &17.73$\pm$1.80 \\
2007qx & - &15.09$\pm$2.76 \\
2007ru & 10.24$\pm$0.05 & - \\
2007sj & - &14.54$\pm$1.66 \\
2007uy & 19.08$\pm$0.28 & - \\
2008D & 19.29$\pm$0.23 & 20.95$\pm$1.66 \\
2008ax & 19.28$\pm$0.13 & 16.05$\pm$1.65 \\
2008hw & 12.31$\pm$0.10 & - \\
2009bb & 12.63$\pm$0.10 & 10.50$\pm$1.65 \\
2009iz & - &21.83$\pm$1.65 \\
2009jf & 21.27$\pm$0.16 & 17.74$\pm$1.71 \\
2010as & 12.44$\pm$0.12 & 15.29$\pm$1.65 \\
2010bh & 12.74$\pm$0.10 & 5.20$\pm$1.68 \\
2010ma & 10.33$\pm$4.34 & - \\
2011bm & 34.59$\pm$0.15 & - \\
2011dh & 15.71$\pm$0.02 & 15.25$\pm$1.65 \\
2011ei & 17.73$\pm$0.03 & 16.22$\pm$1.65 \\
2011hs & 8.59$\pm$0.06 & 12.21$\pm$1.65 \\
2011kl & 15.17$\pm$0.07 & 10.90$\pm$1.66 \\
2012ap & 13.19$\pm$0.31 & 13.69$\pm$2.12 \\
2012bz & 13.49$\pm$0.22 & 15.57$\pm$1.65 \\
2013cq & 13.00$\pm$2.00 & - \\
2013cu & 9.01$\pm$0.08 & 10.90$\pm$1.65 \\
2013df & 21.79$\pm$0.10 & 21.99$\pm$1.65 \\
2013dx & 12.26$\pm$5.48 & - \\
2013ge & - &18.88$\pm$1.70 \\
PTF10vgv & - &10.35$\pm$1.69 \\
PTF11bli & - &20.63$\pm$1.66 \\
PTF11jgj & - &22.07$\pm$2.17 \\
PTF11klg & - &16.39$\pm$1.68 \\
PTF12gzk & 16.35$\pm$0.46 & - \\
iPTF13bvn & 15.95$\pm$0.10 & 14.23$\pm$1.65 \\
iPTF14dby & - &22.46$\pm$1.68 \\
\hline
 \label{tpall}
\end{tabular}
\end{minipage}
\end{table}

\section{Explosion properties}
\subsection{The synthesis of $^{56}$Ni}
With the fully bolometric $L_{\mathrm{peak}}$ and $t_{\mathrm{p}}$, it is now possible to estimate the amount of $^{56}$Ni synthesised in the explosion. To obtain a value for $M_{\mathrm{Ni}}$, we used the formulation from \cite{SL2005} which is based upon ``Arnett's rule'' \citep{Arnett1982}: the approximation that the maximum luminosity of a SN powered by the decay of $^{56}$Ni is equal to the energy released by radioactive decay at that time:

\begin{equation}
\begin{split}
\frac{M_{\mathrm{Ni}}}{{\rm M}_\odot}= & L_{\mathrm{p}}\times\left(10^{43} \textrm{erg s}^{-1} \right)^{-1} \\ & \times \left(6.45\times e^{-t{\mathrm{p}}/8.8}+1.45\times e^{-t{\mathrm{p}}/111.3}\right)^{-1} 
\end{split}
\label{eq:Mni}
\end{equation}

\begin{figure}
\centering
\includegraphics[scale=0.4]{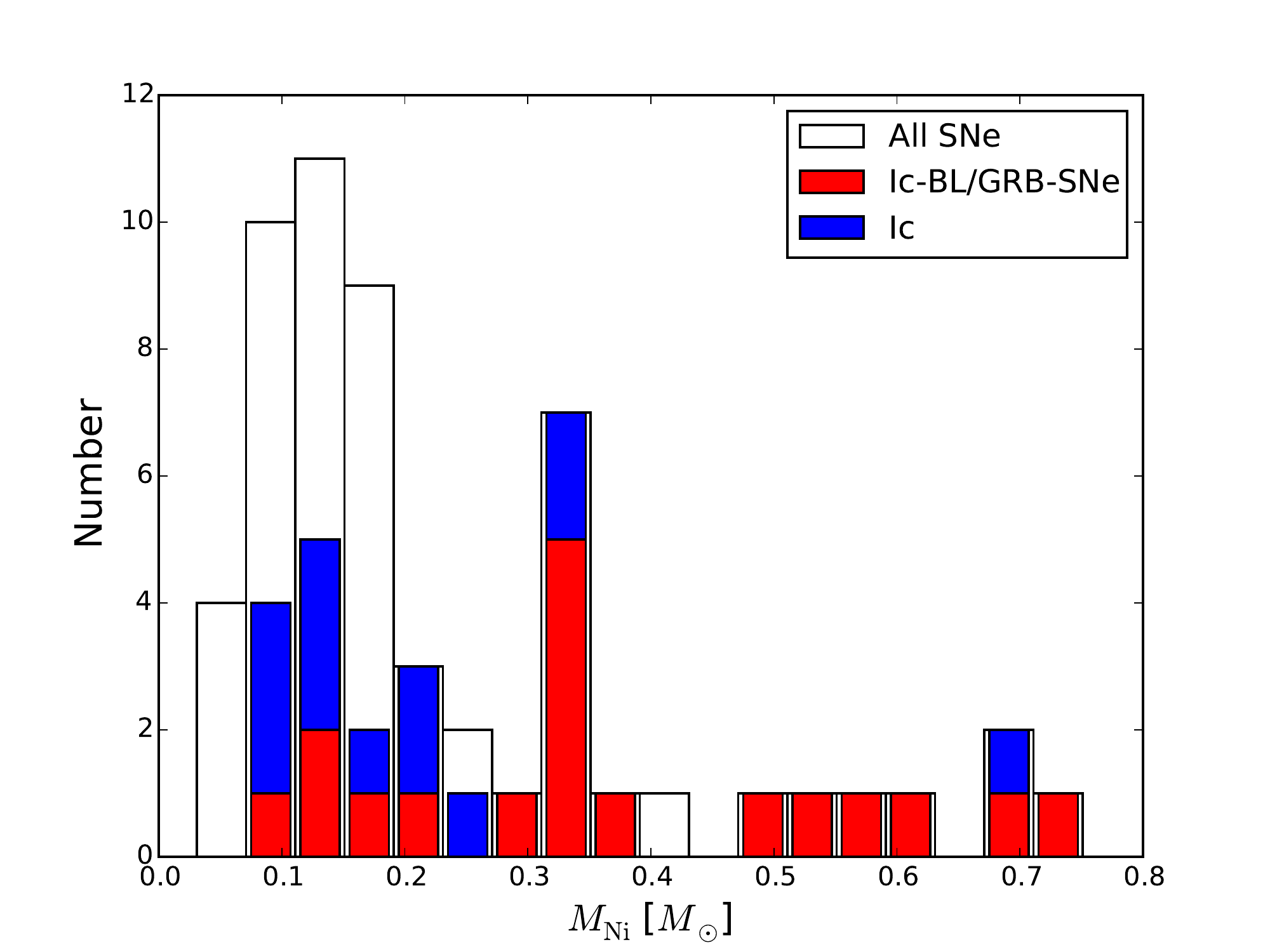}
\caption{The $^{56}$Ni distribution for the fully bolometric sample with the normal SNe~Ic and the SNe~Ic-BL/GRB-SNe in blue and red, respectively. Consistent with our procedure for the bulk analysis, corrections for median host-galaxy extinction have been applied. The bulk nickel-mass distribution, for SNe where the nickel mass can be derived, is shown in white. }
\label{fig:ONIRMNIc}
\end{figure}

\begin{figure}
\centering
\includegraphics[scale=0.4]{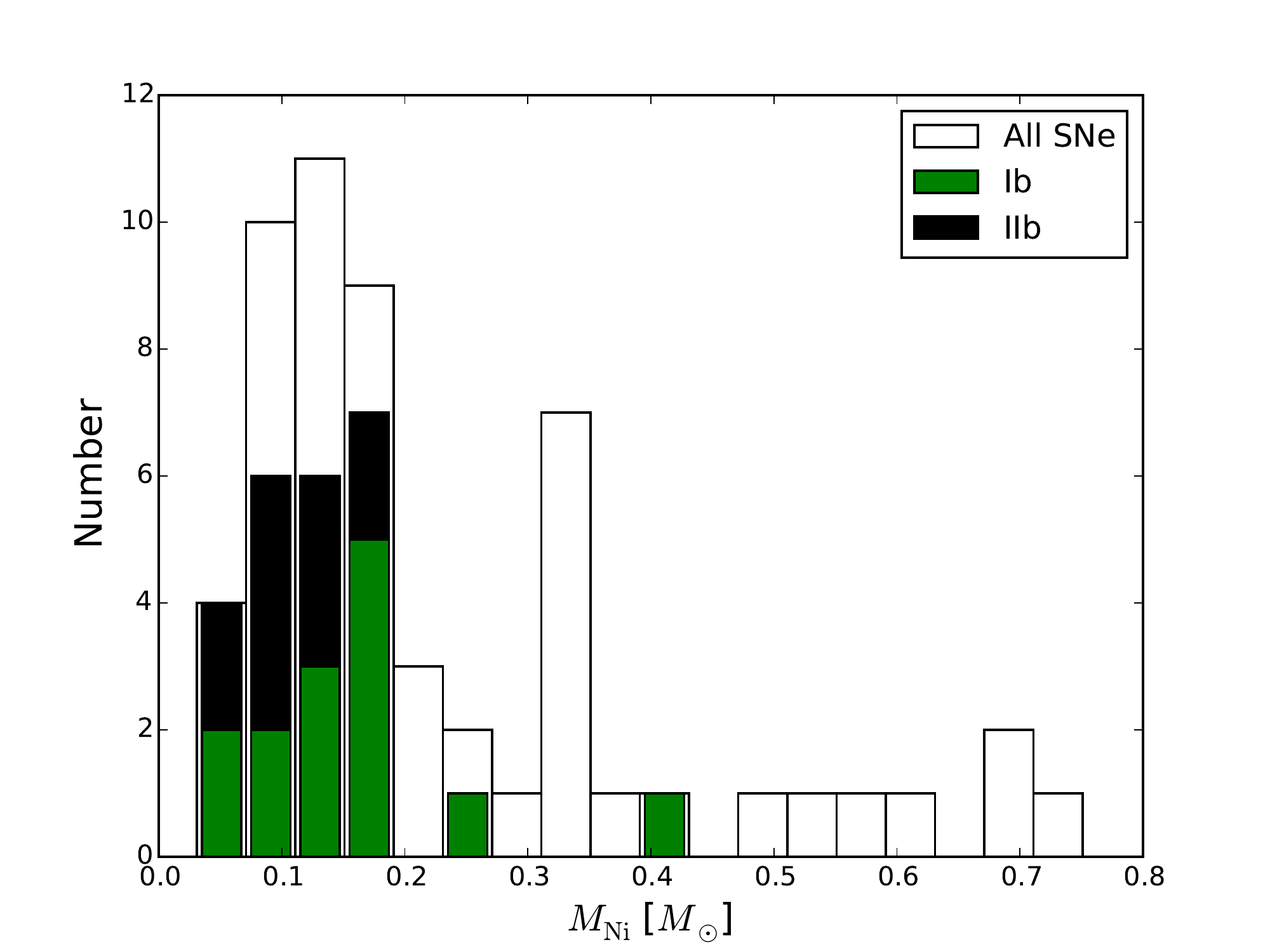}
\caption{The $^{56}$Ni distribution for the fully bolometric sample with the normal SNe~Ib and SNe~IIb in green and black, respectively. Corrections for median host-galaxy extinction have been applied. The white region is the same as in Figure~\ref{fig:ONIRMNIc}.}
\label{fig:ONIRMNIb}
\end{figure}

\begin{figure}
\centering
\includegraphics[scale=0.4]{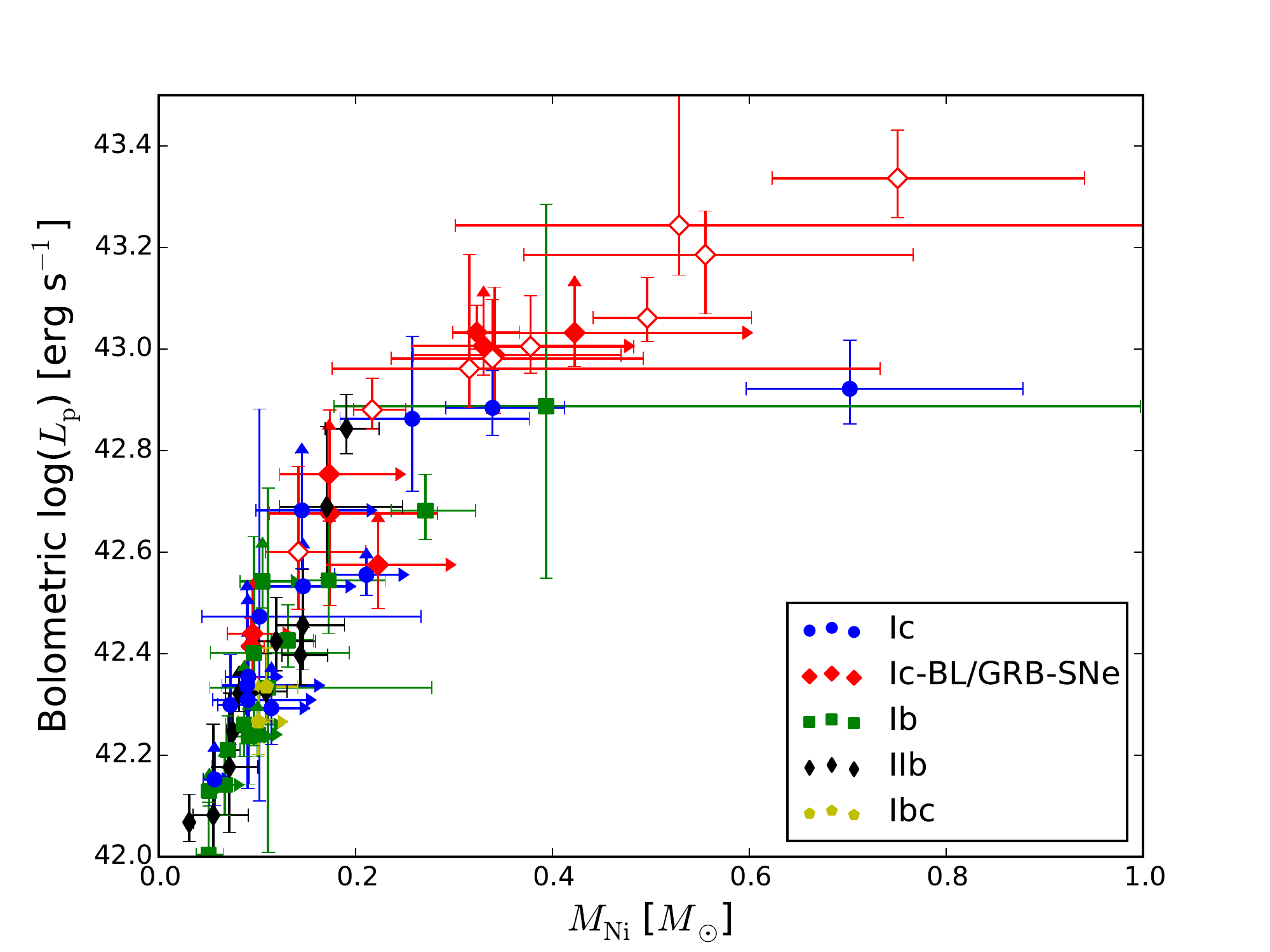}
\caption{Peak bolometric luminosity as a function of nickel mass. Open symbols represent GRB-SNe.}
\label{fig:LpMni}
\end{figure}

We use equation \ref{eq:Mni} to evaluate $M_{\mathrm{Ni}}$ for the \textit{BVRI}, \textit{UBVRI}NIR, and the fully bolometric sample. We adopt the exact value of $t_{\mathrm{p}}$ if known; failing that, we adopt the $t_{-1/2}$ to  $t_{\mathrm{p}}$ conversion (Equation \ref{fit2}) to reduce the propagation of uncertainties. The result of this is given in Table~\ref{fullstats}. The $M_{\mathrm{Ni}}$ distribution for the whole sample, highlighting SNe~Ic/Ic-BL and GRB-SNe, is shown in Figure~\ref{fig:ONIRMNIc}, whereas Figure~\ref{fig:ONIRMNIb} the whole sample emphasises SNe~Ib and SNe~IIb. The spread of synthesised $M_{\mathrm{Ni}}$ varies between SN types. The SN~Ic group shows the largest spread while the SN~Ib and SN~IIb populations cluster toward the lower end of the distribution. The normal (i.e., non-broad-lined) SNe~Ic also include the non-GRB-SN with the largest nickel mass, SN 2011bm, which is a consequence of its exceptionally broad light curve. Figure~\ref{fig:LpMni} gives peak bolometric luminosity as a function of $M_{\mathrm{Ni}}$ for the sample.

% All Mni with Ics

% All Mni with bs

% LP v Mni

\subsection{Characteristic timescales, kinetic energy, and ejecta mass }
The rise time of the SN is linked to the powering mechanism, the spatial distribution of $^{56}$Ni, the opacity of the ejecta ($\kappa$), the mass of the ejecta ($M_{\mathrm{ej}}$), and the photospheric velocity at luminosity peak $v_{\mathrm{ph}}$ via the formulation for the parameter $\tau_m$ given by \cite{Arnett1982},  
\begin{equation}
\tau_m=\sqrt{2}\left(\frac{k}{\beta c}\right)^{\frac{1}{2}}\left(\frac{M_{\mathrm{ej}}}{v_{\mathrm{ph}}}\right)^{\frac{1}{2}}.
\label{tau}
\end{equation}
For ejecta of uniform density undergoing spherically symmetric free expansion ($R(x,t) = v(x)t$), we can convert $v_{\mathrm{ph}}$ to $E_{\mathrm{k}}$ via Equation \ref{Ek},
\begin{equation}
E_{\mathrm{k}}=\frac{3}{10}M_{\mathrm{ej}}v_{\mathrm{ph}}^2,
\label{Ek}
\end{equation}
which leads to
\begin{equation}
\tau_m=\left(\frac{\kappa}{\beta c}\right)^{\frac{1}{2}}\left(\frac{6M_{\mathrm{ej}}^3}{5E_{\mathrm{k}}}\right)^{\frac{1}{4}},
\label{tauEK}
\end{equation}
where $\beta$ is a constant of integration with value of $\sim 13.7$. The parameter $\tau_m$, which defines the scale time of the light curve, is similar to the rise time of the SN but typically lower by $\sim 2$ days. However, by assuming that $t_{\mathrm{p}}\approx \tau_m$, we can estimate a value for $M_{\mathrm{ej}}^3/E_{\mathrm{k}}$ for a range of opacities; see Table~\ref{MoE}. Note that this model assumes that $\kappa$ is independent of time and constant across the ejecta --- but this is not true for any SN, especially SNe~Ib and SNe~IIb. In these cases the opacity of the He shell is extremely low owing to the paucity of He lines in the optical. Consequently, the value for $M_{\mathrm{ej}}$ derived from $\tau_m$ underestimates the total mass of the ejecta because the photons emitted from the photosphere diffuse through this shell with little overall interaction \citep[also see][]{Wheeler2015}. 

\begin{figure}
\centering
\includegraphics[scale=0.4]{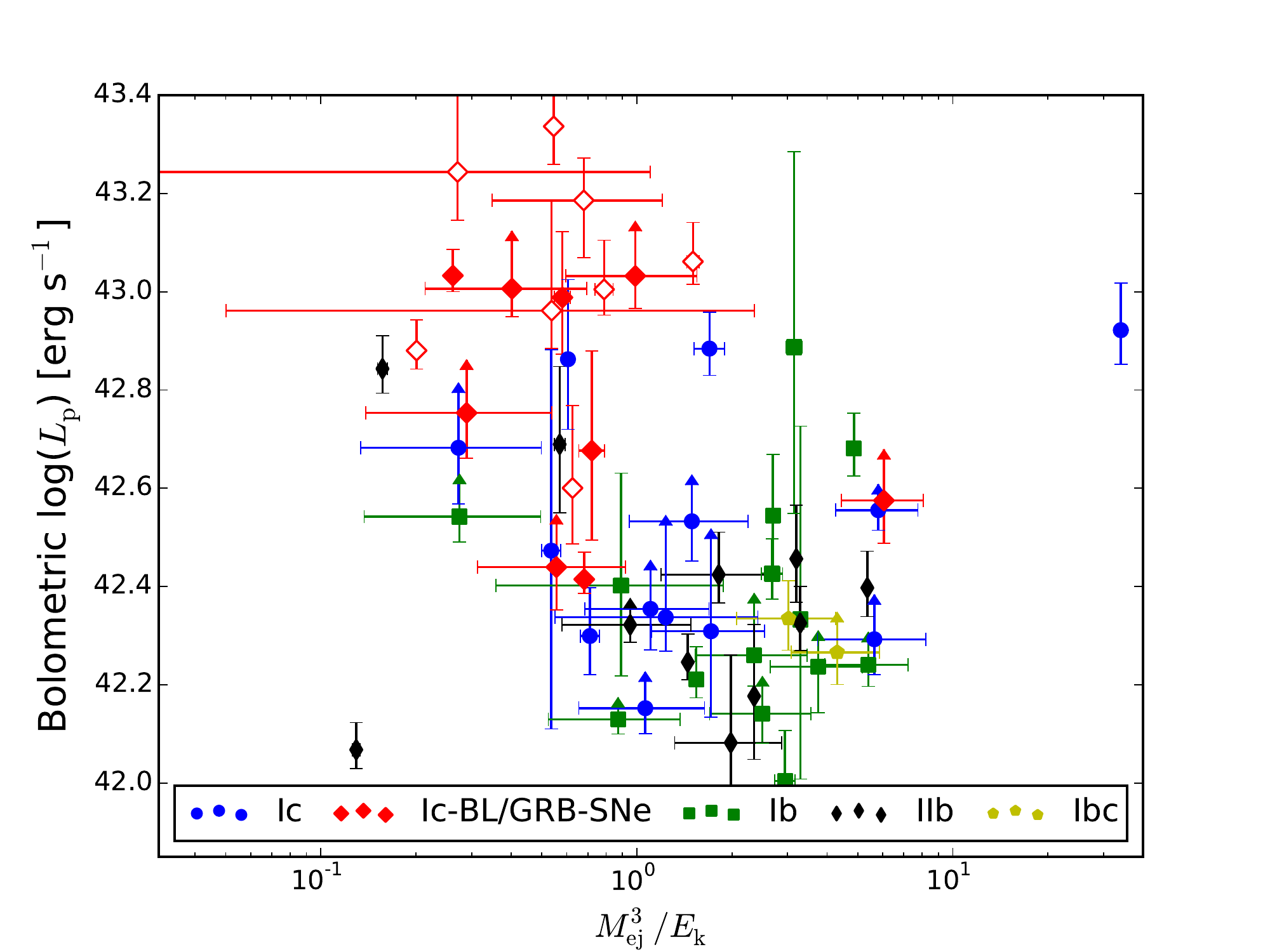}
\caption{Peak bolometric luminosity as a function of $M_{\mathrm{ej}}^3/E_{\mathrm{k}}$. Open symbols represent GRB-SNe.}
\label{fig:LpM3E}
\end{figure}

\begin{figure}
\centering
\includegraphics[scale=0.4]{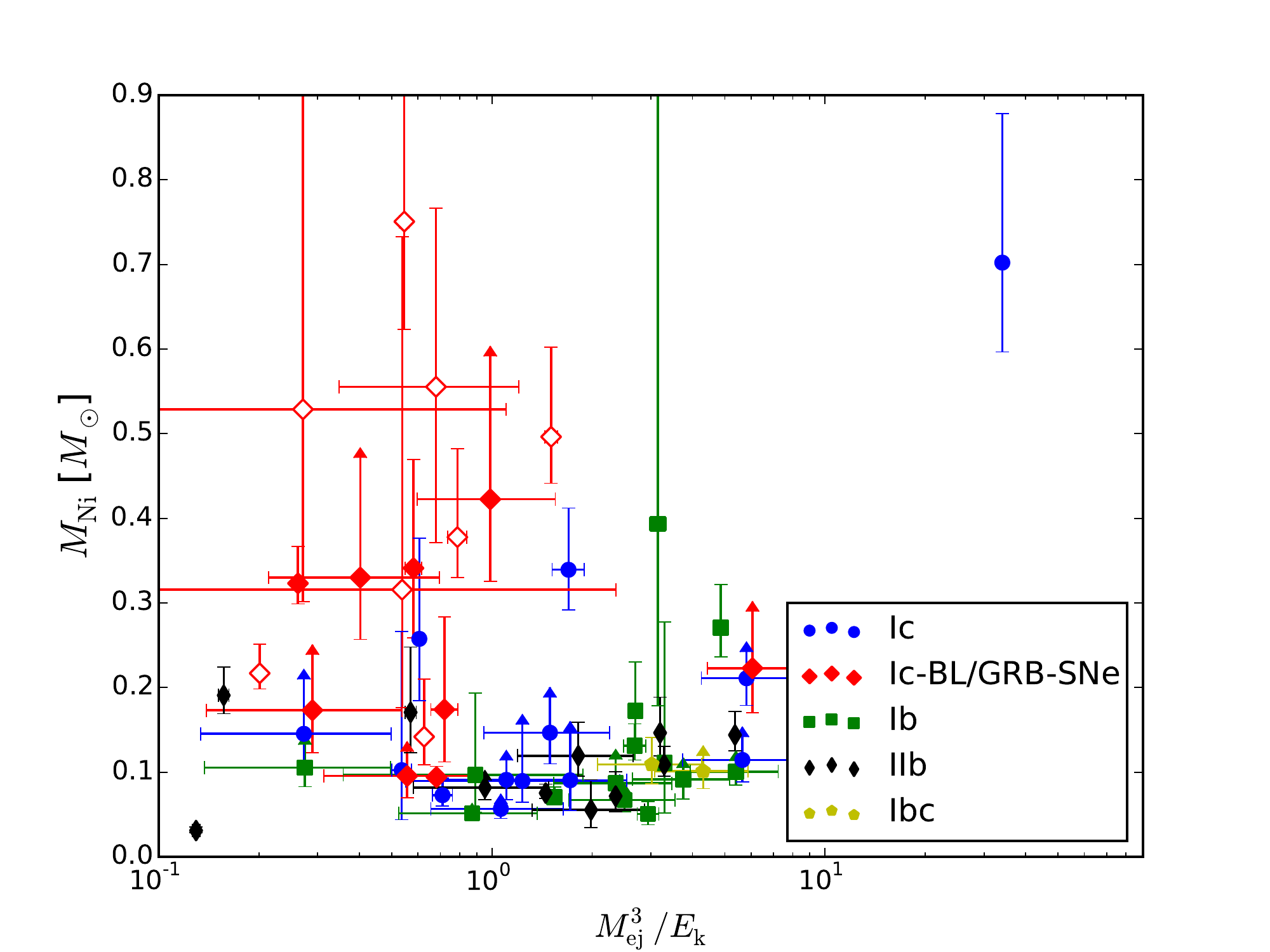}
\caption{Nickel mass as a function of $M_{\mathrm{ej}}^3/E_{\mathrm{k}}$. Open symbols represent GRB-SNe.}
\label{fig:MniM3E}
\end{figure}

We plot $L_{\mathrm{p}}$ and $M_{\mathrm{Ni}}$ as a function of $M_{\mathrm{ej}}^3/E_{\mathrm{k}}$ in Figures~\ref{fig:LpM3E} and \ref{fig:MniM3E}. There is considerable scatter in each plot, but for a given value of $L_{\mathrm{p}}$ or $M_{\mathrm{Ni}}$ the value of $M_{\mathrm{ej}}^3/E_{\mathrm{k}}$ tends to be higher for SNe~Ib and SNe~IIb compared with SNe~Ic. The width of SN 2011bm leads to a large value for $M_{\mathrm{ej}}^3/E_{\mathrm{k}}$, which in turn necessitated the use of a logarithmic abscissa.

% LP v M3/E

% Mni v M3/E

\section{Comparison with multiband photometry}
\subsection{Deriving multiband parameters}
The multiband photometry of each SN was fit using the same process as described in Section 5.3. The raw photometry was dereddened for Galactic extinction in the observer frame and host-galaxy extinction in the rest frame, where possible. 

\subsection{Colour curves}
The diversity of filters used in the observations implies that there is no single colour that can be used to represent every SN in the sample as is; this requires the use of some filter conversion to homogenise the sample. We utilise the colour corrections of \cite{Jordi2006} to convert \textit{B} and \textit{V} to $g'$ and \textit{R} to $r'$, so as to present the $g'-r'$ colour for all of the SNe. This is the most sensible choice for conversion because $g'$ lies between \textit{B} and \textit{V} while $r'$ lies blueward of \textit{R}, so the conversions fall between the available photometry. We note that this process is not perfect and stellar colour transformations can be poor for emission-line objects. However, it should be less of an issue during the photospheric phase ($\lesssim 40$ days) when the spectrum can be more closely approximated by a blackbody. 

As a test of the accuracy of this conversion we utilise SN PTF12gzk, which has photometry in \textit{UBVRIgriz}. We find that the transformed colour falls within the uncertainties of that determined directly from the photometry, giving confidence that this method is sufficiently accurate within the bounds of the photometric uncertainties.

We take the approach of plotting the colours of all SNe at $z<0.05$ with extinction corrections applied only when the reddening is known. The limit on distance is chosen so as to minimise the need for K-corrections, which become more important at $z>0.05$. This importance is amplified by the behaviour of $g'$ and $r'$, as they lie on opposite sides of the SED peak for most SNe until times much later than the peak of the bolometric LC. The behaviour of the K-correction is dependent on the underlying spectrum, the epoch, and the redshift, which can have a dramatic effect on the colours as $g'$ and $r'$ can be corrected in opposite directions. For example, at $\sim 10$ days past bolometric maximum and in the $z < 0.05$ regime, the K-correction in the $g'$ band is typically $\sim 0$ mag, but it begins to rise rapidly to $\sim 0.1$ mag as $z$ tends to 0.05. In the $r'$ band the behaviour is more complex, with the low-$z$ K-correction taking small positive or negative values but tending toward lower values as $z$ increases. At other times the behaviour can be different and more extreme, especially with increasing redshift. Hence, we limit this effect by restricting the sample to low $z$.  
%Hence, the general assumption that, around ten days past bolometric maximum, the flux at the effective wavelength of \textit{g'} will increase while at the effective wavelength of \textit{r'} it will decrease for increasing redshift, but not so large so as to push the peak of the SED blue-ward of $\lambda_{g'}$, is valid in most cases. 

\begin{figure*}
\centering
\includegraphics[scale=0.7]{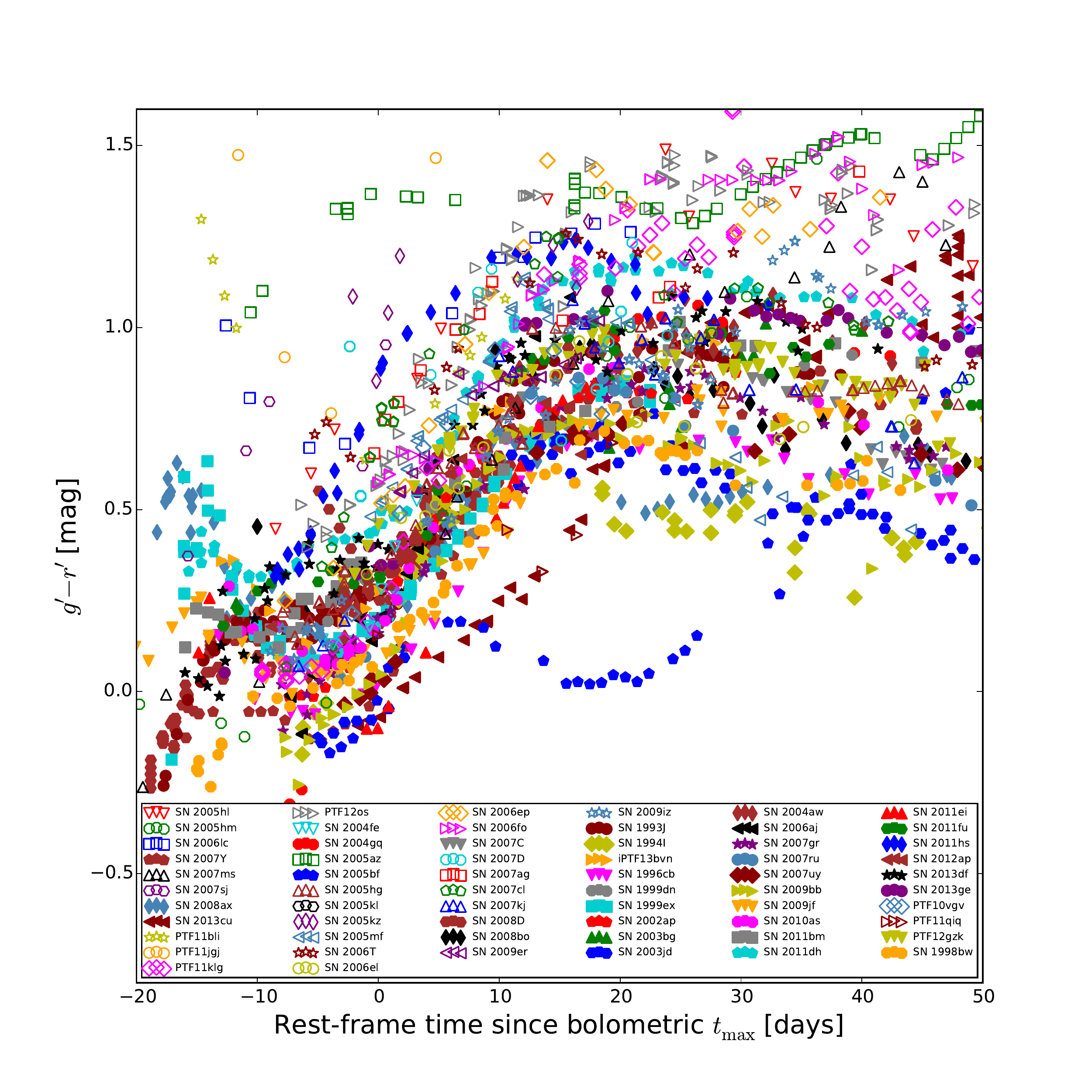}
\caption{$g'-r'$ colour of the SNe in the sample at $z<0.05$. Open markers indicate SNe where $E(B-V)_{\mathrm{host}}$ is not known. To see how correcting the colour curves for host extinction affects the spread, see Figure~\ref{fig:cc}.}
\label{fig:colourwith}
\end{figure*}

\begin{figure*}
\centering
\includegraphics[scale=0.7]{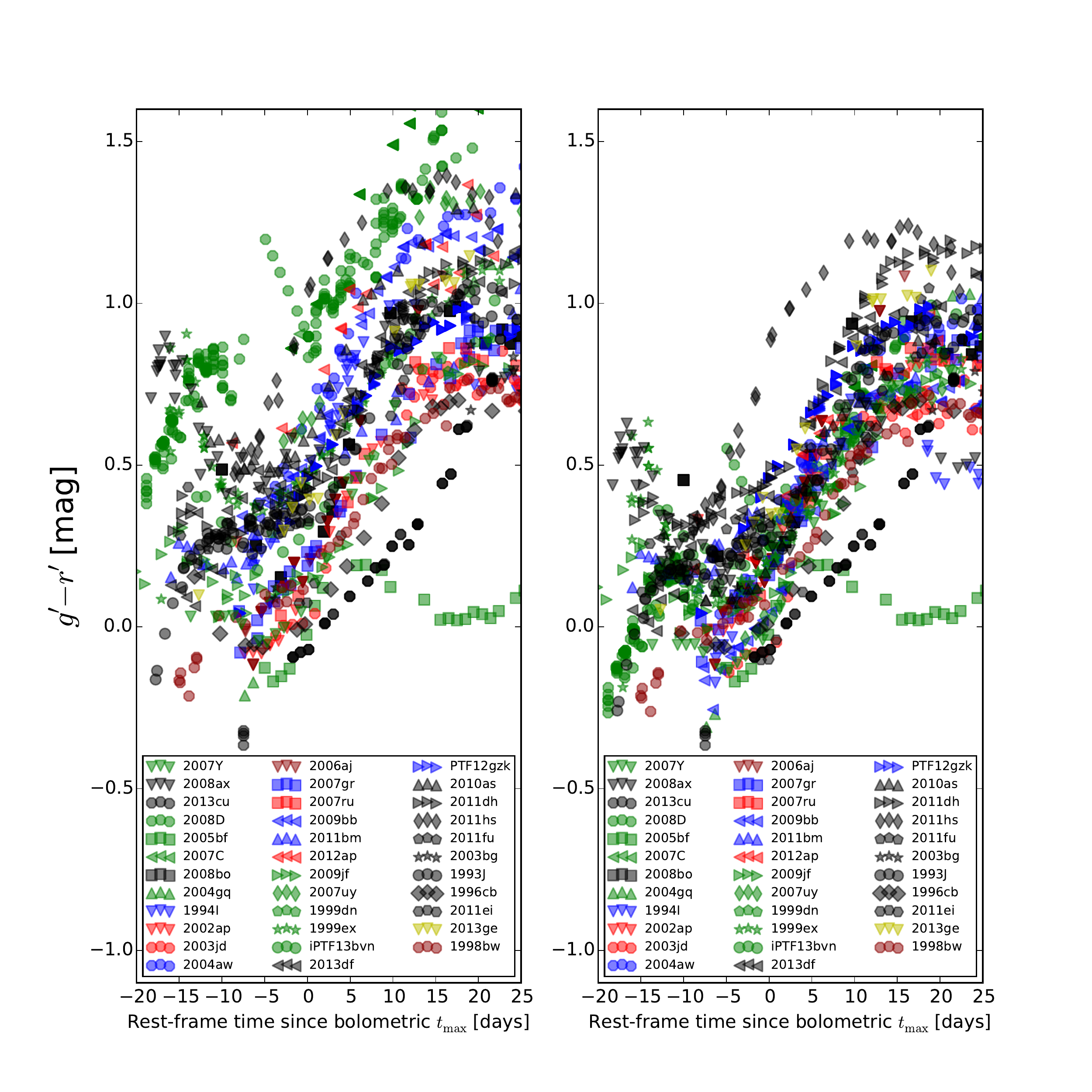}
\caption{$g'-r'$ colour curves of the SNe in the sample at $z<0.05$, sorted by spectral type, showing before (left) and after (right) correction for host extinction. SNe of Types Ic, Ic-BL, Ib, IIb, Ibc, and GRB-SNe are shown in blue, red, green, black, yellow, and dark red, respectively.}
\label{fig:cc}
\end{figure*}

Figure~\ref{fig:colourwith} shows the $g'-r'$ colour evolution over the photospheric phase; SNe without known host-galaxy extinction are plotted in transparent markers to separate them. When host extinction is applied the spread of colours narrows, and at $\sim 10$ days past bolometric peak it forms a ``bottleneck''; see Figure~\ref{fig:cc}. This behaviour is similar to that seen by \cite{Drout2011} in the $V-R$ colours of their sample, and was similarly observed for SNe with $g-r$ by \cite{Taddia2015}. The colour evolution of SN 2005bf is unique and appears to reflect the light curve, with the blue dips being approximately coincident with the luminosity peaks. If one was to take the second peak as the bolometric maximum, then the colour curve would shift to the left by 20 days and line up with the other SNe. Type IIb SN 2011hs, $E_{\mathrm{host}}(B-V) =0.16 \pm0.07$ mag \citep{Bufano2014}, is a notable outlier in Figure~\ref{fig:cc}. \cite{Bufano2014} find that $g'-r'$ is $\sim 0.6$ mag around the time of $g'$ maximum and there is a $\sim 0.5$ mag difference between the colours of SN 2011hs and SN 2008ax. The result here is comparable, with $g'-r' \approx 0.6$ mag in the few days before bolometric maximum and a similar difference between the two SNe, which is consistent with the period around $g'$ maximum. It may be that the host extinction has been underestimated in this case, or it could be indicating that not all SNe sit within the main distribution. 

\begin{figure*}
\centering
\includegraphics[scale=0.7]{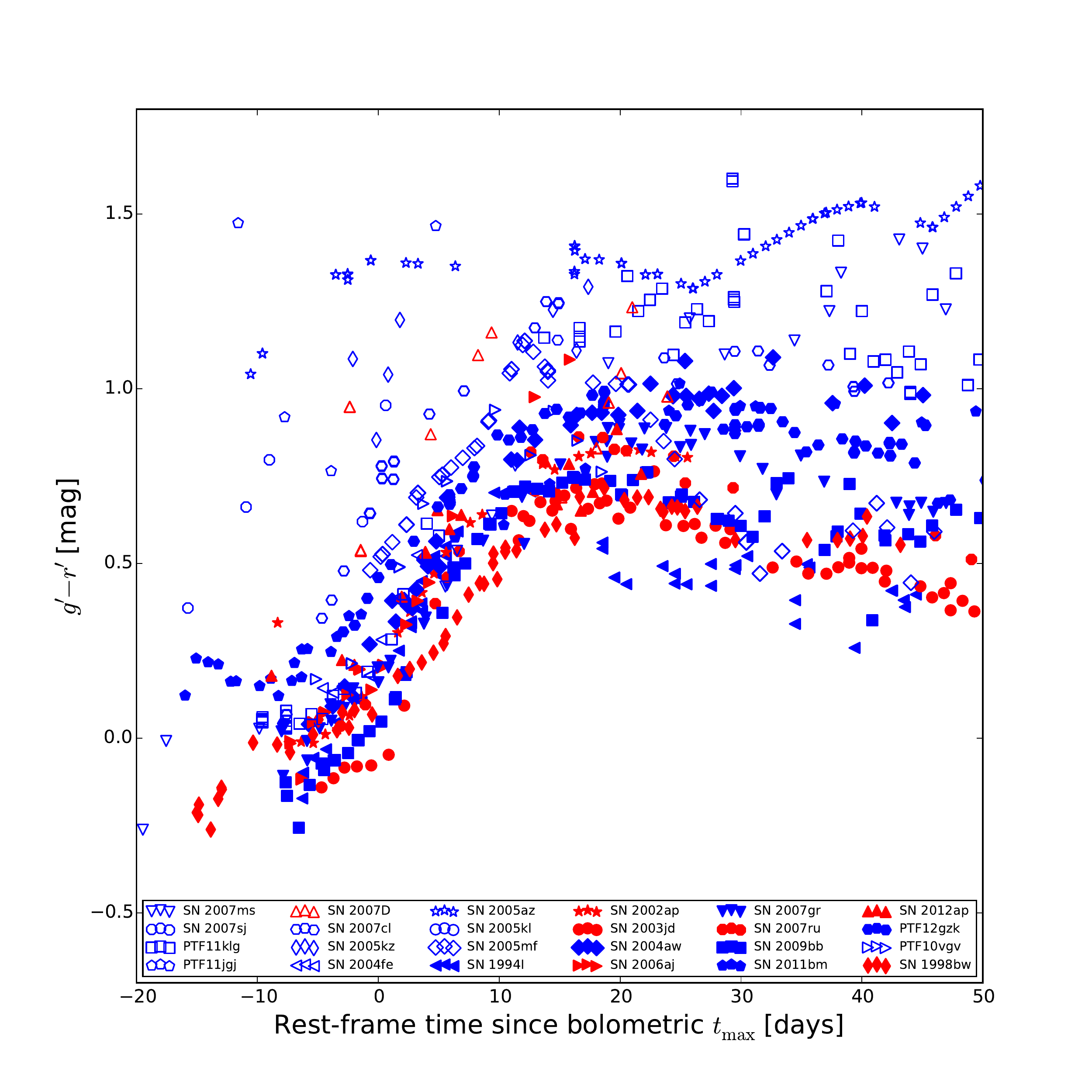}
\caption{$g'-r'$ colour of the SNe~Ic in the sample at $z<0.05$. Open markers indicate SNe where $E(B-V)_{\mathrm{host}}$ is not known.}
\label{fig:colourc}
\end{figure*}

\begin{figure*}
\centering
\includegraphics[scale=0.7]{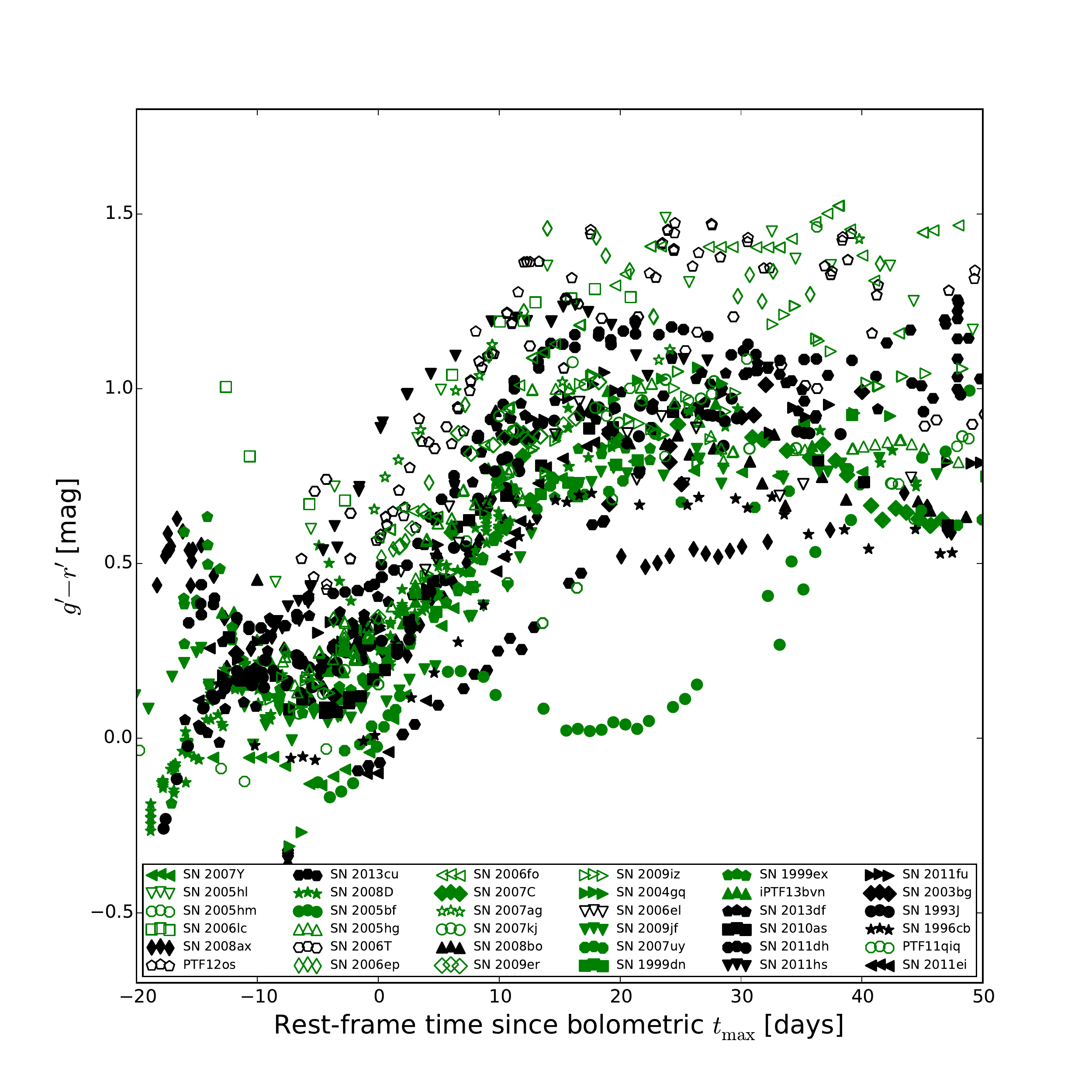}
\caption{$g'-r'$ colour of the SNe~Ib and SNe~IIb in the sample at $z<0.05$. Open markers indicate SNe where $E(B-V)_{\mathrm{host}}$ is not known.}
\label{fig:colourb}
\end{figure*}

Figures~\ref{fig:colourc} and \ref{fig:colourb} show the colour evolution of SNe split by spectral type for all SNe~Ic and SNe~Ib/IIb, respectively. The ``bottleneck'' is more readily apparent in the SN~Ic/Ic-BL/GRB-SN population than it is for SNe~Ib/IIb, providing further caution in attempting to use colours as a basis of determining host extinction. Note that for clarity, uncertainties are not shown in these figures.

% colour curves with host extinction

% colour curves with host extinction

% colour curves Ic

% colour curves Ib

\subsection{Comparing the multiband and bolometric peak time}
The time of peak of the photometry is a function of the effective wavelength of the band, with blue bands peaking earlier than red bands \citep{Taddia2015}. This can be explained by the evolution of the underlying spectrum and the cooling of the photosphere, although absorption features can affect the temporal evolution of the photometry. By using the photometry and interpolating the time of peak between the bands we can estimate the wavelength peak that coincides with $t_{\mathrm{p}}$. Table~\ref{timebandbol} gives the average wavelength where the photometry would peak at $t_{\mathrm{p}}$; it can be seen that the values are around that of the {\it V} band, $\lambda_{\mathrm{eff}}=5505$~\AA. For GRB-SNe the SED can be ``contaminated'' by afterglow flux, which is preferentially blue, leading to a bluer SED peak. Unfortunately, it is hard to account for GRB afterglow beyond simple empirical methods (see Section 3.2). For comparison, the spectra of GRB-SN 1998bw are red, broad-lined, and almost featureless owing to the comprehensive reprocessing of optical photons to lower frequencies by the high-energy ejecta. SN 1998bw is significant in this regard because its SED peak is redder than the median for GRB-SNe ($\sim 5200$~\AA) and the afterglow was negligible, so the photometry and spectrum are not contaminated in the way that (for example) GRB-SN 2013dx is.

% BC stats
\begin{table}
 \centering
 \begin{minipage}{80mm}
  \caption{Effective wavelength when photometry peak matches bolometric peak.}
 \begin{tabular}{lccc}
  \hline
 SN Type& Median $\mathrm{\lambda}$ (\AA) & $\left\langle\mathrm{\lambda}\right\rangle$ (\AA) & $\sigma$ (\AA)\\
  \hline
Ic-BL & 5449 &  $5369$ & $536$\\
GRB-SNe & 4869 & 5065 & 419\\
Ic & 5616 & 5452 & 826\\
Ib & 5416 & 5454 & 500\\
IIb & 5308 & 5277 & 334\\
\hline
 \label{timebandbol}
\end{tabular}
\end{minipage}
\end{table}

\subsection{Bolometric corrections}
In the era of large-scale surveys [e.g., PTF, iPTF, the Panoramic Survey Telescope and Rapid Response System (Pan-STARRS); the future Large Synoptic Survey Telescope (LSST) and the Zwicky Transient Facility (ZTF)], where the number of transients discovered has increased (and will continue to increase) by orders of magnitude, it is apparent that with a limited number of telescopes available, long-term multiband follow-up observations of single objects become more time consuming. Bolometric corrections are a useful way of approximating the bolometric luminosity from a single band or colour \citep{Lyman2014}, which in turn enables the determinations of the bolometric properties of the SN. To this end, we investigate the bolometric corrections (BCs) in single bands for our database. 

The BC is defined as 
\begin{equation}
M_{\mathrm{band}}-{\rm BC}=M_{\mathrm{bol}}.
\label{eq:BC}
\end{equation}
To estimate the BC, we subtract $M_{\mathrm{bol}}$ at the time of bolometric peak from $M_{\mathrm{band}}$ at the time of peak in that band and calculate a distribution.
The mean and standard deviation of each distribution is given in Table~\ref{BC}; the standard deviation is taken to be the error of the bolometric correction $\delta$BC. It is apparent that the band with the least scatter is \textit{R}, where $\delta BC=0.13$ mag, followed by $I$ ($\delta BC=0.18$ mag); $r'$ and $i'$ are less constrained, with $\delta BC=0.28$ mag and $0.27$ mag (respectively). This is an issue for calculating BCs via this method because it relies on the spectral differences of each SN being outside the band in question. If {\it R} ($\lambda_{\mathrm{eff}}=6580$~\AA) and $I$ ($\lambda_{\mathrm{eff}}=8060$~\AA) define such a region, then it should be expected that $i'$ ($\lambda_{\mathrm{eff}}=7630$~\AA) would also be a tracer for the bolometric luminosity, but the scatter is actually considerably greater for this band. The reason behind this could possibly be systematics (e.g., relatively poor photometry), but this would have to be applicable to a large number of observations across many years to avoid being lost in the noise of better observations.
 
There is insufficient evidence to suggest that any particular band is a superior tracer of the bolometric light curve at peak. Additionally, if one were to apply the BC to some SNe, the returned luminosity would be poorly constrained. This is because the standard deviation of the spread of residuals from the BC, which is taken to be the uncertainty in the resulting luminosity, corresponds to approximately a quarter of the value returned. Here we have yet to even consider errors in the input photometry. Given that the spread of SN luminosities typically ranges between $10^{42}$ erg s$^{-1}$ and $10^{43}$ erg s$^{-1}$, this uncertainty represents a significant fraction of the parameter space; hence, the BCs are not applied here.
% BC stats
\begin{table}
 \centering
 \begin{minipage}{70mm}
  \caption{Bolometric corrections derived from single-band photometry. }
 \begin{tabular}{@{}lccccccclcc@{}}
  \hline
 Band & Mean BC (mag) & $\sigma$ (mag) & $N$ \\
  \hline
$B$ & 0.19 & 0.27 & 40\\
$g'$ &  $-0.058$ & 0.33 & 12\\
$V$ & $-0.34$ & 0.21 & 40 \\
$r'$ & $-0.43$ & 0.28 & 27\\
$R$ & $-0.45$ & 0.13 & 25\\
$i'$ & $-0.38$ & 0.29 & 27\\
$I$ & $-0.51$ & 0.18 & 24\\

\hline
 \label{BC}
\end{tabular}
\end{minipage}
\end{table}

\section{Discussion}
\subsection{Biases}
It would be remiss to fail to appreciate the biases involved in the data used for this study. SNe were rare discoveries until the early/mid 1990s. Typically fewer than 20 per year were discovered, but as technology improved and more large-scale surveys were initiated, the number count increased from a few tens per year to the few hundred discovered each year now \citep[e.g.,][]{Galyam2013}. However, telescope time and funding are of limited supply; thus, the sample of SNe is affected in numerous ways. For example, many of the SNe are objects of interest; they display some unusual property that makes them targets for follow-up observations when they are spectroscopically confirmed, or they are nearby and bright. This is especially true of GRB-SNe and XRF-SNe, which are discovered only as a result of the detection of the high-energy transient event with which they are associated. Consequently, GRB-SNe and XRF-SNe are found at much greater distances that other SE-SNe and in different host environments. Some SNe are serendipitous discoveries in surveys designed for research in other areas, typically searches for SNe~Ia, so unless the SE-SN is an object of interest it is not monitored. Furthermore, if a SE-SN is discovered as a result of a targeted survey in which particular galaxy types are observed at regular intervals because they display a propensity for SNe, then we will miss events that occur in less optimal environments. The cost of observing transients over numerous bands has resulted in an increase of single-band observations of SNe. If spectra are available it may be possible to estimate the bolometric properties of the the SN by assuming its evolution is the same as that of a SN with a similar spectrum. Finally, observations favour SNe in less dusty regions of the host galaxy and those that are intrinsically luminous because their apparent brightness is greater. This leads to sampling of a larger comoving volume for more luminous SNe (Malmquist bias).

There are several consequences for our study. First, the luminosity functions are effectively luminosity functions for $z<0.1$ because only the GRB-SNe and some SNe~Ic-BL are sampled at higher redshifts. Second, the luminosity functions themselves may be overestimating the median luminosity if the nondetection of low-luminosity SNe is a significant issue. This is unquantifiable because rates of intrinsically dim SNe are unknown; however, a low-luminosity, nearby SN would be an object of interest (provided this is not a consequence of reddening). Our study shows that there appears to be a clear reduction in the number of SNe with log$(L_{\mathrm{p}}) \la 41.7$, yet there are clearly well-sampled SNe significantly below this luminosity. Given that the majority of these SNe are found at low redshift, and so would be observable at these luminosities, it appears that we are not missing intrinsically dim SNe in large numbers in the comoving volume we are typically sampling. It must also be considered that we do see SNe with relatively low ejecta and nickel masses; for example, SN 1994I was found to have $M_{\mathrm{ej}} <1 {\rm M}_{\odot}$ \citep{Iwamoto1994}, while SN 2007Y synthesised just $0.051 {\rm M}_{\odot}$ of $^{56}$Ni. This raises interesting questions of the SN mechanism and whether there is a lower limit on ejecta mass and nickel synthesis. Finally, we consider the effect of peculiar objects on our sample. This is generally minimised because peculiar objects (e.g., SN 2003bg, \citealt{Mazzali2009}; PTF12gzk, \citealt{BenAmi2012}; SN 2011bm, \citealt{Valenti2012}) do not constitute a significant proportion of each subtype, and if there are enough of them they become subtypes in their own right (e.g., SNe~Ic-BL, GRB-SNe).  We note that a good example of observational bias toward interesting objects is seen with the number of GRB-SNe that could be included in this study. We have 10 GRB-SNe that fulfil the criteria for inclusion, and this represents a significant fraction of those discovered.

\subsection{Explosion characteristics}
% full ONIR stats
\begin{table}
 \centering
 \renewcommand{\arraystretch}{1.5}
 \begin{minipage}{90mm}
  \caption{Median values for the fully bolometric sample. }
 \begin{tabular}{lcc}
  \hline
 SN Type & log$(L{\mathrm{p}})$ & $M_{\mathrm{Ni}}$ (M$_{\odot}$) \\
  \hline
Ic-BL/GRB-SNe & 43.00$\pm^{0.21}_{0.21}$ & 0.34$\pm^{0.13}_{0.19}$\\
Ic & 42.51$\pm^{0.06}_{0.36}$ & 0.16$\pm^{0.03}_{0.10}$\\
Ib & 42.50$\pm^{0.10}_{0.20}$ & 0.14$\pm^{0.04}_{0.04}$\\
IIb & 42.36$\pm^{0.26}_{0.11}$& 0.11$\pm^{0.04}_{0.04}$\\
\hline
 \label{meanstats}
\end{tabular}
\end{minipage}
\end{table}

The median population statistics and relevant 1$\sigma$ uncertainties for the fully bolometric sample are given in Table~\ref{meanstats}. As per Table~\ref{bollumfunc}, a hierarchy of peak luminosities is evident, with SNe~Ic-BL/GRB-SNe generally more luminous than the SN~Ic, IIb, and Ib populations. In terms of $M_{\mathrm{Ni}}$, we find that most types of SNe synthesise similar amounts of $^{56}$Ni with the exception of the broad-lined SNe~Ic, where the median value for $M_{\mathrm{Ni}}$ is more than double that of the others. Additionally, the median $t_{-1/2}$ is found to be shorter for these SNe compared to their less-energetic cousins. The degeneracy between $M_{\mathrm{ej}}$ and $E_{\mathrm{k}}$ in Equation~\ref{tauEK} means that without photospheric velocities, found from spectral modeling, it is not possible to tell whether the typically narrower light curves of the SNe~Ic is a consequence of smaller ejecta masses, more energetic explosions, or some amount of $^{56}$Ni in the outer ejecta. Ejecta with significant amounts of $^{56}$Ni mixed into the outer layers will rise more quickly as the diffusion time for the photons emitted in these regions will be less than that emitted more centrally. The models of \cite{Arnett1982} assume that all light-curve powering sources are located centrally; if $^{56}$Ni is located further out, then this model will return less-accurate values for the light-curve parameters.
\subsection{Photometry}
The scatter of light-curve colours at peak is unsurprising; there is no homogeneous explosion mechanism for SE-SNe as there is for SNe~Ia. The reason for the narrowing of the colour curves after $\sim 10$ days is harder to explain. Attempts have been made to use this as a way of deriving host-galaxy extinction \citep[e.g.,][]{Drout2011,Taddia2015}, but this method is sensitive to the underlying spectrum, K-corrections, the quality of the photometry, the colours used, and the type of SN. It may have most applicability for low-redshift SNe. 

The SNe~Ic-BL appear to have bolometric peak coincident with the SED peak to the blue side of \textit{V}, comparable to SNe~IIb which are known for their blue spectra. This cannot be accounted for in the scatter of the distribution, as it falls short of covering the effective wavelength of \textit{V}. A thorough investigation of this result cannot be undertaken with photometry and will require the use of spectra.

\section{Conclusions}
We have taken 85 SE-SNe from the literature and used the available photometry to build a set of optical pseudobolometric light curves and optical/NIR bolometric light curves. By using the same method and the same cosmological model, the database is as self-consistent as possible. The photometry was corrected for Galactic and host-galaxy extinction where such information was available. We found that fewer than 50\% of the SNe had known host extinction values, so we searched the literature for SNe~Ib/c and SNe~II with this information known. From these data a series of distributions was constructed, and the median host extinction for each SN type was found and used as an estimate to correct for reddening when deriving bulk statistics.
 
The light curves were then analysed to reveal the peak luminosity and various temporal properties which were used to investigate the characteristics of SE-SNe. The analysis revealed that $t_{-1/2}$ is very loosely correlated with $t_{+1/2}$ but showed significant scatter around this value. K-S tests of the cumulative distribution functions for the decay times of the SNe revealed that most were likely drawn from the same population with the exception of SNe~Ic/Ic-BL and SNe~Ic/IIb. However, the presence of the Type Ic SNe PTF11rka and 2011bm, which display broad light curves, and a relatively small sample size skew the SN~Ic CDF. Conversely, it is found that the SNe~Ib and IIb generally take longer to rise than the SN~Ic population. Using the equations of \cite{Arnett1982}, we can estimate the mass to kinetic energy ratio of the ejecta, which is related to the rise time. These properties are degenerate and can only be fully determined via spectral modeling; consequently, we make no comment on the size, mass, or structure of the progenitor star.

It was also found that, in the absence of a known explosion time, the rise time of a light curve could be estimated from $t_{-1/2}$. A comparison between the peak values of the \textit{UBVRI} pseudobolometric light curves and \textit{UBVRI}NIR light curves of those SNe having enough data to build both revealed that a tight correlation forms in the $L_{\mathrm{p,}UBVRI}$--$L_{\mathrm{p,}UBVRI\mathrm{NIR}}$ parameter space that is independent of SN type, allowing the conversion of one value to the other. These relationships were then used to produce bolometric statistics for nearly all of the SN database.

It is shown that SNe~Ic-BL and GRB-SNe occupy the upper part of the SE-SN luminosity function, with SNe~IIb at the bottom. SNe~Ic and SNe~Ib show similar median peak luminosities. Using an approximation of ``Arnett's law,'' the amount of $^{56}$Ni synthesised in the core collapse of the stars that eventually go on to form the SN~Ic-BL population is on average twice that of SNe~Ic, Ib, and IIb. This is partly driven by the fact that all GRB-SNe are broad-lined and luminous. 

The colour curves of the multiband photometry in the sample were analysed. The peak magnitude, time to peak, $t_{-1/2}$, and $t_{+1/2}$ were calculated for each band and compared with those of the bolometric light curve. To determine the colour curves and the colour at maximum, for each SN \textit{BgVRr} photometry was converted to $g'$ and $r'$ where necessary. Our results confirm that there is evidence of a narrowing in the spread of $g'-r'$ for SNe with Milky Way and host-galaxy extinction corrections applied at $\sim 10$ days past bolometric maximum, though there is still a large range of possible values in this region. It is shown that the approximate wavelength of temporally coincident peaks between the multiband photometry and the bolometric LC occurs around the peak of the $V$ band ($\lambda_{\mathrm{eff}}=5505$~\AA) for all but GRB-SNe, which are blueward of this. The photometry allowed us to investigate the possibility of using single-band observations and a BC to derive the bolometric parameters. The smallest spread in values was found for the \textit{R} and \textit{I} bands, but this was not replicated with $i'$, which has an effective wavelength between these two Johnsons-Cousins filters. The uncertainties involved would lead to poorly constrained luminosities; thus, we reject the notion of using a single band as a proxy for the bolometric light curve at peak. 

The importance of knowing the host-galaxy extinction cannot be understated. The peak luminosity of a SN, and all of the subsequent characteristics derived from that, depend on knowing this property. As such, we suggest that future work involving SE-SNe place a high priority on calculating or estimating the extinction at the source. This could be done through medium/high-resolution spectra and analysis of the \ion{Na}{I}~D absorption lines, via some method involving the colour evolution, or preferably both. If the colour method is used it is imperative that K-corrections are included. It is appreciated that these methods are not without their problems, however, and further analysis of the extinction in the environments of SE-SNe is needed.

Finally, each subtype presented here suffers from small-number statistics; hence, we identify the need for well-sampled SE-SN follow-up observations, particularly in volume-limited surveys, in order to improve the statistics and enhance our understanding of the evolutionary paths of massive stars.

\section{Appendix}

\section*{Acknowledgements}
This research used resources of the National Energy Research Scientific Computing Center, a DOE Office of Science User Facility supported by the Office of Science of the U.S. Department of Energy under Contract No. DE-AC02-05CH11231. 
A.V.F.'s research was funded by NSF grant AST-1211916, the TABASGO Foundation, and the Christopher R. Redlich Fund. A.G.Y. is supported by the EU/FP7 via ERC grant no. 307260, "The Quantum Universe" I-Core program by the Israeli Committee for planning and budgeting and the ISF; by Minerva and ISF grants; by the Weizmann-UK “making connections” program; and by Kimmel and YeS awards.
%The Acknowledgements section is not numbered. Here you can thank helpful
%colleagues, acknowledge funding agencies, telescopes and facilities used etc.

%%%%%%%%%%%%%%%%%%%%%%%%%%%%%%%%%%%%%%%%%%%%%%%%%%



\bibliographystyle{mnras}
\bibliography{Bol_bib} % if your bibtex file is called example.bib

%\begin{thebibliography}{99}
%\end{thebibliography}

%%%%%%%%%%%%%%%%%%%%%%%%%%%%%%%%%%%%%%%%%%%%%%%%%%

%%%%%%%%%%%%%%%%% APPENDICES %%%%%%%%%%%%%%%%%%%%%

\appendix

\section{Host-galaxy extinction}

%%%%%%%%%%%%%%%%%%%%%%%%%%%%%%%%%%%%%%%%%%%%%%%%%%

We searched the literature for core-collapse SNe with defined host-galaxy extinction in order to determine the distribution for the sample as a whole or individually. In addition to the SNe used in Table~\ref{database}, we have also included SNe from \cite{Cano2013}, \cite{Pritchard2014}, \cite{Cano2014A}, \cite{Richardson2009}, \cite{Richardson2006}, and other individual SNe from the literature which failed the criteria described in Section 2.1. We also include all SNe~II on the basis that, while the evolution of these objects may be different from that of SE-SNe, their position in the host is likely to be similar given their lifetimes (i.e., in star-forming regions). The total number used was 110.
\subsubsection{K-S tests}
We performed K-S tests on the sample on a type-by-type basis. If the K-S test revealed that the distribution was drawn from the same population by having $P > 0.05$, then the populations were combined. SNe~IIP and SNe~IIL were combined into a single population owing to their small numbers. The interplay between the populations means that combining similar datasets is a risky task. For example, the SN~IIb versus SN~Ib K-S test returns $P = 0.055$, which is on the edge of the $5\sigma$ limit, yet SN~IIb against SN~Ic returns $P = 0.38$, and SN~Ic against SN~Ib returns $P = 0.487$; thus, populations were combined together only if the chance of the two being drawn from the same distribution was greater than 40\% for all permutations of the individuals within the group. This limit is high but necessary to reduce the risk in combining datasets into a larger group that may have a low $P$ value between some of its constituents. 

The clear outlier is the distribution given by GRB-SNe, where K-S tests give $P<0.05$ for every population. This does not imply that GRB-SN hosts are different from those of other SNe; the distribution of extinction values can be explained as a combination of distance and the luminosity function of SNe. We find that GRB-SNe, while more luminous on average than other SN types, are still limited to $L_{\mathrm{peak}} \approx 10^{43}$ erg s$^{-1}$ in bolometric luminosity and are typically observed at redshifts much greater than other SNe. As a result, the sample is biased in two ways. First, these SNe are observed at larger distances because we recover them from GRBs; without this added high-energy component few, if any, of these SNe would be seen. Second, the limiting luminosity and distance mean that only the least-attenuated SN light will be observed, so if a GRB exploded in a dusty environment and had a corresponding SN, we would not see it as it would be below current detection limits. Furthermore, if we consider the entire sample, the non-Gaussian nature of all the distributions may be expected but their observed shape is determined by similar constraints. The probability that a SN is detected is reduced for higher local extinctions and lower intrinsic luminosities because these lower its apparent brightness. Hence, the distribution is more likely to show SNe with low host extinction, which in turn biases the result in favour of SNe in ``clean'' environments. 

Thus, we return four populations:
\begin{itemize}
	\item{SNe~IIP, IIL, II, IIb}
	\item{SNe~Ib}
	\item{SNe~Ic, Ic-BL}
	\item{GRB-SNe}.
\end{itemize}

% Population stats
\begin{table}
 \centering
 \renewcommand{\arraystretch}{1.5}
 \begin{minipage}{100mm}
  \caption{Reddening [$E(B-V)$, mag] statistics of the populations. }
 \begin{tabular}{@{}lccccccclcc@{}}
  \hline
 SN Population & $N$ & median & mean  \\
  \hline
IIP, IIL, II, IIb, IIn& 25 & 0.081 $\pm_{0.026}^{0.1298}$& 0.130\\
Ib & 22 & 0.2235	 $\pm_{0.157}^{0.305}$ & 0.30  \\
Ic, Ic-BL & 27 & 0.174 $\pm_{0.116}^{0.187}$ &	0.244 \\
GRB-SN & 22& 0.0325 $\pm_{0.024}^{0.00635}$ & 0.040  \\
\hline
 \label{redstats}
\end{tabular}
\end{minipage}
\end{table}

\begin{figure}
\centering
\includegraphics[scale=0.4]{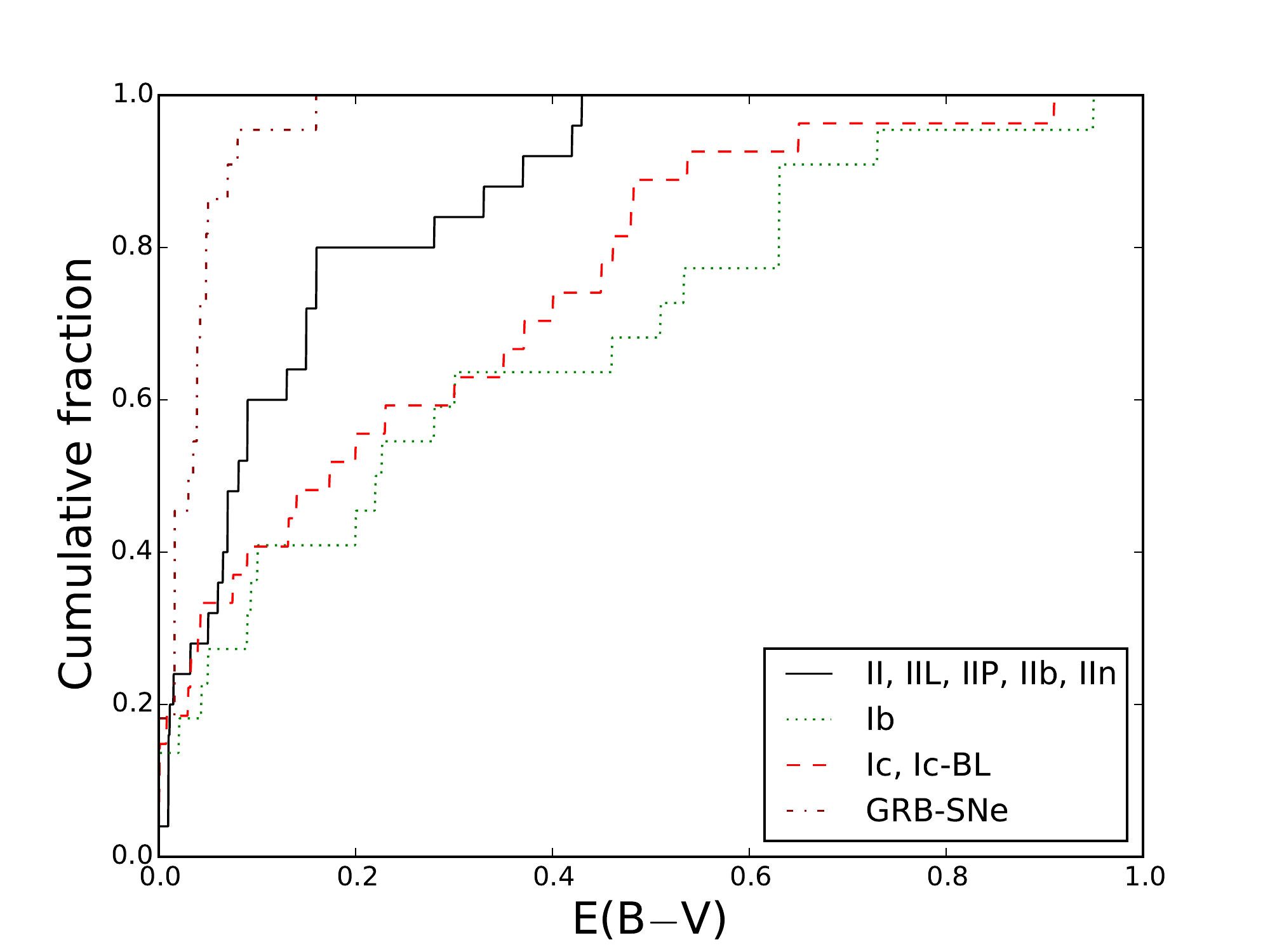}
\caption{Cumulative distribution of reddening values for SNe sorted by population.}
\label{fig:cumdist1}
\end{figure}

The median, upper and lower 34\%, mean, and standard deviation for each population are given in Table~\ref{redstats}. The discrepancy in median and mean is caused by the non-Gaussian distribution, so we use the median extinction as a typical extinction for SNe of that type and the upper and lower 34\% as the $1\sigma$ boundaries. This median extinction and associated uncertainty is then applied to any SN without determined host extinction. For an individual SN such a correction could be deemed to be misleading; however, for a large dataset with several SNe utilising this method, we rely on ``regression to the mean'' --- that is, for each SN where the extinction is underestimated, there will be a corresponding SN with extinction overestimated, and the uncertainties in this range provide a balance in terms of numbers centred on the average (the median in this case). The cumulative distributions are shown in Figure~\ref{fig:cumdist1}.

\begin{figure}
\centering
\includegraphics[scale=0.4]{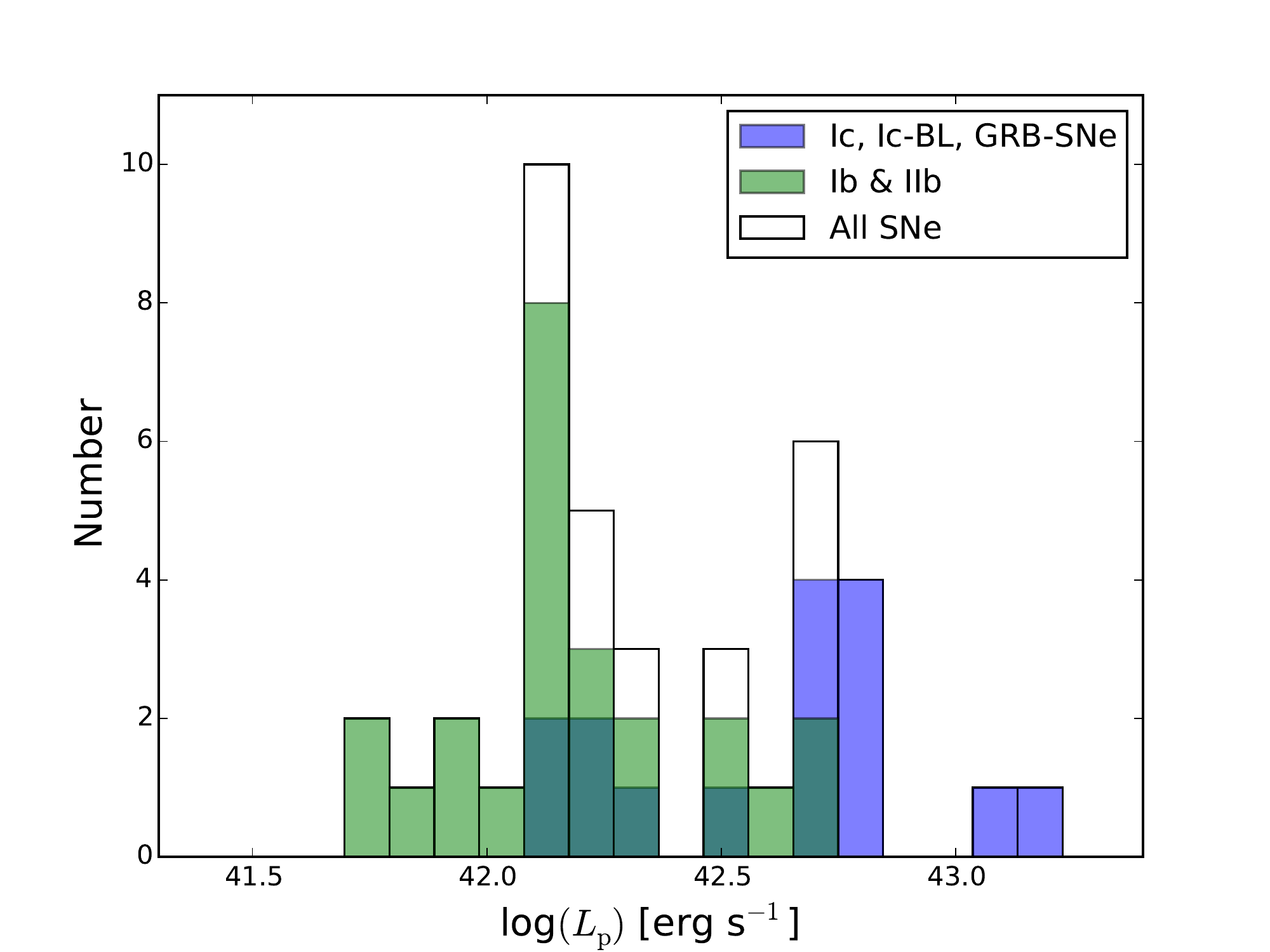}
\caption{\textit{BVRI} luminosity function for SNe with their host extinction correction applied using the values given in the literature. Colours are as described in Figure~\ref{fig:BVRIlumfunc}.}
\label{fig:test1}
\end{figure}

\begin{figure}
\centering
\includegraphics[scale=0.4]{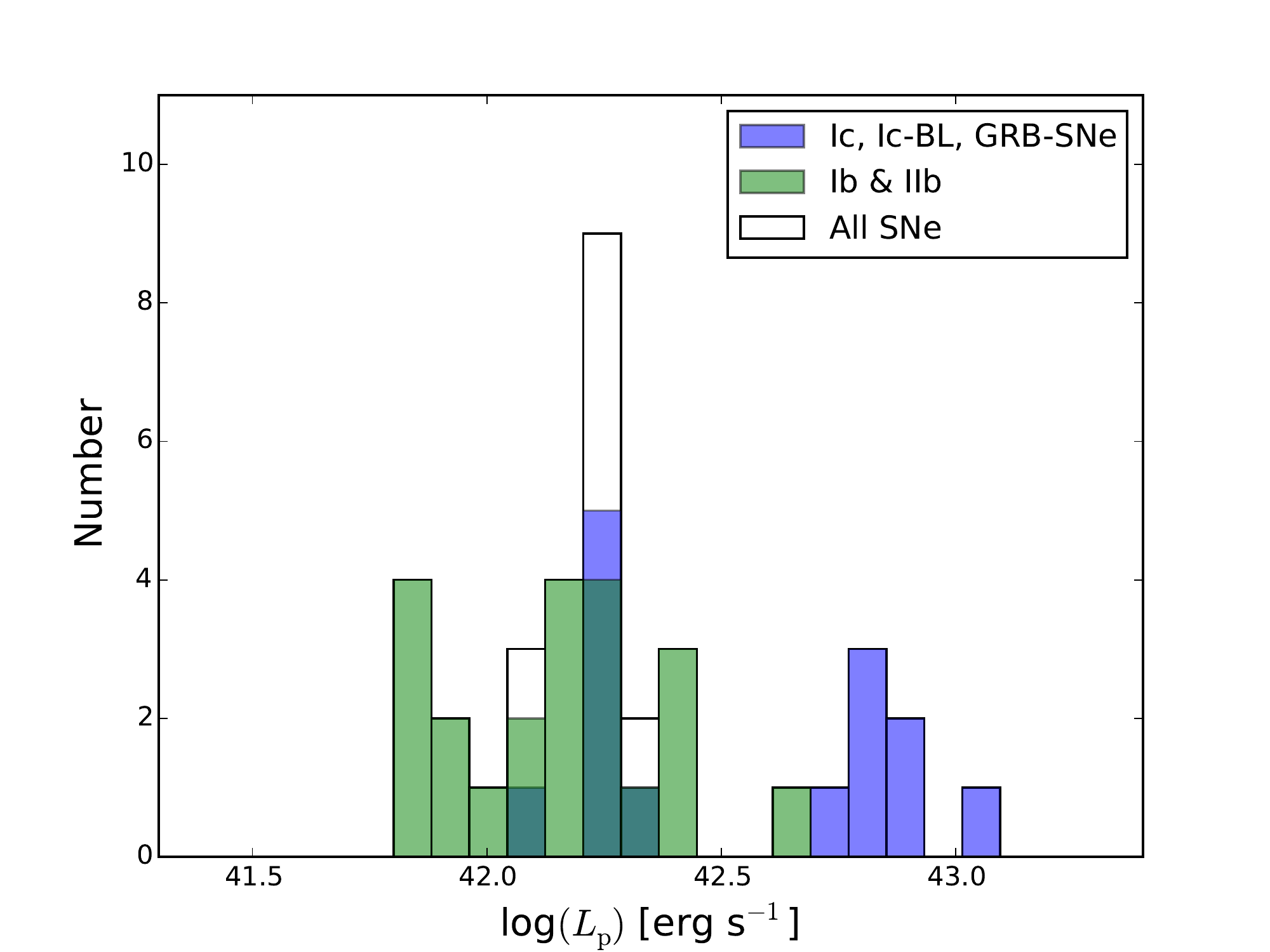}
\caption{\textit{BVRI} luminosity function of the same SNe as in Figure~\ref{fig:test1} but with a type-dependent host extinction correction applied. Colours are as described in Figure~\ref{fig:BVRIlumfunc}.}
\label{fig:test2}
\end{figure}

\begin{figure}
\centering
\includegraphics[scale=0.4]{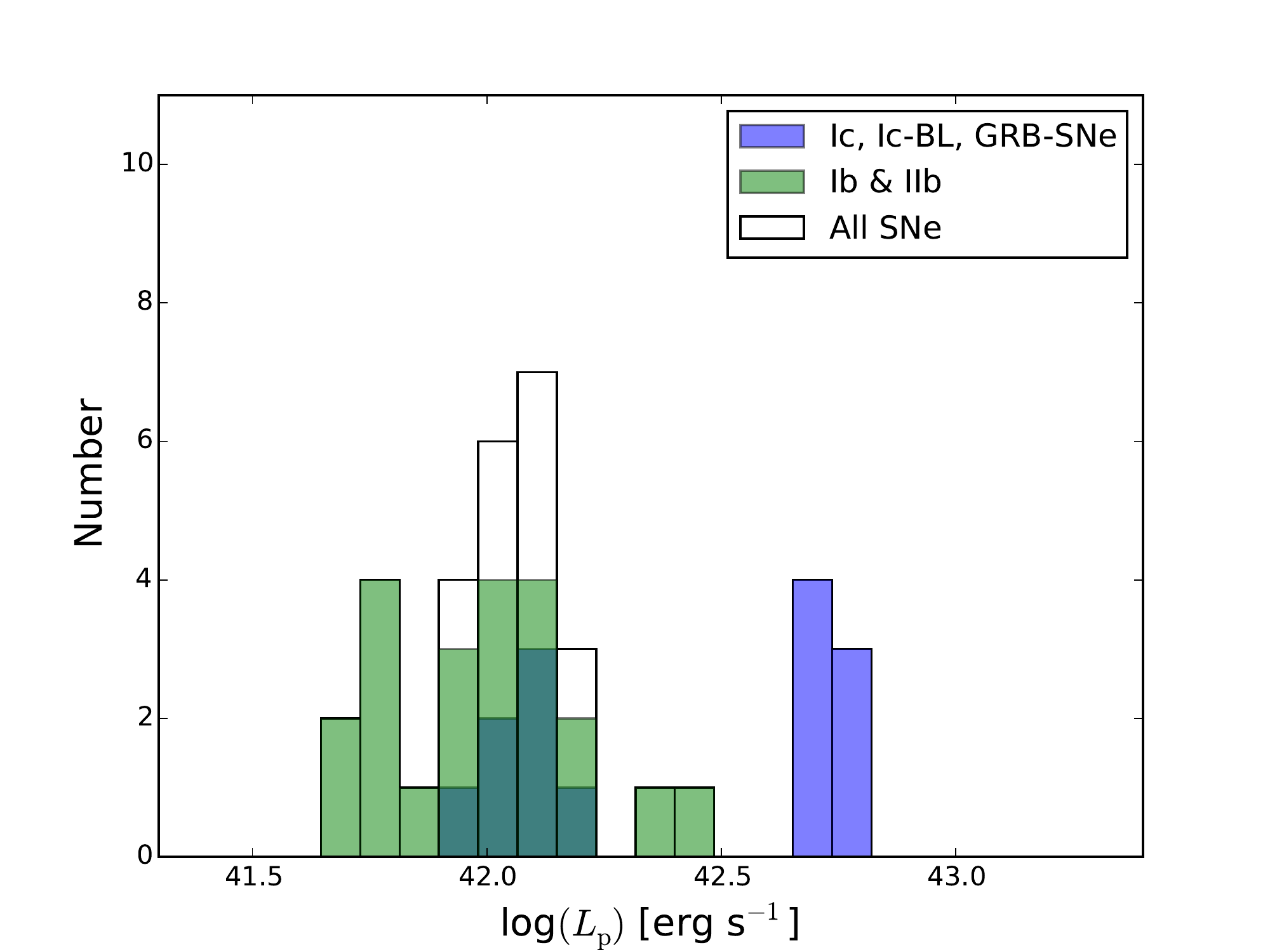}
\caption{\textit{BVRI} luminosity function of the same SNe as in Figure~\ref{fig:test1} but with no host extinction correction applied. Colours are as described in Figure~\ref{fig:BVRIlumfunc}.}
\label{fig:test3}
\end{figure}

To test the accuracy of using this method, we take the SNe from the database with known host-galaxy extinction (SNe~Ic/Ic-BL/GRB-SNe = 15, SNe~Ib/IIb = 22) and apply the median extinction correction when constructing the bolometric light curve. This is then compared with the case when no host extinction correction is applied and when the actual extinction is applied. The results are shown in Figure~\ref{fig:test2} for the application of median host extinction, Figure~\ref{fig:test1} for the actual values, and Figure~\ref{fig:test3} for no correction. It is apparent that the character of the luminosity function changes when median $E(B-V)_{\mathrm{host}}$ is used in place of the literature value; this is to be expected around the extremes of the distribution because no SN is being corrected for a large host extinction but all are being corrected. The statistics returned in Table~\ref{comparison} show that the mean and median of the luminosity functions remain similar, giving confidence in the conversion method to return bulk statistics. However, in the absence of $E(B-V)_{\mathrm{host}}$ the statistics returned are generally lower than the corrected values; again, this would be expected as they represent a lower limit. It is interesting to note that the SN~Ic median values are very similar for no extinction and the known $E(B-V)_{\mathrm{host}}$, although this is not reflected in the mean value.

%\begin{figure}
%\centering
%\includegraphics[scale=0.4]{cum_dist0.pdf}
%\caption{Cumulative distribution of reddening values for SNe sorted by population.}
%\label{fig:cumdist2}
%\end{figure}
%
%\begin{figure}
%\centering
%\includegraphics[scale=0.4]{cum_dist1.pdf}
%\caption{Cumulative distribution of reddening values for SNe sorted by population.}
%\label{fig:cumdist3}
%\end{figure}

\begin{table}
	\centering
	\caption{Host extinction tests on peak-luminosity statistics.}
	\begin{tabular}{lccc}
	\hline
	\multicolumn{4}{c}{Literature extinction values}\\
	SN Type & median & mean & $\sigma$ \\
	\hline
	All Ic & 42.69 & 42.59 & 0.32 \\
	Ib/IIb & 42.15 & 42.17 & 0.24 \\
    bulk & 42.23 & 42.34 & 0.34 \\
    \hline
    \multicolumn{4}{c}{Median extinction values}\\
	SN Type & median & mean & $\sigma$ \\
	\hline
	All Ic & 42.72 & 42.61 & 0.38 \\
	Ib/IIb & 42.17 & 42.15 & 0.23 \\
    bulk & 42.24 & 42.34 & 0.38 \\
    \hline
    \multicolumn{4}{c}{No extinction correction}\\
	SN Type & median & mean & $\sigma$ \\
	\hline
	All Ic & 42.67 & 42.47 & 0.4 \\
	Ib/IIb & 42.01 & 41.99 & 0.2 \\
    bulk & 42.11 & 42.12 & 0.38 \\
    \hline
	\end{tabular}
	\label{comparison}
\end{table}

\section{Tables}

\begin{table*}
\renewcommand{\arraystretch}{1.5}
 \centering
 \begin{minipage}{160mm}
  \caption{\textit{BVRI} pseudobolometric light curve statistics.}
 \begin{tabular}{lccccccc}
  \hline
 SN & Type & log$(L{\mathrm{p}})$ &$M_{\mathrm{Ni}}$ (M$_{\odot}$) &$t_{\mathrm{p}}$ (days) & $t_{-1/2}$ (days) & $t_{+1/2}$ (days) & Width (days)  \\
  \hline

1993J & IIb & 42.231$\pm^{0.068}_{0.066}$ & 0.087$\pm^{0.015}_{0.012}$ & 19.148$\pm{0.033}$ & 10.313$\pm 0.017$ & 12.808$\pm 0.116$ & 23.121$\pm 0.118$\\
1994I & Ic & 42.271$\pm^{0.346}_{0.328}$ & 0.064$\pm^{0.080}_{0.034}$ & 12.252$\pm{0.211}$ & 5.541$\pm 0.137$ & 8.522$\pm 0.509$ & 14.063$\pm 0.527$\\
1996cb & IIb & 41.871$\pm^{0.123}_{0.123}$ & 0.034$\pm^{0.015}_{0.011}$ & 16.993$\pm{1.650}$ & 10.737$\pm 0.000$ & 15.444$\pm 0.017$ & 26.181$\pm 0.017$\\
1998bw & GRB-SN & 42.780$\pm^{0.031}_{0.023}$ & 0.260$\pm^{0.022}_{0.016}$ & 15.861$\pm{0.177}$ & 9.734$\pm 0.372$ & 15.858$\pm 0.779$ & 25.593$\pm 0.863$\\
1999dn & Ib & 42.134$\pm^{0.084}_{0.061}$ & 0.052$\pm^{0.023}_{0.015}$ & 13.916$\pm{2.841}$ & - & 21.827$\pm 3.638$ & -\\
1999ex & Ib & 42.335$\pm^{0.082}_{0.082}$ & 0.106$\pm^{0.022}_{0.019}$ & 18.349$\pm{0.036}$ & 9.226$\pm 0.053$ & 15.958$\pm 0.606$ & 25.183$\pm 0.608$\\
2002ap & Ic-BL & 42.217$\pm^{0.010}_{0.006}$ & 0.060$\pm^{0.001}_{0.001}$ & 13.011$\pm{0.000}$ & 6.466$\pm 0.060$ & 15.506$\pm 0.165$ & 21.972$\pm 0.175$\\
2003bg & IIb & 42.226$\pm^{0.027}_{0.014}$ & 0.076$\pm^{0.012}_{0.009}$ & 16.624$\pm{1.656}$ & 10.504$\pm 0.146$ & 28.374$\pm 1.318$ & 38.878$\pm 1.326$\\
2003jd & Ic-BL & 42.795$\pm^{0.090}_{0.089}$ & 0.219$\pm^{0.054}_{0.042}$ & 12.504$\pm{0.179}$ & 9.372$\pm 0.505$ & 14.067$\pm 0.317$ & 23.439$\pm 0.596$\\
2004aw & Ic & 42.448$\pm^{0.030}_{0.044}$ & - & - & - & 21.273$\pm 1.340$ & -\\
2004fe* & Ic & 42.152$\pm^{0.031}_{0.030}$ & 0.057$\pm^{0.010}_{0.009}$ & 14.680$\pm{1.653}$ & 9.276$\pm 0.101$ & - & -\\
2004gq & Ib & 42.087$\pm^{0.010}_{0.006}$ & - & - & - & 18.582$\pm 0.465$ & -\\
2005az* & Ic & 41.981$\pm^{0.013}_{0.015}$ & - & - & - & 27.282$\pm 0.146$ & -\\
2005bf & Ib & 42.170$\pm^{0.025}_{0.023}$ & 0.073$\pm^{0.006}_{0.005}$ & 18.325$\pm{0.351}$ & - & - & -\\
2005hg* & Ib & 42.233$\pm^{0.010}_{0.013}$ & - & - & - & 17.943$\pm 0.237$ & -\\
2005hl* & Ib & 41.993$\pm^{0.008}_{0.005}$ & - & - & - & 21.656$\pm 0.667$ & -\\
2005hm* & Ib & 41.958$\pm^{-0.013}_{0.041}$ & 0.048$\pm^{0.002}_{0.008}$ & 19.927$\pm{1.669}$ & 12.591$\pm 0.250$ & 14.682$\pm 1.455$ & 27.274$\pm 1.476$\\
2005kl* & Ic & 41.485$\pm^{0.008}_{0.008}$ & - & - & - & 20.115$\pm 0.027$ & -\\
2005kr* & Ic-BL & 42.759$\pm^{0.072}_{0.038}$ & 0.187$\pm^{0.059}_{0.035}$ & 11.405$\pm{1.669}$ & 7.207$\pm 0.252$ & - & -\\
2005ks* & Ic-BL & 42.291$\pm^{0.066}_{0.078}$ & 0.068$\pm^{0.020}_{0.017}$ & 12.368$\pm{1.656}$ & 7.815$\pm 0.142$ & - & -\\
2005kz* & Ic & 42.004$\pm^{0.034}_{0.031}$ & - & - & - & - & -\\
2005mf* & Ic & 42.121$\pm^{0.033}_{0.024}$ & - & - & - & 17.281$\pm 0.493$ & -\\
2006T* & IIb & 42.145$\pm^{0.011}_{0.011}$ & 0.054$\pm^{0.007}_{0.007}$ & 14.148$\pm{1.651}$ & 8.940$\pm 0.051$ & 14.433$\pm 0.031$ & 23.372$\pm 0.059$\\
2006aj & GRB-SN & 42.685$\pm^{0.020}_{0.019}$ & 0.138$\pm^{0.007}_{0.006}$ & 9.587$\pm{0.040}$ & 6.779$\pm 0.155$ & - & -\\
2006el* & IIb & 42.041$\pm^{0.024}_{0.016}$ & - & - & - & 15.594$\pm 0.660$ & -\\
2006ep* & Ib & 41.942$\pm^{0.012}_{0.013}$ & 0.042$\pm^{0.005}_{0.005}$ & 18.002$\pm{1.653}$ & 11.375$\pm 0.101$ & 15.034$\pm 0.013$ & 26.409$\pm 0.102$\\
2006fe* & Ic & 42.263$\pm^{0.025}_{0.026}$ & 0.079$\pm^{0.013}_{0.012}$ & 15.830$\pm{1.711}$ & 10.002$\pm 0.455$ & 19.197$\pm 1.431$ & 29.199$\pm 1.502$\\
2006fo* & Ib & 42.164$\pm^{0.019}_{0.019}$ & - & - & - & 19.017$\pm 0.124$ & -\\
14475* & Ic-BL & 42.558$\pm^{0.047}_{0.048}$ & 0.110$\pm^{0.029}_{0.024}$ & 10.501$\pm{1.760}$ & 6.635$\pm 0.613$ & 18.037$\pm 1.403$ & 24.672$\pm 1.531$\\
2006jo* & Ib & 42.371$\pm^{0.020}_{0.020}$ & 0.071$\pm^{0.013}_{0.011}$ & 10.366$\pm{1.651}$ & 6.550$\pm 0.067$ & 10.081$\pm 0.472$ & 16.630$\pm 0.476$\\
2006lc* & Ib & 41.954$\pm^{0.005}_{0.008}$ & 0.034$\pm^{0.004}_{0.004}$ & 13.837$\pm{1.651}$ & 8.743$\pm 0.050$ & 14.036$\pm 0.050$ & 22.779$\pm 0.071$\\
2006nx* & Ic-BL & 42.847$\pm^{0.051}_{0.021}$ & 0.276$\pm^{0.067}_{0.040}$ & 14.279$\pm{1.698}$ & 9.022$\pm 0.401$ & - & -\\
2007C & Ib & 42.588$\pm^{0.085}_{0.084}$ & - & - & - & 12.373$\pm 0.186$ & -\\
2007D* & Ic-BL & 42.121$\pm^{0.085}_{0.082}$ & - & - & - & - & -\\
2007Y & Ib & 41.750$\pm^{0.064}_{0.072}$ & 0.028$\pm^{0.005}_{0.005}$ & 18.757$\pm{0.350}$ & 9.284$\pm 0.569$ & 15.327$\pm 0.285$ & 24.610$\pm 0.636$\\
2007ag* & Ib & 41.884$\pm^{0.015}_{0.014}$ & - & - & - & 20.052$\pm 0.673$ & -\\
2007cl* & Ic & 42.077$\pm^{0.010}_{0.010}$ & - & - & - & 15.616$\pm 0.259$ & -\\
2007gr & Ic & 42.087$\pm^{0.042}_{0.043}$ & 0.045$\pm^{0.005}_{0.005}$ & 13.146$\pm{0.229}$ & 8.530$\pm 0.025$ & 15.270$\pm 0.081$ & 23.800$\pm 0.085$\\
2007kj* & Ib & 42.017$\pm^{0.009}_{0.008}$ & - & - & - & - & -\\
2007ms* & Ic & 42.319$\pm^{0.008}_{0.017}$ & 0.122$\pm^{0.011}_{0.013}$ & 22.231$\pm{1.666}$ & 14.047$\pm 0.231$ & 24.625$\pm 0.434$ & 38.672$\pm 0.491$\\

\hline
 \label{BVRIstats}
\end{tabular}
\end{minipage}
\end{table*}
\begin{table*}
\renewcommand{\arraystretch}{1.5}
 \centering
 \begin{minipage}{160mm}
 \contcaption{\textit{BVRI} pseudobolometric light curve statistics.}
 \begin{tabular}{lccccccc}
  \hline
  SN & Type & log$(L{\mathrm{p}})$ &$M_{\mathrm{Ni}}$ (M$_{\odot}$) &$t_{\mathrm{p}}$ (days) & $t_{-1/2}$ (days) & $t_{+1/2}$ (days) & Width (days) \\
  \hline
2007nc* & Ib & 42.065$\pm^{0.064}_{0.016}$ & 0.055$\pm^{0.015}_{0.007}$ & 17.734$\pm{1.800}$ & 11.205$\pm 0.718$ & 19.059$\pm 4.382$ & 30.264$\pm 4.440$\\  
2007qv* & Ic & 42.514$\pm^{0.023}_{0.022}$ & - & - & - & - & -\\
2007qx* & Ic & 42.118$\pm^{0.197}_{0.041}$ & 0.054$\pm^{0.045}_{0.013}$ & 15.094$\pm{2.758}$ & 9.537$\pm 2.211$ & - & -\\
2007ru & Ic-BL & 42.810$\pm^{0.038}_{0.009}$ & 0.194$\pm^{0.018}_{0.005}$ & 10.242$\pm{0.047}$ & - & 14.640$\pm 0.485$ & -\\
2007sj* & Ic & 41.961$\pm^{0.022}_{0.021}$ & 0.036$\pm^{0.006}_{0.005}$ & 14.542$\pm{1.658}$ & 9.188$\pm 0.164$ & - & -\\
2007uy & Ib & 42.675$\pm^{0.270}_{0.260}$ & 0.241$\pm^{0.214}_{0.110}$ & 19.075$\pm{0.284}$ & - & 15.360$\pm 1.160$ & -\\
2008D & Ib & 42.125$\pm^{0.252}_{0.240}$ & 0.069$\pm^{0.055}_{0.030}$ & 19.293$\pm{0.229}$ & 13.239$\pm 0.194$ & 17.042$\pm 0.722$ & 30.281$\pm 0.747$\\
2008ax & IIb & 42.099$\pm^{0.025}_{0.024}$ & 0.065$\pm^{0.004}_{0.004}$ & 19.283$\pm{0.127}$ & 10.144$\pm 0.085$ & 15.279$\pm 0.318$ & 25.423$\pm 0.329$\\
2008bo & IIb & 41.757$\pm^{0.011}_{0.011}$ & - & - & - & 9.240$\pm 0.010$ & -\\
2008hw & GRB-SN & 43.156$\pm^{0.050}_{0.050}$ & 0.496$\pm^{0.064}_{0.057}$ & 12.307$\pm{0.100}$ & - & - & -\\
2009bb & Ic & 42.685$\pm^{0.123}_{0.121}$ & 0.171$\pm^{0.057}_{0.042}$ & 12.631$\pm{0.095}$ & 6.637$\pm 0.059$ & 13.060$\pm 0.357$ & 19.697$\pm 0.362$\\
2009er* & Ib & 42.724$\pm^{0.013}_{0.012}$ & - & - & - & 14.225$\pm 0.049$ & -\\
2009iz* & Ib & 42.085$\pm^{0.010}_{0.010}$ & 0.070$\pm^{0.006}_{0.006}$ & 21.828$\pm{1.650}$ & 13.793$\pm 0.000$ & 23.220$\pm 0.169$ & 37.013$\pm 0.169$\\
2009jf & Ib & 42.478$\pm^{0.022}_{0.027}$ & 0.169$\pm^{0.010}_{0.011}$ & 21.267$\pm{0.164}$ & 11.212$\pm 0.444$ & 19.901$\pm 0.631$ & 31.113$\pm 0.771$\\
2010as & IIb & 42.524$\pm^{0.088}_{0.087}$ & 0.117$\pm^{0.027}_{0.022}$ & 12.442$\pm{0.122}$ & 9.661$\pm 0.024$ & 16.541$\pm 0.415$ & 26.202$\pm 0.415$\\
2010bh & GRB-SN & 42.365$\pm^{0.053}_{0.053}$ & 0.082$\pm^{0.011}_{0.010}$ & 12.737$\pm{0.099}$ & 3.287$\pm 0.303$ & 9.286$\pm 0.884$ & 12.573$\pm 0.934$\\
2010ma & GRB-SN & 43.060$\pm^{0.234}_{0.071}$ & 0.346$\pm^{0.444}_{0.136}$ & 10.331$\pm{4.338}$ & - & 12.450$\pm 0.482$ & -\\
2011bm & Ic & 42.705$\pm^{0.031}_{0.037}$ & 0.427$\pm^{0.033}_{0.036}$ & 34.586$\pm{0.151}$ & - & 43.446$\pm 1.544$ & -\\
2011dh & IIb & 42.084$\pm^{0.008}_{0.012}$ & 0.052$\pm^{0.001}_{0.001}$ & 15.712$\pm{0.017}$ & 9.638$\pm 0.034$ & 14.525$\pm 0.119$ & 24.163$\pm 0.124$\\
2011ei & IIb & 41.969$\pm^{0.089}_{0.088}$ & 0.044$\pm^{0.010}_{0.008}$ & 17.732$\pm{0.033}$ & 10.247$\pm 0.033$ & 16.893$\pm 0.360$ & 27.140$\pm 0.362$\\
2011fu & IIb & 42.338$\pm^{0.008}_{0.012}$ & - & - & - & 18.016$\pm 0.283$ & -\\
2011hs & IIb & 41.903$\pm^{0.011}_{0.012}$ & 0.021$\pm^{0.001}_{0.001}$ & 8.588$\pm{0.057}$ & 7.718$\pm 0.023$ & 11.199$\pm 0.263$ & 18.917$\pm 0.264$\\
2011kl$^{\mathrm{a}}$ & GRB-SN & 43.324$\pm^{0.166}_{0.157}$ &-& 15.169$\pm{0.071}$ & 6.889$\pm 0.142$ & 12.712$\pm 0.355$ & 19.601$\pm 0.382$\\
2012ap & Ic-BL & 42.472$\pm^{0.162}_{0.158}$ & 0.109$\pm^{0.052}_{0.035}$ & 13.192$\pm{0.308}$ & 8.648$\pm 1.337$ & 15.546$\pm 0.174$ & 24.193$\pm 1.348$\\
2012bz & GRB-SN & 42.795$\pm^{0.029}_{0.032}$ & 0.240$\pm^{0.026}_{0.013}$ & 13.491$\pm{0.224}$ & 9.838$\pm 0.066$ & 17.110$\pm 0.177$ & 26.947$\pm 0.189$\\
2013cq & GRB-SN & 42.960$\pm^{0.050}_{0.100}$ & - & - & - & - & -\\
2013cu & IIb & 42.661$\pm^{0.007}_{0.006}$ & 0.125$\pm^{0.003}_{0.002}$ & 9.009$\pm{0.082}$ & 6.890$\pm 0.031$ & 12.516$\pm 0.163$ & 19.406$\pm 0.166$\\
2013df & IIb & 42.198$\pm^{0.034}_{0.037}$ & 0.091$\pm^{0.008}_{0.008}$ & 21.793$\pm{0.103}$ & 13.897$\pm 0.101$ & 11.727$\pm 0.310$ & 25.624$\pm 0.326$\\
2013dx & GRB-SN & 42.831$\pm^{0.027}_{0.032}$ & - & - & 11.697$\pm 1.768$ & 14.389$\pm 0.841$ & 26.086$\pm 1.958$\\
2013ge & Ibc & 42.132$\pm^{0.019}_{0.024}$ & 0.068$\pm^{0.009}_{0.009}$ & 18.878$\pm{1.697}$ & 11.928$\pm 0.395$ & 19.953$\pm 0.285$ & 31.881$\pm 0.487$\\
PTF09dh/2009dr* & Ic-BL &42.892$\pm^{0.035}_{0.03}$& - & - & - & 18.546$\pm 0.247$ & - \\
PTF10gvb* & Ic-BL & 42.775$\pm^{0.067}_{0.067}$ & - & - & 18.179$\pm 19.629$ & 14.344$\pm 0.483$ & 32.522$\pm 19.635$\\
PTF10inj* & Ib & 42.438$\pm^{0.157}_{0.143}$ & - & - & - & 36.035$\pm 0.490$ & -\\
PTF10qif* & Ib & 42.455$\pm^{0.123}_{0.118}$ & - & - & - & 13.452$\pm 1.764$ & -\\
PTF10vgv* & Ic & 42.493$\pm^{0.065}_{0.063}$ & 0.094$\pm^{0.029}_{0.022}$ & 10.347$\pm{1.687}$ & 6.538$\pm 0.354$ & 9.969$\pm 0.000$ & 16.507$\pm 0.354$\\
PTF11bli* & Ibc & 42.071$\pm^{0.018}_{0.018}$ & 0.065$\pm^{0.008}_{0.007}$ & 20.627$\pm{1.658}$ & 13.033$\pm 0.164$ & - & -\\
PTF11jgj* & Ic & 42.088$\pm^{0.022}_{0.018}$ & 0.072$\pm^{0.010}_{0.009}$ & 22.072$\pm{2.170}$ & 13.947$\pm 1.409$ & - & -\\
PTF11klg* & Ic & 42.105$\pm^{0.144}_{0.125}$ & 0.057$\pm^{0.030}_{0.018}$ & 16.393$\pm{1.678}$ & 10.358$\pm 0.303$ & 14.530$\pm 0.039$ & 24.887$\pm 0.306$\\
PTF11qiq* & Ib & 42.161$\pm^{0.500}_{0.010}$ & - & - & - & 17.000$\pm 1.000$ & -\\
PTF11rka* & Ic & 42.723$\pm^{0.050}_{0.045}$ & - & - & - & 45.781$\pm 3.909$ & -\\
\hline
\end{tabular}
\end{minipage}
\end{table*}
\begin{table*}
\renewcommand{\arraystretch}{1.5}
 \centering
 \begin{minipage}{160mm}
 \contcaption{\textit{BVRI} pseudobolometric light curve statistics.}
 \begin{tabular}{lccccccc}
  \hline

PTF12gzk & Ic & 42.708$\pm^{0.031}_{0.029}$ & 0.226$\pm^{0.023}_{0.020}$ & 16.345$\pm{0.455}$ & - & 23.643$\pm 0.065$ & -\\
PTF12os* & IIb & 41.409$\pm^{0.007}_{0.010}$ & - & - & - & 17.095$\pm 0.492$ & -\\
iPTF13bvn & Ib & 41.997$\pm^{0.012}_{0.011}$ & 0.043$\pm^{0.001}_{0.001}$ & 15.953$\pm{0.098}$ & 8.990$\pm 0.024$ & 13.582$\pm 0.098$ & 22.572$\pm 0.101$\\
iPTF14dby* & Ic-BL & 42.373$\pm^{0.043}_{0.042}$ & 0.140$\pm^{0.025}_{0.022}$ & 22.462$\pm{1.678}$ & 14.193$\pm 0.303$ & 20.615$\pm 0.884$ & 34.808$\pm 0.934$\\
 \hline
 \multicolumn{8}{p{\textwidth}}{* SN has not been corrected for host extinction. $L_{\mathrm{p}}$ and $M_{\mathrm{Ni}}$ values are lower limits}\\
  \multicolumn{8}{p{\textwidth}}{$^{\mathrm{a}}$SN 2011kl was primarily powered by a magnetar. The mass of $^{56}$Ni synthesised in the explosion was negligible \citep{Greiner2015}}
%\label{BVRIstats}
\end{tabular}
\end{minipage}
\end{table*}

\begin{table*}
\renewcommand{\arraystretch}{1.5}
 \centering
 \begin{minipage}{160mm}
  \caption{\textit{UBVRI}NIR pseudobolometric light curve statistics.}
 \begin{tabular}{lccccccc}
  \hline
  SN & Type & log$(L{\mathrm{p}})$ &$M_{\mathrm{Ni}}$ (M$_{\odot}$) &$t_{\mathrm{p}}$ (days) & $t_{-1/2}$ (days) & $t_{+1/2}$ (days) & Width (days)  \\
  \hline
1998bw & GRB-SN & 42.953$\pm^{0.042}_{0.026}$ & 0.386$\pm^{0.043}_{0.026}$ & 15.861$\pm{0.177}$ & 9.445$\pm 0.607$ & 14.639$\pm 0.894$ & 24.084$\pm 1.081$\\
1999dn & Ib & 42.360$\pm^{0.191}_{0.163}$ & 0.088$\pm^{0.073}_{0.038}$ & 13.916$\pm{2.841}$ & - & 19.790$\pm 4.210$ & -\\
2002ap & Ic-BL & 42.373$\pm^{0.018}_{0.008}$ & 0.086$\pm^{0.004}_{0.002}$ & 13.011$\pm{0.000}$ & 6.076$\pm 0.130$ & 13.629$\pm 0.362$ & 19.705$\pm 0.384$\\
2003dh & GRB-SN & 42.94$\pm^{0.08}_{0.09}$ & 0.308$\pm^{0.102}_{0.083}$ & 12.65$\pm{1.66}$ & - & 15.845$\pm 0.590$ & -\\
2005bf & Ib & 42.385$\pm^{0.033}_{0.032}$ & 0.119$\pm^{0.012}_{0.010}$ & 18.325$\pm{0.351}$ & - & - & -\\
2005hg* & Ib & 42.377$\pm^{0.029}_{0.027}$ & - & - & 12.215$\pm 0.051$ & 19.900$\pm 0.727$ & 32.115$\pm 0.728$\\
2005mf* & Ic & 42.255$\pm^{0.095}_{0.051}$ & - & - & - & 18.696$\pm 0.899$ & -\\
2006aj & GRB-SN & 42.839$\pm^{0.025}_{0.017}$ & 0.197$\pm^{0.012}_{0.008}$ & 9.587$\pm{0.040}$ & 6.227$\pm 0.526$ & 11.059$\pm 4.141$ & 17.285$\pm 4.175$\\
2007Y & Ib & 41.963$\pm^{0.065}_{0.099}$ & 0.046$\pm^{0.008}_{0.010}$ & 18.757$\pm{0.350}$ & 7.406$\pm 0.916$ & 11.507$\pm 0.916$ & 18.912$\pm 1.295$\\
2007gr & Ic & 42.258$\pm^{0.061}_{0.059}$ & 0.066$\pm^{0.011}_{0.009}$ & 13.146$\pm{0.229}$ & 7.951$\pm 0.051$ & 14.300$\pm 0.067$ & 22.251$\pm 0.084$\\
2007uy & Ib & 42.846$\pm^{0.360}_{0.318}$ & 0.358$\pm^{0.474}_{0.188}$ & 19.075$\pm{0.284}$ & - & 12.944$\pm 1.512$ & -\\
2008D & Ib & 42.292$\pm^{0.355}_{0.305}$ & 0.101$\pm^{0.130}_{0.051}$ & 19.293$\pm{0.229}$ & 13.697$\pm 0.106$ & 16.897$\pm 0.606$ & 30.594$\pm 0.615$\\
2008ax & IIb & 42.284$\pm^{0.037}_{0.036}$ & 0.099$\pm^{0.009}_{0.008}$ & 19.283$\pm{0.127}$ & 9.259$\pm 0.243$ & 13.262$\pm 0.526$ & 22.522$\pm 0.579$\\
2009iz* & Ib & 42.200$\pm^{0.028}_{0.024}$ & 0.091$\pm^{0.014}_{0.012}$ & 21.750$\pm{1.936}$ & 12.959$\pm 0.278$ & 25.824$\pm 0.893$ & 38.782$\pm 0.936$\\
2009jf & Ib & 42.640$\pm^{0.034}_{0.036}$ & 0.246$\pm^{0.022}_{0.021}$ & 21.267$\pm{0.164}$ & 10.633$\pm 0.652$ & 18.821$\pm 0.285$ & 29.453$\pm 0.711$\\
2011bm & Ic & 42.880$\pm^{0.058}_{0.049}$ & 0.638$\pm^{0.094}_{0.070}$ & 34.586$\pm{0.151}$ & - & 43.461$\pm 1.449$ & -\\
2011dh & IIb & 42.205$\pm^{0.019}_{0.016}$ & 0.068$\pm^{0.003}_{0.003}$ & 15.712$\pm{0.017}$ & 9.837$\pm 0.075$ & 13.346$\pm 0.045$ & 23.182$\pm 0.087$\\
2011hs & IIb & 42.027$\pm^{0.017}_{0.018}$ & 0.028$\pm^{0.001}_{0.001}$ & 8.588$\pm{0.057}$ & 8.303$\pm 0.000$ & 12.072$\pm 0.227$ & 20.375$\pm 0.227$\\
\hline
\multicolumn{8}{p{\textwidth}}{* SN has not been corrected for host extinction. $L_{\mathrm{p}}$ and $M_{\mathrm{Ni}}$ values are lower limits}
 \label{intONIRstats}
\end{tabular}
\end{minipage}
\end{table*}

\begin{table*}
\renewcommand{\arraystretch}{1.5}
 \centering
 \begin{minipage}{110mm}
  \caption{Parameters derived from the fully bolometric $L_{\mathrm{p}}$ values.}
 \begin{tabular}{lcccc}
  \hline
 SN & Type & log\,$(L{\mathrm{p}})$  & $t_{\mathrm{p}}$ (days) &  $M_{\mathrm{Ni}}$ (M$_{\odot}$)  \\
  \hline

1993J & IIb & 42.456$\pm^{0.109}_{0.088}$ &19.148$\pm{0.033}$ &0.146$\pm^{0.042}_{0.027}$ \\
1994I & Ic & 42.473$\pm^{0.409}_{0.363}$ &12.252$\pm{0.211}$ &0.102$\pm^{0.164}_{0.059}$ \\
1996cb & IIb & 42.082$\pm^{0.179}_{0.163}$ &16.993$\pm{1.650}$ &0.055$\pm^{0.036}_{0.021}$ \\
1998bw & GRB-SN & 43.061$\pm^{0.080}_{0.046}$ &15.861$\pm{0.177}$ &0.496$\pm^{0.091}_{0.047}$ \\
1999dn & Ib & 42.402$\pm^{0.229}_{0.183}$ &13.916$\pm{2.841}$ &0.097$\pm^{0.097}_{0.044}$ \\
1999ex & Ib & 42.544$\pm^{0.124}_{0.105}$ &18.349$\pm{0.036}$ &0.172$\pm^{0.058}_{0.037}$ \\
2002ap & Ic-BL & 42.415$\pm^{0.055}_{0.028}$ &13.011$\pm{0.000}$ &0.094$\pm^{0.013}_{0.006}$ \\
2003bg & IIb & 42.424$\pm^{0.086}_{0.058}$ &16.624$\pm{1.656}$ &0.119$\pm^{0.040}_{0.024}$ \\
2003dh & GRB-SN & 42.981$\pm^{0.116}_{0.107}$ &12.65$\pm{1.66}$ &0.339$\pm^{0.153}_{0.102}$ \\
2003jd & Ic-BL & 42.988$\pm^{0.134}_{0.115}$ &12.504$\pm{0.179}$ &0.341$\pm^{0.128}_{0.082}$ \\
2004aw & Ic & 42.659$\pm^{0.063}_{0.056}$ &- &- \\
2004fe* & Ic & 42.354$\pm^{0.099}_{0.083}$ &14.680$\pm{1.653}$ &0.091$\pm^{0.035}_{0.023}$ \\
2004gq & Ib & 42.388$\pm^{0.046}_{0.028}$ &- &- \\
2005az* & Ic & 42.173$\pm^{0.086}_{0.070}$ &- &- \\
2005bf & Ib & 42.426$\pm^{0.071}_{0.052}$ &18.325$\pm{0.351}$ &0.131$\pm^{0.026}_{0.017}$ \\
2005hg* & Ib & 42.418$\pm^{0.067}_{0.047}$ &- &- \\
2005hl* & Ib & 42.167$\pm^{0.060}_{0.027}$ &- &- \\
2005hm* & Ib & 42.237$\pm^{0.073}_{0.093}$ &19.927$\pm{1.669}$ &0.092$\pm^{0.025}_{0.023}$ \\
2005kl* & Ic & 41.979$\pm^{0.080}_{0.064}$ &- &- \\
2005kr* & Ic-BL & 43.006$\pm^{0.117}_{0.057}$ &11.405$\pm{1.669}$ &0.330$\pm^{0.153}_{0.073}$ \\
2005ks* & Ic-BL & 42.440$\pm^{0.107}_{0.087}$ &12.368$\pm{1.656}$ &0.096$\pm^{0.040}_{0.026}$ \\
2005kz* & Ic & 42.212$\pm^{0.101}_{0.084}$ &- &- \\
2005mf* & Ic & 42.296$\pm^{0.133}_{0.071}$ &- &- \\
2006T* & IIb & 42.322$\pm^{0.054}_{0.036}$ &14.148$\pm{1.651}$ &0.082$\pm^{0.020}_{0.014}$ \\
2006aj & GRB-SN & 42.880$\pm^{0.062}_{0.038}$ &9.587$\pm{0.040}$ &0.217$\pm^{0.034}_{0.019}$ \\
2006el* & IIb & 42.237$\pm^{0.086}_{0.063}$ &- &- \\
2006ep* & Ib & 42.142$\pm^{0.075}_{0.060}$ &18.002$\pm{1.653}$ &0.067$\pm^{0.019}_{0.014}$ \\
2006fe* & Ic & 42.533$\pm^{0.094}_{0.081}$ &15.830$\pm{1.711}$ &0.147$\pm^{0.053}_{0.037}$ \\
2006fo* & Ib & 42.372$\pm^{0.099}_{0.080}$ &- &- \\
14475* & Ic-BL & 42.753$\pm^{0.108}_{0.092}$ &10.501$\pm{1.760}$ &0.173$\pm^{0.077}_{0.050}$ \\
2006jo* & Ib & 42.542$\pm^{0.086}_{0.052}$ &10.366$\pm{1.651}$ &0.105$\pm^{0.039}_{0.023}$ \\
2006lc* & Ib & 42.130$\pm^{0.044}_{0.030}$ &13.837$\pm{1.651}$ &0.051$\pm^{0.011}_{0.008}$ \\
2006nx* & Ic-BL & 43.032$\pm^{0.111}_{0.066}$ &14.279$\pm{1.698}$ &0.422$\pm^{0.181}_{0.097}$ \\
2007C & Ib & 42.749$\pm^{0.155}_{0.136}$ &- &- \\
2007D* & Ic-BL & 42.282$\pm^{0.156}_{0.135}$ &- &- \\
2007Y & Ib & 42.005$\pm^{0.103}_{0.119}$ &18.757$\pm{0.350}$ &0.051$\pm^{0.015}_{0.013}$ \\
2007ag* & Ib & 42.085$\pm^{0.078}_{0.061}$ &- &- \\
2007cl* & Ic & 42.282$\pm^{0.078}_{0.064}$ &- &- \\
2007gr & Ic & 42.299$\pm^{0.099}_{0.079}$ &13.146$\pm{0.229}$ &0.073$\pm^{0.020}_{0.013}$ \\
2007kj* & Ib & 42.214$\pm^{0.072}_{0.055}$ &- &- \\
2007ms* & Ic & 42.556$\pm^{0.052}_{0.041}$ &22.231$\pm{1.666}$ &0.211$\pm^{0.043}_{0.032}$ \\
\hline
 \label{fullstats}
\end{tabular}
\end{minipage}
\end{table*}
\begin{table*}
\renewcommand{\arraystretch}{1.5}
 \centering
 \contcaption{Parameters derived from the fully bolometric $L_{\mathrm{p}}$ values.}
 \begin{minipage}{110mm}
 \begin{tabular}{lcccc}
  \hline
 SN & Type & log$(L{\mathrm{p}})$  & $t_{\mathrm{p}}$ (days) & $M_{\mathrm{Ni}}$ (M$_{\odot}$)   \\
  \hline
2007nc* & Ib & 42.260$\pm^{0.125}_{0.063}$ &17.734$\pm{1.800}$ &0.087$\pm^{0.040}_{0.019}$ \\  
2007qv* & Ic & 42.783$\pm^{0.062}_{0.043}$ &- &- \\  
2007qx* & Ic & 42.337$\pm^{0.207}_{0.069}$ &15.094$\pm{2.758}$ &0.090$\pm^{0.078}_{0.025}$ \\
2007ru & Ic-BL & 43.033$\pm^{0.054}_{0.032}$ &10.242$\pm{0.047}$ &0.323$\pm^{0.044}_{0.024}$ \\
2007sj* & Ic & 42.152$\pm^{0.074}_{0.052}$ &14.542$\pm{1.658}$ &0.057$\pm^{0.017}_{0.011}$ \\
2007uy & Ib & 42.887$\pm^{0.398}_{0.338}$ &19.075$\pm{0.284}$ &0.394$\pm^{0.604}_{0.215}$ \\
2008D & Ib & 42.333$\pm^{0.393}_{0.325}$ &19.293$\pm{0.229}$ &0.111$\pm^{0.167}_{0.059}$ \\
2008ax & IIb & 42.326$\pm^{0.075}_{0.056}$ &19.283$\pm{0.127}$ &0.109$\pm^{0.021}_{0.014}$ \\
2008bo & IIb & 41.972$\pm^{0.071}_{0.055}$ &- &- \\
2008hw & GRB-SN & 43.337$\pm^{0.095}_{0.078}$ &12.307$\pm{0.100}$ &0.751$\pm^{0.190}_{0.127}$ \\
2009bb & Ic & 42.863$\pm^{0.162}_{0.143}$ &12.631$\pm{0.095}$ &0.258$\pm^{0.119}_{0.073}$ \\
2009er* & Ib & 42.916$\pm^{0.089}_{0.072}$ &- &- \\
2009iz* & Ib & 42.241$\pm^{0.066}_{0.044}$ &21.828$\pm{1.650}$ &0.101$\pm^{0.024}_{0.016}$ \\
2009jf & Ib & 42.682$\pm^{0.072}_{0.057}$ &21.267$\pm{0.164}$ &0.271$\pm^{0.051}_{0.035}$ \\
2010as & IIb & 42.689$\pm^{0.158}_{0.139}$ &12.442$\pm{0.122}$ &0.171$\pm^{0.077}_{0.048}$ \\
2010bh & GRB-SN & 42.600$\pm^{0.168}_{0.113}$ &12.737$\pm{0.099}$ &0.142$\pm^{0.068}_{0.033}$ \\
2010ma & GRB-SN & 43.244$\pm^{0.272}_{0.098}$ &10.331$\pm{4.338}$ &0.529$\pm^{0.789}_{0.227}$ \\
2011bm & Ic & 42.922$\pm^{0.096}_{0.069}$ &34.586$\pm{0.151}$ &0.702$\pm^{0.176}_{0.106}$ \\
2011dh & IIb & 42.246$\pm^{0.057}_{0.036}$ &15.712$\pm{0.017}$ &0.075$\pm^{0.011}_{0.006}$ \\
2011ei & IIb & 42.177$\pm^{0.146}_{0.129}$ &17.732$\pm{0.033}$ &0.072$\pm^{0.029}_{0.018}$ \\
2011fu & IIb & 42.538$\pm^{0.045}_{0.034}$ &- &- \\
2011hs & IIb & 42.068$\pm^{0.055}_{0.038}$ &8.588$\pm{0.057}$ &0.031$\pm^{0.004}_{0.003}$ \\
2011kl$^{\mathrm{a}}$ & GRB-SN & 43.529$\pm^{0.174}_{0.148}$ &15.169$\pm{0.071}$ &- \\
2012ap & Ic-BL & 42.676$\pm^{0.203}_{0.182}$ &13.192$\pm{0.308}$ &0.174$\pm^{0.109}_{0.062}$ \\
2012bz & GRB-SN & 43.005$\pm^{0.100}_{0.052}$ &13.491$\pm{0.224}$ &0.378$\pm^{0.105}_{0.048}$ \\
2013cq & GRB-SN & 43.186$\pm^{0.086}_{0.117}$ &13.000$\pm{2.000}$ &0.555$\pm^{0.211}_{0.184}$ \\
2013cu & IIb & 42.843$\pm^{0.067}_{0.050}$ &9.009$\pm{0.082}$ &0.191$\pm^{0.033}_{0.022}$ \\
2013df & IIb & 42.397$\pm^{0.074}_{0.058}$ &21.793$\pm{0.103}$ &0.144$\pm^{0.028}_{0.019}$ \\
2013dx & GRB-SN & 42.962$\pm^{0.225}_{0.077}$ &12.261$\pm{5.484}$ &0.316$\pm^{0.417}_{0.139}$ \\
2013ge & Ibc & 42.335$\pm^{0.077}_{0.064}$ &18.878$\pm{1.697}$ &0.109$\pm^{0.032}_{0.023}$ \\
PTF09dh/2009dr* & Ic-BL & 43.082$\pm^{0.078}_{0.059}$&-&-\\
PTF10inj* & Ib & 42.620$\pm^{0.215}_{0.186}$ &- &- \\
PTF10qif* & Ib & 42.636$\pm^{0.182}_{0.162}$ &- &- \\
PTF10vgv* & Ic & 42.682$\pm^{0.132}_{0.115}$ &10.347$\pm{1.687}$ &0.145$\pm^{0.076}_{0.047}$ \\
PTF11bli* & Ibc & 42.266$\pm^{0.081}_{0.065}$ &20.627$\pm{1.658}$ &0.101$\pm^{0.030}_{0.021}$ \\
PTF11jgj* & Ic & 42.293$\pm^{0.090}_{0.072}$ &22.072$\pm{2.170}$ &0.114$\pm^{0.039}_{0.026}$ \\
PTF11klg* & Ic & 42.309$\pm^{0.208}_{0.175}$ &16.393$\pm{1.678}$ &0.090$\pm^{0.069}_{0.036}$ \\
PTF11qiq* & Ib & 42.352$\pm^{0.545}_{0.057}$ &- &- \\
PTF11rka* & Ic & 42.905$\pm^{0.117}_{0.097}$ &- &- \\
\hline
\end{tabular}
\end{minipage}
\end{table*}
\begin{table*}
\renewcommand{\arraystretch}{1.5}
 \centering
 \contcaption{Parameters derived from the fully bolometric $L_{\mathrm{p}}$ values.}
 \begin{minipage}{110mm}
 \begin{tabular}{lcccc}
  \hline
 SN & Type & log$(L{\mathrm{p}})$  & $t_{\mathrm{p}}$ (days) & $M_{\mathrm{Ni}}$ (M$_{\odot}$)   \\
  \hline
PTF12gzk & Ic & 42.904$\pm^{0.077}_{0.058}$ &16.345$\pm{0.455}$ &0.355$\pm^{0.079}_{0.052}$ \\
PTF12os* & IIb & 41.588$\pm^{0.047}_{0.030}$ &- &- \\
iPTF13bvn & Ib & 42.211$\pm^{0.067}_{0.037}$ &15.953$\pm{0.098}$ &0.070$\pm^{0.012}_{0.006}$ \\
iPTF14dby* & Ic-BL & 42.575$\pm^{0.104}_{0.087}$ &22.462$\pm{1.678}$ &0.223$\pm^{0.079}_{0.053}$ \\
 \hline
 \multicolumn{5}{p{\textwidth}}{* SN has not been corrected for host extinction. $L_{\mathrm{p}}$ and $M_{\mathrm{Ni}}$ values are lower limits}\\
 \multicolumn{5}{p{\textwidth}}{$^{\mathrm{a}}$SN 2011kl was primarily powered by a magnetar. The mass of $^{56}$Ni synthesised in the explosion was negligible \citep{Greiner2015}}
\end{tabular}
\end{minipage}
\end{table*}

\begin{table*}
\renewcommand{\arraystretch}{1.5}
 \centering
 \begin{minipage}{140mm}
  \caption{$M_{\mathrm{ej}}^3/E_{\mathrm{k}}$ values for the sample at different opacities.}
 \begin{tabular}{@{}llcccccclcc@{}}
  \hline
 SN & Type & $\frac{M_{\mathrm{ej}}^3}{[{\rm M}_\odot]}/ \frac{E_{\mathrm{k}}}{[10^{51}\mathrm{ erg}]}$  & $\frac{M_{\mathrm{ej}}^3}{[{\rm M}_\odot]}/ \frac{E_{\mathrm{k}}}{[10^{51}\mathrm{ erg}]}$ & $\frac{M_{\mathrm{ej}}^3}{[{\rm M}_\odot]}/ \frac{E_{\mathrm{k}}}{[10^{51}\mathrm{ erg}]}$ &  \\
 & & $\kappa = 0.05$ g cm$^{-2}$ & $\kappa = 0.07$ g cm$^{-2}$ & $\kappa = 0.1$ g cm$^{-2}$ \\
  \hline
14475 & Ic-BL & 0.567$ \pm^{0.487}_{0.295}$ &0.289$ \pm^{0.248}_{0.150}$ &0.142$ \pm^{0.122}_{0.074}$ \\
1993J & IIb & 6.267$ \pm^{0.044}_{0.043}$ &3.198$ \pm^{0.022}_{0.022}$ &1.567$ \pm^{0.011}_{0.011}$ \\
1994I & Ic & 1.051$ \pm^{0.074}_{0.071}$ &0.536$ \pm^{0.038}_{0.036}$ &0.263$ \pm^{0.019}_{0.018}$ \\
1996cb & IIb & 3.888$ \pm^{1.745}_{1.304}$ &1.984$ \pm^{0.890}_{0.665}$ &0.972$ \pm^{0.436}_{0.326}$ \\
1998bw & GRB-SN & 2.951$ \pm^{0.134}_{0.130}$ &1.506$ \pm^{0.068}_{0.066}$ &0.738$ \pm^{0.033}_{0.032}$ \\
1999dn & Ib & 1.749$ \pm^{1.928}_{1.047}$ &0.892$ \pm^{0.984}_{0.534}$ &0.437$ \pm^{0.482}_{0.262}$ \\
1999ex & Ib & 5.286$ \pm^{0.041}_{0.041}$ &2.697$ \pm^{0.021}_{0.021}$ &1.321$ \pm^{0.010}_{0.010}$ \\
2002ap & Ic-BL & 1.336$ \pm^{0.000}_{0.000}$ &0.682$ \pm^{0.000}_{0.000}$ &0.334$ \pm^{0.000}_{0.000}$ \\
2003bg & IIb & 3.561$ \pm^{1.646}_{1.221}$ &1.817$ \pm^{0.840}_{0.623}$ &0.890$ \pm^{0.411}_{0.305}$ \\
2003jd & Ic-BL & 1.140$ \pm^{0.067}_{0.064}$ &0.582$ \pm^{0.034}_{0.033}$ &0.285$ \pm^{0.017}_{0.016}$ \\
2004fe & Ic & 2.165$ \pm^{1.153}_{0.823}$ &1.105$ \pm^{0.588}_{0.420}$ &0.541$ \pm^{0.288}_{0.206}$ \\
2005bf & Ib & 5.257$ \pm^{0.415}_{0.392}$ &2.682$ \pm^{0.212}_{0.200}$ &1.314$ \pm^{0.104}_{0.098}$ \\
2005hm & Ib & 7.352$ \pm^{2.790}_{2.170}$ &3.751$ \pm^{1.423}_{1.107}$ &1.838$ \pm^{0.697}_{0.543}$ \\
2005kr & Ic-BL & 0.789$ \pm^{0.573}_{0.370}$ &0.403$ \pm^{0.293}_{0.189}$ &0.197$ \pm^{0.143}_{0.092}$ \\
2005ks & Ic-BL & 1.091$ \pm^{0.713}_{0.477}$ &0.557$ \pm^{0.364}_{0.243}$ &0.273$ \pm^{0.178}_{0.119}$ \\
2006T & IIb & 1.868$ \pm^{1.037}_{0.731}$ &0.953$ \pm^{0.529}_{0.373}$ &0.467$ \pm^{0.259}_{0.183}$ \\
2006aj & GRB-SN & 0.394$ \pm^{0.007}_{0.007}$ &0.201$ \pm^{0.003}_{0.003}$ &0.098$ \pm^{0.002}_{0.002}$ \\
2006ep & Ib & 4.896$ \pm^{2.062}_{1.566}$ &2.498$ \pm^{1.052}_{0.799}$ &1.224$ \pm^{0.515}_{0.391}$ \\
2006fe & Ic & 2.928$ \pm^{1.487}_{1.075}$ &1.494$ \pm^{0.759}_{0.549}$ &0.732$ \pm^{0.372}_{0.269}$ \\
2006jo & Ib & 0.538$ \pm^{0.434}_{0.269}$ &0.275$ \pm^{0.221}_{0.137}$ &0.135$ \pm^{0.109}_{0.067}$ \\
2006lc & Ib & 1.709$ \pm^{0.974}_{0.681}$ &0.872$ \pm^{0.497}_{0.347}$ &0.427$ \pm^{0.243}_{0.170}$ \\
2006nx & Ic-BL & 1.938$ \pm^{1.100}_{0.770}$ &0.989$ \pm^{0.561}_{0.393}$ &0.485$ \pm^{0.275}_{0.193}$ \\
2007Y & Ib & 5.771$ \pm^{0.443}_{0.419}$ &2.945$ \pm^{0.226}_{0.214}$ &1.443$ \pm^{0.111}_{0.105}$ \\
2007gr & Ic & 1.393$ \pm^{0.100}_{0.094}$ &0.711$ \pm^{0.051}_{0.048}$ &0.348$ \pm^{0.025}_{0.024}$ \\
2007ms & Ic & 11.388$ \pm^{3.817}_{3.049}$ &5.810$ \pm^{1.948}_{1.556}$ &2.847$ \pm^{0.954}_{0.762}$ \\
2007nc & Ib & 4.611$ \pm^{2.176}_{1.606}$ &2.353$ \pm^{1.110}_{0.819}$ &1.153$ \pm^{0.544}_{0.401}$ \\
2007qx & Ic & 2.420$ \pm^{2.316}_{1.341}$ &1.235$ \pm^{1.182}_{0.684}$ &0.605$ \pm^{0.579}_{0.335}$ \\
2007ru & Ic-BL & 0.513$ \pm^{0.010}_{0.009}$ &0.262$ \pm^{0.005}_{0.005}$ &0.128$ \pm^{0.002}_{0.002}$ \\
2007sj & Ic & 2.085$ \pm^{1.126}_{0.800}$ &1.064$ \pm^{0.575}_{0.408}$ &0.521$ \pm^{0.282}_{0.200}$ \\
2007uy & Ib & 6.173$ \pm^{0.377}_{0.360}$ &3.149$ \pm^{0.192}_{0.184}$ &1.543$ \pm^{0.094}_{0.090}$ \\
2008D & Ib & 6.460$ \pm^{0.312}_{0.301}$ &3.296$ \pm^{0.159}_{0.154}$ &1.615$ \pm^{0.078}_{0.075}$ \\
2008ax & IIb & 6.447$ \pm^{0.172}_{0.169}$ &3.289$ \pm^{0.088}_{0.086}$ &1.612$ \pm^{0.043}_{0.042}$ \\
2008hw & GRB-SN & 1.070$ \pm^{0.035}_{0.034}$ &0.546$ \pm^{0.018}_{0.018}$ &0.267$ \pm^{0.009}_{0.009}$ \\
2009bb & Ic & 1.187$ \pm^{0.036}_{0.035}$ &0.606$ \pm^{0.018}_{0.018}$ &0.297$ \pm^{0.009}_{0.009}$ \\
2009iz & Ib & 10.585$ \pm^{3.582}_{2.856}$ &5.401$ \pm^{1.828}_{1.457}$ &2.646$ \pm^{0.896}_{0.714}$ \\
2009jf & Ib & 9.537$ \pm^{0.297}_{0.290}$ &4.866$ \pm^{0.151}_{0.148}$ &2.384$ \pm^{0.074}_{0.072}$ \\
2010as & IIb & 1.117$ \pm^{0.044}_{0.043}$ &0.570$ \pm^{0.023}_{0.022}$ &0.279$ \pm^{0.011}_{0.011}$ \\
2010bh & GRB-SN & 1.227$ \pm^{0.039}_{0.038}$ &0.626$ \pm^{0.020}_{0.019}$ &0.307$ \pm^{0.010}_{0.009}$ \\

\hline

 \label{MoE}
\end{tabular}
\end{minipage}
\end{table*}
\begin{table*}
\renewcommand{\arraystretch}{1.5}
 \centering
 \begin{minipage}{140mm}
 \contcaption{$M_{\mathrm{ej}}^3/E_{\mathrm{k}}$ values for the sample at different opacities.}
 \begin{tabular}{@{}llcccccclcc@{}}
  \hline
  SN & Type & $\frac{M_{\mathrm{ej}}^3}{[{\rm M}_\odot]}/ \frac{E_{\mathrm{k}}}{[10^{51}\mathrm{ erg}]}$  & $\frac{M_{\mathrm{ej}}^3}{[{\rm M}_\odot]}/ \frac{E_{\mathrm{k}}}{[10^{51}\mathrm{ erg}]}$ & $\frac{M_{\mathrm{ej}}^3}{[{\rm M}_\odot]}/ \frac{E_{\mathrm{k}}}{[10^{51}\mathrm{ erg}]}$ &  \\
 & & $\kappa = 0.05$ g cm$^{-2}$ & $\kappa = 0.07$ g cm$^{-2}$ & $\kappa = 0.1$ g cm$^{-2}$ \\
  \hline
2010ma & GRB-SN & 0.531$ \pm^{1.628}_{0.471}$ &0.271$ \pm^{0.830}_{0.240}$ &0.133$ \pm^{0.407}_{0.118}$ \\
2011bm & Ic & 66.714$ \pm^{1.170}_{1.154}$ &34.038$ \pm^{0.597}_{0.589}$ &16.678$ \pm^{0.292}_{0.289}$ \\
2011dh & IIb & 2.842$ \pm^{0.012}_{0.012}$ &1.450$ \pm^{0.006}_{0.006}$ &0.710$ \pm^{0.003}_{0.003}$ \\
2011ei & IIb & 4.609$ \pm^{0.034}_{0.034}$ &2.352$ \pm^{0.017}_{0.017}$ &1.152$ \pm^{0.009}_{0.008}$ \\
2011hs & IIb & 0.254$ \pm^{0.007}_{0.007}$ &0.129$ \pm^{0.003}_{0.003}$ &0.063$ \pm^{0.002}_{0.002}$ \\
2011kl & GRB-SN & 2.468$ \pm^{0.047}_{0.046}$ &1.259$ \pm^{0.024}_{0.023}$ &0.617$ \pm^{0.012}_{0.011}$ \\
2012ap & Ic-BL & 1.412$ \pm^{0.137}_{0.127}$ &0.720$ \pm^{0.070}_{0.065}$ &0.353$ \pm^{0.034}_{0.032}$ \\
2012bz & GRB-SN & 1.544$ \pm^{0.105}_{0.100}$ &0.788$ \pm^{0.054}_{0.051}$ &0.386$ \pm^{0.026}_{0.025}$ \\
2013cq & GRB-SN & 1.332$ \pm^{1.029}_{0.649}$ &0.679$ \pm^{0.525}_{0.331}$ &0.333$ \pm^{0.257}_{0.162}$ \\
2013cu & IIb & 0.307$ \pm^{0.011}_{0.011}$ &0.157$ \pm^{0.006}_{0.006}$ &0.077$ \pm^{0.003}_{0.003}$ \\
2013df & IIb & 10.517$ \pm^{0.201}_{0.198}$ &5.366$ \pm^{0.103}_{0.101}$ &2.629$ \pm^{0.050}_{0.050}$ \\
2013dx & GRB-SN & 1.054$ \pm^{3.570}_{0.955}$ &0.538$ \pm^{1.821}_{0.487}$ &0.263$ \pm^{0.892}_{0.239}$ \\
2013ge & Ibc & 5.921$ \pm^{2.433}_{1.858}$ &3.021$ \pm^{1.241}_{0.948}$ &1.480$ \pm^{0.608}_{0.465}$ \\
PTF10vgv & Ic & 0.534$ \pm^{0.443}_{0.272}$ &0.273$ \pm^{0.226}_{0.139}$ &0.134$ \pm^{0.111}_{0.068}$ \\
PTF11bli & Ibc & 8.440$ \pm^{3.059}_{2.404}$ &4.306$ \pm^{1.561}_{1.226}$ &2.110$ \pm^{0.765}_{0.601}$ \\
PTF11jgj & Ic & 11.066$ \pm^{5.037}_{3.751}$ &5.646$ \pm^{2.570}_{1.914}$ &2.767$ \pm^{1.259}_{0.938}$ \\
PTF11klg & Ic & 3.367$ \pm^{1.605}_{1.181}$ &1.718$ \pm^{0.819}_{0.602}$ &0.842$ \pm^{0.401}_{0.295}$ \\
PTF12gzk & Ic & 3.328$ \pm^{0.387}_{0.356}$ &1.698$ \pm^{0.197}_{0.181}$ &0.832$ \pm^{0.097}_{0.089}$ \\
iPTF13bvn & Ib & 3.020$ \pm^{0.075}_{0.073}$ &1.541$ \pm^{0.038}_{0.037}$ &0.755$ \pm^{0.019}_{0.018}$ \\
iPTF14dby & Ic-BL & 11.869$ \pm^{3.963}_{3.168}$ &6.056$ \pm^{2.022}_{1.616}$ &2.967$ \pm^{0.991}_{0.792}$ \\

 \hline
\end{tabular}
\end{minipage}
\end{table*}

% Don't change these lines
\bsp	% typesetting comment
\label{lastpage}
\end{document}